\documentclass[11pt]{article}
\usepackage{jheppub}
\usepackage[utf8]{inputenc}\usepackage[T1]{fontenc}
\usepackage{amsmath}
\usepackage{amssymb}
\usepackage{graphicx}
\usepackage{slashed}
\usepackage{cases}
\usepackage[toc,page]{appendix}
\usepackage{empheq}
\usepackage[normalem]{ulem} 
\usepackage{xcolor} 
\usepackage{afterpage}
\usepackage{float}
\usepackage{bbold}
\usepackage{hypbmsec}
\usepackage{gensymb}

\allowdisplaybreaks
\usepackage[export]{adjustbox}

\makeatletter
\newcommand\fake@math{}
\def\fake@math#1\){[math]}
\makeatother

\preprint{MIT-CTP/5520}

\begin{document}

\title{Gravity-improved metastability bounds for the Type-I Seesaw Mechanism}

\author[a,b]{Garv Chauhan,}
\author[c,d]{Thomas Steingasser}
\affiliation[a]{Center for Neutrino Physics, Department of Physics, Virginia Tech, Blacksburg, VA 24061, USA}
\affiliation[b]{Centre for Cosmology, Particle Physics and Phenomenology (CP3), 
Universit\'{e} catholique de Louvain, Chemin du Cyclotron 2, 
B-1348 Louvain-la-Neuve, Belgium}
\affiliation[c]{Department of Physics, Massachusetts Institute of Technology, Cambridge, MA 02139, USA}
\affiliation[d]{Black Hole Initiative at Harvard University, 20 Garden Street, Cambridge, MA 02138, USA}
\emailAdd{gchauhan@vt.edu}\emailAdd{tstngssr@mit.edu}


\abstract{ 
Right-handed neutrinos (RHN) destabilize the electroweak vacuum by increasing its decay rate. In the SM, the latter is dominated by physics at the RG scale at which $\lambda$ reaches its minimum, $\mu_*^{\text{SM}} \sim 10^{17}$~GeV. For large neutrino Yukawa coupling $Y_\nu$,  RHNs can push $\mu_*$ beyond the Planck scale, implying that gravitational effects need to be taken into account. In this work, we perform the first comprehensive study of electroweak vacuum metastability in the type-I seesaw mechanism including these effects. Our analysis covers both low- and high-scale seesaw models, with two as well as three RHNs and for multiple values of the Higgs' non-minimal coupling to gravity. We find that gravitational effects can significantly stabilize the vacuum, leading to weaker metastability bounds. We show that metastability sets the strongest bounds for low-scale seesaws with $M_N>1$ TeV. For high-scale seesaws, we find upper bounds on the allowed masses for the RHNs, which are relevant for high-scale leptogenesis. We also point out that $\text{Tr}(Y_\nu^\dagger Y_\nu)$, which is commonly used to express these metastability bounds, cannot be used for all of parameter space. Instead, we argue that bounds can always be expressed reliably through $\text{Tr}(Y_\nu^\dagger Y_\nu\,Y_\nu^\dagger Y_\nu)$. Lastly, we use this insight to develop a new technique for an easier RG analysis applicable to scenarios with degenerate RHN masses.
}


\maketitle

\section{Introduction}
The absence of neutrino masses remains one of the biggest shortcomings of the Standard Model (SM). The arguably simplest solution to this problem is given by the \textit{Type-I Seesaw Mechanism}, in which the SM-neutrinos are coupled to \textit{right-handed neutrinos} (RHNs) $N_i$~\cite{Minkowski:1977sc, Mohapatra:1979ia, Yanagida:1979as, GellMann:1980vs, Glashow:1979nm, Schechter:1980gr}. Given that their masses could be as large as~$10^{15}$~GeV, it is not surprising that they have so far evaded both direct and indirect experimental detection. Thus, there has been growing interest in strategies to constrain their parameter space through theoretical arguments.

One of the most successful attempts to do so relies on the stability of the electroweak vacuum. As we will review in Sec.~\ref{EW vacuum decay}, current measurements suggest that the latter is not a global minimum of the Higgs potential, but can instead decay into lower-lying states through bubble nucleation. The rate of such nucleation events is increased through the effects of additional fermions with a Yukawa-coupling to the Higgs, such as RHNs~\cite{Grzadkowski:1987tf,Casas:1999tg,Antusch:2002rr,Pirogov:1998tj}. Meanwhile our ongoing existence provides an upper bound on the decay rate of our vacuum, which can thus be translated to bounds on the RHNs' parameter space.

This strategy and its consequences have been explored extensively in the literature, covering the type-I seesaw with degenerate RHN masses,~\cite{Rodejohann:2012px,Lindner:2015qva,Bambhaniya:2016rbb,Mandal:2019ndp}, non-degenerate RHN masses (restricted to two heavy generations)~\cite{Ipek:2018sai}, the linear~\cite{Khan:2012zw} and inverse seesaw mechanism~\cite{DelleRose:2015bms} as well as type-III~\cite{Lindner:2015qva}. Restricting ourselves to the type-I model for simplicity and concreteness, we revisit these computations with a particular focus on an important, yet previously neglected aspect of it: The Higgs' coupling to gravity. 

On the first look, it might appear counter-intuitive that the parameter space of RHNs, whose masses are usually well below the Planck scale, could be so sensitive to gravitational effects. This feature can be understood by realizing that, at high energies, the Higgs potential's only scale dependence is through the RG-running of the quartic coupling~$\lambda$. This seems to cause the tunneling process to be determined by the value of $\lambda$ at its minimum~\cite{Andreassen:2017rzq}. Assuming only SM particle content, this value lies at $\mu_*\sim 10^{17}$~GeV. However, introducing additional fermions, and in particular RHNs, drives this scale to even larger values, in many cases beyond the Planck scale. Thus, it is evident that gravitational effects need to be taken into account.

In the context of vacuum decay, they manifest to leading order through their influence on the instanton mediating the decay, which has been thoroughly investigated in the literature~\cite{Salvio:2016mvj,Espinosa:2020qtq,Khoury:2021zao}. One of the most important consequences of the Higgs' interaction with gravity is that it gives rise to corrections to the Euclidean action, which stabilize the vacuum. 

The strength of this stabilization depends not only on the RG-trajectory of the quartic coupling, but also on the Higgs' \textit{non-minimal coupling} to gravity, characterized by the coupling constant $\xi$. It is well-established that the most general theory for the Higgs on a curved background should not only include its minimal coupling to gravity, but also an interaction term $\propto \xi R h^2$, where $R$ is curvature scalar and $h$ is the Higgs field. Such a term is not only allowed by the theory's symmetries, but would in any case be generated radiatively~\cite{Callan:1970ze,Bunch:1980bs,Birrell:1982ix,Odintsov:1990mt,Buchbinder:1992rb,Parker:2009uva,Markkanen:2013nwa,Kaiser:2015usz,Markkanen:2018bfx}. As the gravitational correction to the decay rate is controlled by the combination $(1+6 \xi)^2$, we find that already relatively small values of $\xi$ notably enhance the stabilizing effect of gravity, and even more so for larger values. 

As an additional result, we provide a systematic analysis of the decay rate's dependence on various combinations of the neutrinos' Yukawa couplings. We find in particular that it is most sensitive to the combination $\text{Tr}( Y_\nu^\dagger Y_\nu Y_\nu^\dagger Y_\nu )$ rather than $\text{Tr}( Y_\nu^\dagger Y_\nu)$, which is often used in the literature. While these give equivalent results for all previously considered cases, we find significant deviations for others. 

We perform all our stability computations at one-loop, using the consistent scheme developed in~\cite{Andreassen:2016cvx,Andreassen:2017rzq,Khoury:2021zao,Steingasser:2022yqx}. Given the very high scales we are interested in, we run all couplings using their three-loop (two-loop) beta functions for the SM (neutrino) contributions. At the RHN mass scales we also take into account all relevant threshold corrections, as recently derived in~\cite{Zhang:2021jdf}.

The rest of this article is organized as follows. In Sec.~\ref{EW vacuum decay}, we discuss the consistent calculation of vacuum decay probabilities at NLO including gravitational effects. In Sec.~\ref{sec:seesawmech}, we review the most important features of the type-I seesaw mechanism and discuss the effects of adding RHNs on the stability of the electroweak vacuum. In Sec.~\ref{sec:lowscale}, we present our results for the metastability bounds in the case of a low-scale seesaw. We find that the parameters used by previous works become unreliable at intermediate scales, for which we provide a suitable set of parameters. We furthermore show how this parametrization can be used to construct an approximate RG running scheme for scenarios with degenerate RHN masses. We conclude our discussion of low-scale seesaw models with a discussion of their impact on the instability scale. In Sec.~\ref{sec:highscale}, we present our results for a high-scale seesaw, covering both the possibility of two and three heavy RHNs, with important implications for high-scale leptogenesis. In the appendix, we provide threshold corrections~(\ref{RGappendix}), the beta functions~(\ref{beta functions append}), and corrections to the decay rate~(\ref{Instantonloops}) used in our calculations.

\section{Electroweak vacuum decay}\label{EW vacuum decay}
\subsection{Standard Model at NLO}\label{MSSM}
At energies far above the electroweak scale, the SM Higgs' effective potential takes the simple form
\begin{equation}
V_{\text{eff}}(\mu,h)\simeq  \frac{1}{4} \lambda_{\text{eff}} (\mu ,h) h^4.
\label{Veff4}
\end{equation}
Assuming ongoing validity of the SM, the quartic coupling $\lambda$ becomes negative at energies larger than the so-called \textit{instability scale}~$h\sim \mu_I \sim 10^{11}$~GeV. This behavior can be understood through the quartic coupling's beta function $\beta_\lambda$, which is rendered negative by a large contribution from the top Yukawa coupling, $\Delta \beta_\lambda \propto - 6 y_t^4$. As a consequence, the potential becomes unbounded from below at large field values. This allows the Higgs to tunnel out of the current (false) electroweak vacuum into the region of negative $\lambda$, with dramatic consequences.

Such a tunneling would occur through bubble nucleation, i.e. the formation of a "bubble" in some point in space-time populated with field values far beyond the instability scale. This process can be described by an \textit{instanton}, whose Euclidean action determines the decay rate to leading order. Due to the classical scale invariance of the potential Eq.~\eqref{Veff4}, the bounce is not unique. Instead there exists a one-parameter family of solutions characterized by the bubble size $R$,
\begin{equation}
h_R(r)= \sqrt{\frac{8}{|\lambda|}} \frac{R^2}{R^2+r^2} \cdot \frac{1}{R}.
\label{bounce}
\end{equation}
The total decay rate is given by the integral over the contributions from all these solutions, which has been shown to imply a divergent decay rate~\cite{Andreassen:2017rzq}. However, leading-order loop corrections to the decay rate break the scale invariance, yielding the finite expression
\begin{equation}
\frac{\Gamma}{V}=  \int \frac{\text{d}R}{R^5} \ {\rm e}^{-S_E [h_R]} D \big(R^{-1}\big)\,, \  \text{where} \ S_E[h_R] =  \frac{8 \pi^2}{3  \left\vert\lambda \left(R^{-1}\right)\right\vert}.
\label{DecayR}
\end{equation}
Here,~$D$ summarizes all but the gravitational corrections at NLO as derived in~\cite{Andreassen:2017rzq}, which we present in the appendix for the convenience of the reader.\footnote{This scheme allows for a complete incorporation of one-loop effects, unlike the common procedure of simply using the one-loop corrected effective action when finding the instanton. As pointed out in~\cite{Andreassen:2016cvx}, doing so would induce an error comparable to the remaining one-loop terms, as the instanton prevents a momentum-expansion of corrections to the Higgs' gradient term.} In principle, the integral over~$R$ can be performed analytically by resumming loop corrections to the Euclidean action, ultimately leading to a result of the form
\begin{equation}\label{GammaV}
    \frac{\Gamma}{V} \propto \mu_*^4 {\rm e}^{- \frac{8 \pi^2}{3  \left\vert\lambda \left(\mu_* \right)\right\vert}},
\end{equation}
where $\mu_*$ satisfies $\beta_\lambda (\mu_*)=0$. In other words, the leading order decay rate is determined by the value of the quartic coupling at its minimum.

On a conceptual level, this result can be understood as the decay rate being dominated by the bounce whose size minimizes the Euclidean action, i.e., maximizes $\vert \lambda \vert$. A similar result could have been obtained by applying a saddle point approximation to Eq.~\eqref{DecayR}. Using the central values of the most recent world averages for the running parameters' value at the top mass scale~\cite{Huang:2020hdv}, we find that $\mu_*=4.34 \cdot 10^{17}$~GeV and $\lambda (\mu_*)=-0.01$.\footnote{This value differs by almost $30\%$ from that found by the authors of~\cite{Andreassen:2017rzq}. This can be traced back to a shift in the central value of the top mass together the quartic coupling's near-criticality and thus sensitivity to such a correction.}

\subsection{Effect of gravity and the Higgs' non-minimal coupling}\label{VDG}
The SM decay rate is dominated by the contribution from the instanton linked to the RG scale $\mu_* \sim 4.34 \cdot 10^{17}$~GeV, where the quartic coupling reaches its minimum. Introducing additional fermions with sizeable Yukawa couplings to the Higgs drives this scale to even higher values, oftentimes beyond the Planck scale, where gravitational corrections should be taken into account. One of the most important results we review in the remainder of this section is that doing so automatically keeps the relevant scale sub-Planckian for all scenarios of interest in this article.

As a scalar, the Higgs can interact with gravity not only in the "standard" (\textit{minimal}) way, but also through a \textit{non-minimal coupling},
\begin{equation}
    S= \int \text{d}^4x \sqrt{-g} \ \left( \mathcal{L}_{\text{minimal}} + \frac{\xi}{2} R h^2 \right), 
\end{equation}
where $R$ is the usual curvature scalar and $\xi$ characterizes the strength of the coupling. 

To appreciate the importance of this term, it is helpful to note that, even if one were to set $\xi$ to zero at some scale, e.g. the Planck scale, a non-vanishing value would nevertheless be generated radiatively at all other scales. This can be seen from its beta function~\cite{Markkanen:2018bfx}, which is, in the regime of interest for our analysis, given by
\begin{equation}
    \beta_\xi^{(1)} = \frac{1}{(4 \pi)^2} \left( \xi + \frac{1}{6} \right) \left( 12 \lambda +6 y_t^2 + 2 \text{Tr}(Y_\nu^\dagger Y_\nu) - \frac{3}{2} {g^\prime}^2 - \frac{9}{2} g^2 \right),
\end{equation}
where $Y_\nu$ describes the neutrinos Yukawa couplings and will be defined through Eq.~\eqref{eq:lag}. An important special case is $\xi = - \frac{1}{6}$, for which not only the one-loop beta function vanishes, but also, as we will show in this subsection, the leading order gravitational correction to the decay rate. At our level of accuracy this amounts to neglecting gravity entirely and thus, as we will show, leads to a trans-Planckian RG scale. To allow for a comparison of our results to the existing literature, but also to make transparent the effect of our more precise calculations, we nevertheless consistently keep track of it throughout the remainder of this article. When doing so, we will thus also adopt the common procedure of introducing a hard cutoff at $\mu=M_{\text{Pl}}$, evaluating all couplings at the latter whenever $\mu_*>M_{\text{Pl}}$\footnote{Any attempt to obtain precise results from this approach would require further modifications to our underlying scheme. Going forward we will ignore this issue, as making these modifications would prevent a comparison with the existing literature, but also since we estimate that the effects would be far smaller than the error related to the introduction of a cutoff.}. 

Large values of $\xi$ are often considered in the context of Higgs inflation. While early works suggested that the latter can be achieved through values of $\xi$ as large as $\xi=10^4$~\cite{Bezrukov:2007ep}, more recent theoretical developments have shifted the focus to smaller values~$\xi\sim 10^2$~\cite{Barvinsky:2009fy,Barvinsky:2009ii,Bezrukov:2012sa,Schutz:2013fua}, which are also of interest for other applications, such as the production of primordial black holes~\cite{Iacconi:2021ltm,Geller:2022nkr,Braglia:2022phb,Qin:2023lgo} and other scenarios to produce dark matter~\cite{Lebedev:2022vwf}. It is also crucial to note that Higgs inflation requires a positive $\lambda$ in the regime of interest, i.e., that it is not compatible with the scenarios considered in this article. We will thus restrict ourselves to values of $\xi$ smaller than 10, which are more than sufficient to convey our central point and easily avoid any issues with unitarity or naturalness. 

Neglecting the expansion of spacetime, which is justified as the Hubble parameter satisfies $\frac{\dot{a}}{a} \ll \mu_*$~\cite{Joti:2017fwe,Shkerin:2015exa,Kobakhidze:2013tn,Mantziris:2020rzh}, the Higgs' coupling to gravity manifests to leading order in an additional term in the Euclidean action describing the gravitational potential induced by the instanton,
\begin{equation}
\label{SEgrav}
    S_E^{\text{grav}} (\mu_R ) \simeq \frac{8 \pi^2}{3 \left\vert\lambda \left(\mu_R \right)\right\vert}+\frac{256 \pi^3 (1+ 6 \xi)^2}{45 \lambda^2 \left(R^{-1}\right)} \frac{\mu_R^2}{M_{\text{Pl}}^2} + O \left( \left( \frac{\mu_R}{M_{\text{Pl}}}\right)^{4} \right),
\end{equation}
where $\mu_R \equiv R^{-1}$ and the Planck mass is $M_{\text{Pl}}= 1.22 \cdot 10^{19}$~GeV~\cite{Isidori:2007vm,Salvio:2016mvj}\footnote{Note that, throughout this article, we will assume metric gravity. Gravitational corrections to the decay rate in Palatini gravity have recently been investigated in~\cite{Gialamas:2022gxv}, whose results suggest that it should be straightforward to extend our analysis to this model.}. The backreaction of gravity on the instanton only emerges at the next order of $\vert \lambda \vert^{- \frac{1}{2}} (R\cdot M_{\text{Pl}})$. This observation is essential for our RG analysis, which relies on the identification $\mu \sim h_R \sim R^{-1}$. We will discuss the validity of our perturbative treatment and the mechanism that ensures it at the end of this subsection.

On a technical level, the new term prevents a straightforward application of the techniques allowing for an exact evaluation of the $R$-integral outlined in~\cite{Andreassen:2017rzq}. Instead, following~\cite{Khoury:2021zao,Steingasser:2022yqx}, this integral can be evaluated through a saddle point approximation,
\begin{equation}
\frac{\Gamma}{V} \simeq {\rm e}^{-S_{\text{E}}\left(\lambda (\mu_S),\mu_S\right)} \sqrt{\frac{2 \pi}{  \frac{\text{d}^2}{\text{d} \ln \mu^2}S_{\text{E}}\left(\lambda (\mu_S),\mu_S\right)}} \,\mu_S^4 D (\mu_S)\,, \label{ratefull}
\end{equation}
where the Euclidean action is again given by Eq.~\eqref{SEgrav}.

All couplings, as well as the corrections summarized in~$D$, are evaluated at the saddle point~$R_S^{-1} =\mu_S$ minimizing the Euclidean action. This scale, which has been named \textit{instanton scale} in~\cite{Khoury:2021zao,Steingasser:2022yqx}, is the solution of
\begin{equation}
\beta_\lambda (\mu_S) \bigg( \vert  \lambda (\mu_S) \vert +\frac{64  (1+6 \xi)^2 \pi}{15} \frac{\mu_S^2}{M_{\text{Pl}}^2} \bigg) =  - \vert\lambda (\mu_S)  \vert \frac{64 (1+6 \xi)^2 \pi}{15} \frac{\mu_S^2}{M_{\text{Pl}}^2} \left(1 + \frac{\beta_\xi}{1+6 \xi} \right).
\label{eq:saddle}
\end{equation}
In the limit $M_{\text{Pl}}^2 \to \infty$ and keeping in mind that $\lambda (\mu_S)<0$, it reduces to the condition $\beta_\lambda (\mu_S)=0$, i.e., $\mu_S=\mu_*$. The right-hand side is strictly negative, as is the bracket on the left-hand side. Thus, also $\beta_\lambda (\mu_S)$ needs to be negative, and therefore $\mu_S < \mu_*$. However, as $\beta_\lambda$ is loop-suppressed and varies only logarithmically, the effect of the right-hand side is significant already for scales up to two orders of magnitude below the Planck scale even in the scenario of a relatively small non-minimal coupling $\xi \sim 0$. 

The correction to the Euclidean action induced by the gravitational correction is strictly positive, thus increasing the Euclidean action and thereby lowering the decay rate. While the increase in the Euclidean action is partially compensated for by a shift in the instanton scale away from $\mu_*$, it is easy to see that this effect is always subdominant.

The gravitational contribution to the instanton scale Eq.~\eqref{eq:saddle} as well as to the Euclidean action appear multiplied by a factor of $(1+6 \xi)^2$, i.e., the non-minimal coupling can be expected to enhance the gravitational stabilization already for $\xi \gtrsim O(10^{-1})$. However, we will find in later sections that this effect is already significant without a non-minimal coupling, in agreement with our initial motivation and previous discussions.

Finally, we may observe that Eq.~\eqref{eq:saddle} also helps ensure the applicability of our computation in general and our perturbative treatment in particular, as it typically implies that $\mu_S \ll M_{\text{Pl}}$, even for trans-Planckian values of $\mu_*$. We make this manifest in our results through the introduction of a smallness parameter,
\begin{equation}\label{epsgrav}
    \epsilon_{\text{grav}} \equiv \frac{\Delta S_E^{\text{grav}}}{S_E^{\text{Mink}}} =\frac{32 \pi}{15 \vert \lambda (\mu_S) \vert} (1 + 6 \xi)^2 \frac{\mu_S^2}{M_{\text{Pl}}^2},
\end{equation}
which we will demand to be strictly smaller than one. We will find that this is always the case for smaller values of $\xi$ due to the mechanism described above. The only significant exceptions we find are for parameter space scans with $\xi \sim 10$, in which we highlight the corresponding points separately. 

The extension of our discussion beyond the leading order in gravitational corrections is well-understood, albeit not within the framework laid out in~\cite{Andreassen:2016cvx,Andreassen:2017rzq}. It is, however, easy to see that the next order of the perturbative treatment would manifest through corrections to the shape of the instanton.\footnote{The interplay between this effect and the RG running at LO has been discussed in~\cite{Espinosa:2020qtq}.}

\subsection{Decay probability and metastability}\label{sec:msbounds}
The decay of the electroweak vacuum would have drastic consequences for the universe, and our ongoing existence implies that it has not yet occured. We can thus safely conclude that the total probability that such an event has occured within our past lightcone $\mathcal{P}$ must be smaller than $O(1)$. Having at hand an expression for the probability per unit volume, this probability can be obtained by 
\begin{equation}
    P_{\text{decay} }= \int_{\mathcal{P}} \frac{\Gamma}{V}=V_{\mathcal{P}} \cdot \frac{\Gamma}{V} <1,
    \label{eq:metabound}
\end{equation}
where the second equality is true as the decay rate is constant throughout a majority of our past lightcone. The volume of the latter can be determined in terms of the current Hubble constant,
\begin{equation}
    V_{\mathcal{P}}=\frac{0.15}{H_0^4}= 2.2 \cdot 10^{163} (\text{GeV})^{-4},
\end{equation}
with $H_0 \approx 67.4 \frac{km}{s \cdot Mpc}$~\cite{Buttazzo:2013uya}. Thus, the continued existence of the electroweak vacuum requires that 
\begin{equation}\label{eq:mscond}
    \frac{\Gamma}{V}< \left(\frac{\Gamma}{V} \right)_{\text{crit}} \equiv 4.57 \cdot 10^{-164} \text{GeV}^4 =  e^{-376.104} \text{GeV}^4  ,
\end{equation}
which, through Eq.~\eqref{GammaV}, can be understood as a lower bound on the quartic coupling at high energies or, equivalently, an upper bound on the impact of any destabilizing effect.

Many previous works have relied on a related, but not equivalent condition. Setting aside the issue that not specifying the scale of the RG rate would manifest in a divergent decay rate, one can nevertheless obtain a (conservative) bound by demanding that the equivalent of Eq.~\eqref{eq:mscond} is satisfied for instantons at \textbf{all} allowed renormalization scales. To leading order, this amounts to,
\begin{equation}\label{boundnaive}
    \left( \frac{\Gamma}{V} \right)\sim \max_{\mu_I < \mu < M_{\text{Pl}}} \mu^4 e^{- \frac{8 \pi^2}{3 \vert \lambda (\mu) \vert}} < \left(\frac{\Gamma}{V} \right)_{\text{crit}},
\end{equation}
which can be translated to a lower bound on the quartic coupling at its minimum. In the absence of gravity, this inequality agrees to leading order with the one obtained from the more rigorous approach reviewed in Sec.~\ref{MSSM}. While this indeed happens to be the case in the SM, it is easy to see that a modified running, e.g., through the addition of RHNs, could easily drive $\mu_*$ beyond $M_{\text{Pl}}$. As a consequence, one would either have to accept dropping the most important contribution to Eq.~\eqref{boundnaive} or extrapolate the SM couplings beyond the Planck scale.

\section{Type-I Seesaw Mechanism}
\label{sec:seesawmech}
The \textit{Type-I Seesaw Mechanism} is one of the simplest SM extensions capable of explaining light neutrino masses~\cite{Minkowski:1977sc, Mohapatra:1979ia, Yanagida:1979as, GellMann:1980vs, Glashow:1979nm, Schechter:1980gr}. It relies on the introduction of fermionic SM singlets $N_i$, usually referred to as \textit{right handed neutrinos} (RHNs). Their dynamics and interactions with the SM lepton doublets $L_\alpha$ is described by the Lagrangian
\begin{align}
   {\cal L} \ = \ \bar{N}_i \slashed{\partial} N_i- (Y_\ell)_{\alpha\beta} \bar{L}_\alpha H l_R - (Y_{\nu})_{\alpha i}\bar{L}_{\alpha}H^c N_i - \frac{1}{2}(M_N)_{ij}\overline{N_i^c}N_j+{\rm h.c.} \, ,
   \label{eq:lag}
\end{align}
where $Y_\ell$ is charged lepton mass matrix and $L_\alpha$ is the $SU(2)_L$ lepton doublet. In the flavor basis ({$\nu^c,N$}), neutrino mass matrix can be written as
\begin{equation}
m_\nu = \left(
\begin{array}{cc}
0 & M_D \\
M_D^T & M_N 
\end{array}
\right),
\end{equation}
where $M_D=\frac{Y_{\nu}}{\sqrt{2}} v_{EW}$. Upon diagonalization in the limit $M_N \gg M_D$, this implies the famous seesaw relation for the light neutrino mass matrix after spontaneous symmetry breaking, 
\begin{equation}
    m_\nu \simeq -\frac{v_{EW}^2}{2}\: Y_{\nu}\,{M_N}^{-1}\,Y_{\nu}^T.
    \label{eq:seesaw}
\end{equation}
In the basis in which the charged leptons' mass matrix and gauge-interactions are flavor diagonal, the light neutrinos' mass matrix can be diagonalized by the unitary matrix $U_{\text{PMNS}}$ 
\begin{equation}
    U_{\text{PMNS}}^\text{T}m_\nu U_{\text{PMNS}} = \text{Diag}(m_1,m_2,m_3) = m_{\text{Diag}}.
\end{equation}
$U_{\text{PMNS}}$ can be completely parametrized by the the mixing angles $\theta_{ij}$ and the Dirac CP phase $\delta$. It also depends on two currently unknown Majorana phases, which we set to $\alpha_1=\alpha_2=0$ for our analysis. Since only the active neutrino mass splittings $\Delta m_{ij}^2$ are known, two different mass hierarchies are possible: \textit{normal hierarchy} (NH), $m_1<m_2<m_3$, and \textit{inverted hierarchy} (IH), $m_3<m_1<m_2$. Throughout this work, we use the central values of these known parameters derived from a recent global fit to the neutrino oscillation data \cite{Esteban:2020cvm}. 

Using Eq.~\eqref{eq:seesaw}, the Yukawa matrix can be expressed in terms of the~\textit{Casas-Ibarra parametrization}, i.e.,
\begin{equation}
    Y_{\nu}= -\frac{i\sqrt{2}}{v_{EW}} \,U_{\text{PMNS}}^* \:m_{\text{Diag}}^{1/2}\, R\, M_N^{1/2}
\end{equation}
where $R$ is a general complex orthogonal matrix and, together with $M_N$, contains the model's remaining free parameters. 

\subsection{Neutrinos and vacuum stability}\label{NuDecs}
\begin{figure}[t!]
    \centering
    \includegraphics[width=0.51\textwidth]{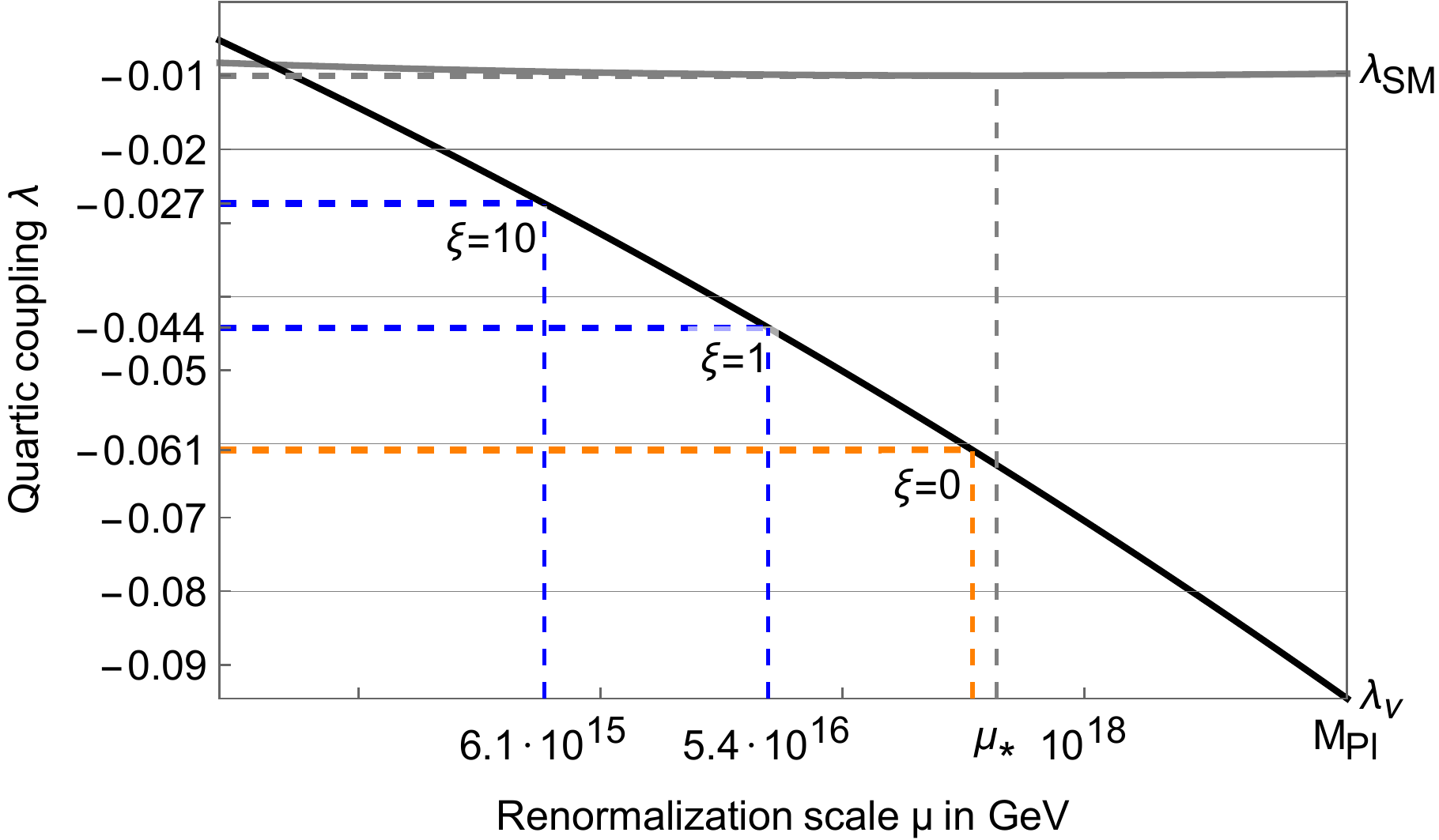}
    \includegraphics[width=0.47\textwidth]{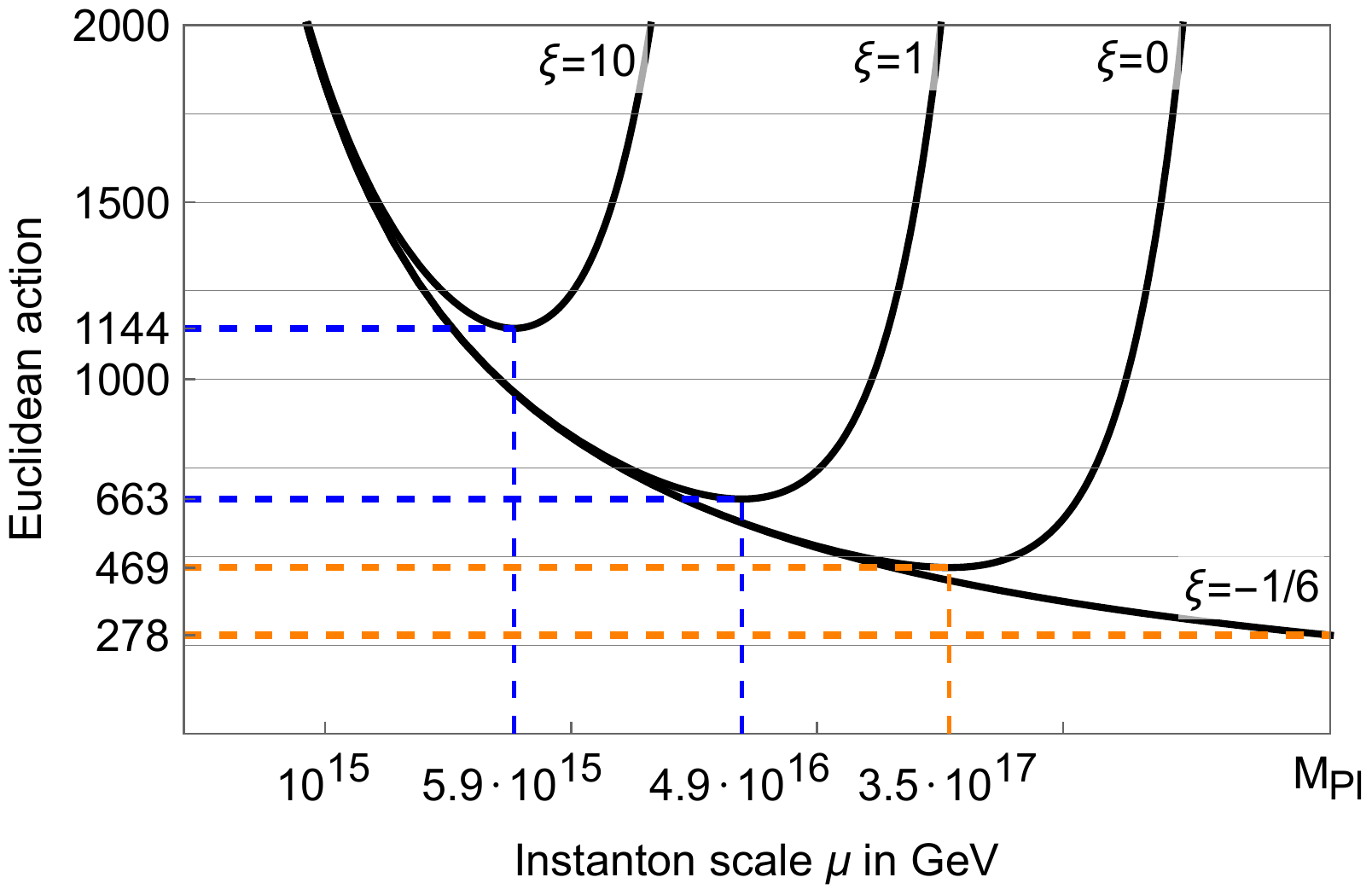}
    \caption{(Left panel) The RG evolution of $\lambda$ with and without the effect of RHNs. The SM trajectory ($Y_\nu=0$) is labelled $\lambda_{\text{SM}}$, while the curved labelled $\lambda_{\nu}$ contains the effects of RHNs with a generic set of parameters near the metastability bound. (Right panel) The corresponding Euclidean action including leading-order gravitational corrections as a function of $\mu \equiv \frac{1}{R}$ - before the saddle point approximation has been performed - for different values of the non-minimal coupling $\xi$. \\
    In both panels, orange lines mark unstable electroweak vacua, while blue lines mark metastable ones.}
    \label{lS}
\end{figure} The effect of RHNs on vacuum stability can be best understood through Eq.~\eqref{GammaV}. Like all other fermions, their inclusion gives rise to a negative contribution to the quartic coupling's beta function, which for a type-I-seesaw is to leading order given by
\begin{align} 
    \Delta \beta_\lambda &=\frac{1}{(4\pi)^2}\left[4 \lambda \text{Tr}(Y^\dagger_\nu Y_\nu)-2\text{Tr}(Y^\dagger_\nu Y_\nu Y^\dagger_\nu Y_\nu) \right].
    \label{eq:betaL}
\end{align}
In terms of Feynman diagrams, the first term can be understood as an insertion of a neutrino loop into the propagator of one of the Higgs particles entering/leaving the vertex. Meanwhile, the second term is generated by a diagram in which the four-Higgs vertex itself gets a correction from a neutrino loop.

Assuming reasonably large Yukawa couplings, these terms drive $\lambda$ to more negative values, thereby increasing the decay rate. Besides their direct negative contribution to the running of $\lambda$, introducing neutrinos also amplifies the destabilizing effect of the top quark by increasing its Yukawa coupling,
\begin{equation}\label{eq:betayt}
    \Delta \beta_{y_t} =\frac{1}{(4\pi)^2} y_t \text{Tr}(Y^\dagger_\nu Y_\nu).
\end{equation}
In addition to lowering the values of $\lambda$ at high energies, Eqs.~\eqref{eq:betaL} and~\eqref{eq:betayt} push $\mu_*$ towards larger values, often even beyond the Planck scale, as illustrated in the left panel of Fig.~\ref{lS} for a generic example. This is evident from the full expression for $\beta_\lambda$, as given, e.g., in Eq.~\eqref{eq:betalambdafull}. In the SM, $\lambda$ reaches a minimum when the negative contribution from the top quark becomes smaller than its positive counterparts from the gauge bosons. If, however, its decline is counteracted as well as compensated for by the neutrino terms, the cancellation characterizing the minimum of $\lambda$ occurs only at much larger scales.
 
Following our discussion in Sec.~\ref{VDG}, taking into account gravitational corrections leads to a "decoupling" of the relevant RG scale, the instanton scale $\mu_S$, from the scale marking the minimum of $\lambda$, $\mu_*$. This is illustrated in Fig.~\ref{lS}, where we plot the running of the quartic coupling $\lambda$ beyond the heaviest RHN mass scale for a generic choice of neutrino parameters near the metastability bound. Due to the non-zero $Y_\nu$, no minimum forms below the Planck scale, and $\lambda$ is driven to significantly smaller values. Using Eq.~\eqref{GammaV} with $\lambda (M_{\text{Pl}})$, this causes an increase in the exponent of the decay rate by a factor of roughly~$10$, rendering the vacuum unstable. However, due to the effect of gravitational corrections described in Sec.~\ref{VDG}, the \textit{instanton scale} is pushed to lower values and the decay rate decreased. For large enough values of the non-minimal coupling, the vacuum becomes metastable.

The Yukawa couplings also contribute directly to the decay rate, namely through the functional determinants summarized in the factor $D$ in Eq.~\eqref{DecayR}. This factor leads to an effective one-loop contribution from each particle coupled to the Higgs, which has been calculated analytically in~\cite{Andreassen:2017rzq}. We present the results extended by the neutrino contribution in Sec.~\ref{Instantonloops}.

\subsection{Relevant parameters for the RG evolution}\label{relpar}
We established in Sec.~\ref{EW vacuum decay} that the neutrinos' influence on vacuum stability is predominantly through their effect on the running of $\lambda$. This allows us to identify suitable quantities for the presentation of the metastability bounds. To leading order, the neutrinos' correction to $\beta_\lambda$, given in Eq.~\eqref{eq:betaL}, is sensitive to the neutrinos' parameters only through the combinations $\text{Tr}(Y_\nu^\dagger Y_\nu )$ and $\text{Tr}(Y_\nu^\dagger Y_\nu Y_\nu^\dagger Y_\nu )$. To simplify the following discussion, we can now introduce a set of new parameters $\{T_{2n}\}_{n \in \mathbb{N}}$ defined as
\begin{equation}
T_{2n}\equiv \text{Tr} \left[\left(Y_\nu^\dagger Y_\nu \right)^n \right],
\end{equation}
to which we will refer as \textit{trace parameters}.

The parameter $T_2$ appears in $\beta_\lambda$ multiplied with $\lambda$, which, for all scenarios of interest to us, causes a significant suppression relative to the term $\propto T_4$. In other words, we can expect that our bounds take the clearest form when expressed in terms of $T_4$, with a remaining dependence on $T_2$. It is well-known that there exist many instances in which these two parameters are related through the relation $T_2^2 \simeq T_4$, which motivated previous works to express their results in terms of $T_2$. However, as we will demonstrate explicitly in Sec.~\ref{sec:mslowscale}, this is not guaranteed. 

To extend this analysis beyond the leading order, we need to take into account NLO contributions to the decay rate as well as higher orders of the loop expansion in the running of the relevant parameters and threshold corrections. An important observation necessary for this analyis is that the beta function for $Y_\nu$ can be used to obtain beta functions for the trace parameters, which we provide in Sec.~\ref{beta functions append}.

To illustrate our point, let us first consider the running of $\lambda$ itself at two-loop and that of the remaining parameters at one-loop accuracy. Since the latter affect the Euclidean action only indirectly through the running of $\lambda$, we estimate these effects to be comparable. This allows us to approximate the effect of the RHNs on the overall running as
\begin{gather}
    \Delta \lambda \sim \left( \beta_\lambda^{(1)}(\lambda T_2,T_4,y_t) + \beta_\lambda^{(2)}(T_2,T_4,T_6)  \right) \ln \left(\frac{\mu_S}{M_N} \right) \label{dl} +...\\
    \Delta y_t \sim \left( \beta_{y_t}^{(1)}(T_2)  \right) \ln \left(\frac{\mu_S}{M_N} \right) + ... \\
    \Delta T_4 \sim \left( \beta_{T_4}^{(1)}(T_2,T_4,T_6) \right) \ln \left(\frac{\mu_S}{M_N} \right) \\
    \Delta T_2 \sim \left( \beta_{T_2}^{(1)}(T_2,T_4) \right) \ln \left(\frac{\mu_S}{M_N} \right) +... \label{dt4}
\end{gather}
Thus, we find that increasing the order of the loop expansion in the running for $\lambda$ by $1$ requires the inclusion of one additional trace parameter. 

A consistent running at two-loop also requires the inclusion of threshold corrections at one-loop, which we provide in Sec.~\ref{thresh}. For scenarios with approximately degenerate RHN masses, these introduce an additional parameter $\text{Tr}(Y_\nu^\dagger Y_\nu Y_\nu^T Y_\nu^* )$, which we estimate to have an effect comparable to that of $T_6$. For non-degenerate RHN masses, these corrections mix different flavors. This prevents a complete, consistent analyis along the lines of this subsection for such cases. Crucially, this does \textbf{not} interfere with our conclusion that $T_4$ can be expected to be the most important parameter, which we concluded solely through leading order arguments, which are insensitive to these threshold corrections.

Lastly, we may note that the decay rate also depends on the Yukawa couplings $Y_\nu$ directly through the neutrinos' functional determinants, which we derive in Sec.~\ref{Instantonloops}. While these cannot strictly be expressed through the trace parameters, they can be related with $T_2$ up to a deviation that has proven negligible in our numerical analysis.

\section{Low-scale Type-I Seesaw}
\label{sec:lowscale}
For RHNs to appreciably affect the RG running of $\lambda$, the entries of $Y_\nu$ need to be of~$\mathcal{O}(0.1-1)$. In the canonical scenario, assuming this to be the case and the light neutrino mass scale to be the observed $0.1$ eV, the required mass scale for the heavy states is around $10^{15}$~GeV. 

In the standard picture described above, trying to lower $M_N$ below $10^{15}$~GeV leads to tiny values for $Y_\nu$, which therefore cannot contribute appreciable to $\beta_\lambda$. For instance, requiring $M_N$ to be around the TeV scale leads to $Y_\nu \sim \mathcal{O}(10^{-6})$, implying a negligible impact on the decay rate. This simple estimate neglects however the general matrix structure of the seesaw relation, which can be made manifest, e.g., through the Casas-Ibarra parametrization discussed in Sec.~\ref{sec:seesawmech}. Due to this structure, it is possible for the entries in the matrix Eq.~\eqref{eq:seesaw} to cancel out in such a way as to yield light neutrino masses and yet retain high values for the entries in $Y_\nu$ (for large complex phases in $R$). The latter feature leads to improved prospects for detection at current/near future experiments due to high values of the mixing parameter $U=Y_\nu\; v \, M^{-1}_N $. Previous studies~\cite{Pilaftsis:1991ug,Gluza:2002vs,Xing:2009in,He:2009ua,Ibarra:2010xw,Mitra:2011qr,Lee:2013htl} have shown that larger mixing angles are possible with various special textures for $Y_D$ and $M_N$. 

In this article, we consider light neutrino mass generation from the seesaw mechanism at tree-level. For low-scale seesaw models, it has been shown that the seesaw relation is radiatively stable under quantum corrections only if $M_N$ has a degenerate spectrum \cite{Kersten:2007vk, Moffat:2017feq}. This also reduces the number of free parameters in the Casas-Ibarra parametrization of $Y_\nu$. In the limit of large entries for $Y_\nu$, this effectively corresponds to an approximate $B-\Tilde{L}$ symmetry, where $\Tilde{L}$ is the generalized lepton number~\cite{Shaposhnikov:2006nn,Kersten:2007vk}.

\subsection{Metastability bounds}\label{sec:mslowscale}

 \begin{figure}[t!]
    \centering
     \includegraphics[width=0.49\textwidth]{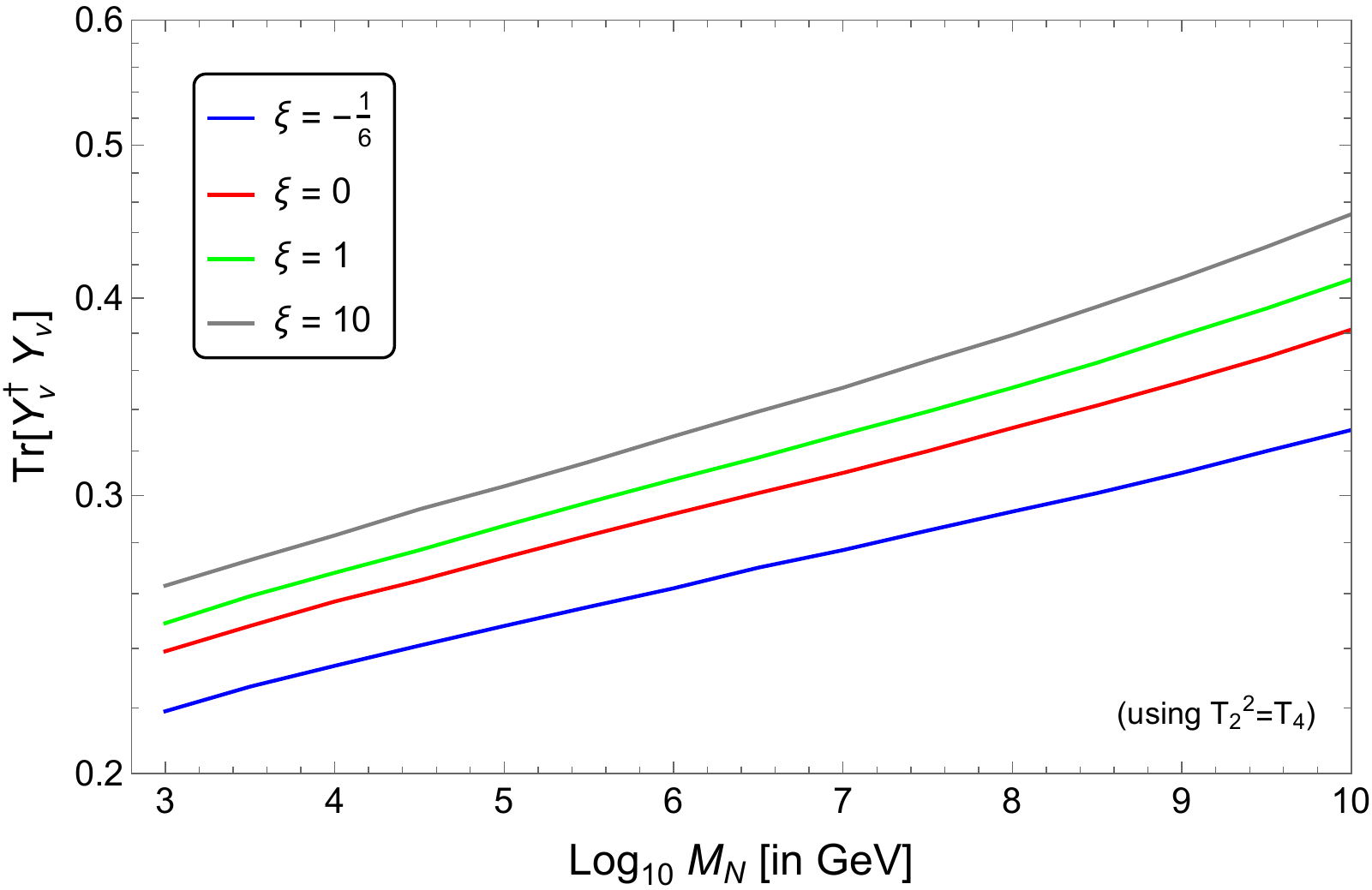}
    \includegraphics[width=0.49\textwidth]{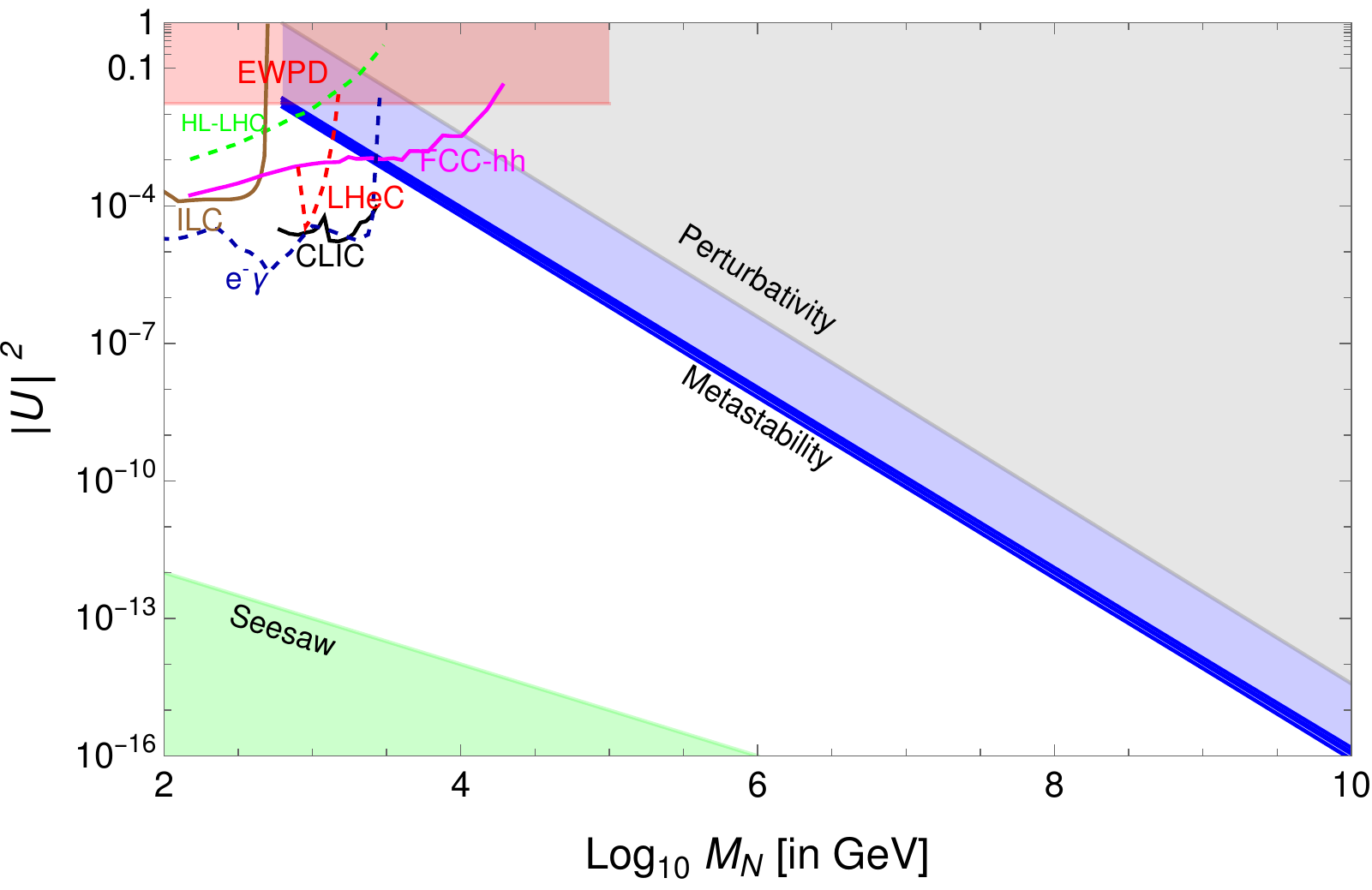}
    \caption{(Left) The upper bound $\text{Tr}(Y_\nu^\dagger\, Y_\nu)$ as a function of degenerate RHN mass scale $M_N$ for non-minimal coupling $\xi = \left(-\frac{1}{6},0,1,10\right)$. (Right) Bounds on mixing parameter $|U|^2$ (shown in blue) as a function of degenerate RHN mass scale $M_N$ for different non-minimal couplings along with current constraints from EWPD~\cite{delAguila:2008pw,Akhmedov:2013hec,deBlas:2013gla,Antusch:2014woa,Blennow:2016jkn,Flieger:2019eor} and future projections for HL-LHC, FCC-hh, ILC, LHeC, CLIC and $e^- \gamma$ colliders~\cite{Banerjee:2015gca,Mondal:2016kof,Antusch:2016ejd,Pascoli:2018heg,Hernandez:2018cgc,Das:2018usr,FCC:2018vvp,Chakraborty:2018khw, Antusch:2019eiz,Bolton:2019pcu,Das:2023tna}. We incorporate the set of values of $\xi$ also highlighted in the left panel (including the uncertainties, see Sec.~\ref{subsec:uncertain}), but the differences between their impacts are too small to be distinguished in this plot (dark blue band).}
    \label{fig:lowSapprox}
\end{figure}

Following our arguments in Sec.~\ref{sec:msbounds}, we may exclude scenarios in which condition Eq.~\eqref{eq:metabound} is violated. To obtain the RG-trajectories necessary to evaluate this relation we first integrate the SM beta functions as given in Sec.~\ref{beta functions append} at three-loop\footnote{The effects of the bottom-quark and tau lepton at one-loop are comparable to those of the top quark at three-loop, and thus included in our analysis.} until the matching scale $M_N$. As matching condition in the IR we use the current central values at the top mass scale as given in~\cite{Huang:2020hdv},
\begin{gather}
    \lambda (M_t)=0.12607\,; \ \ y_t (M_t)=0.9312\,; \ \  \nonumber\\  g_s (M_t)=1.1618\,; \ \ g^\prime (M_t)=0.358545\,; \ \    g(M_t)=0.64765 \,. 
\end{gather}
where $g_s$, $g'$ and $g$ are the $SU(3)_c$, $SU(2)_L$ and $U(1)_Y$ gauge coupling respectively.

Following our discussion in Sec.~\ref{relpar}, the relevant parameters for the running of $\lambda$ are the traces $\text{Tr}(Y_\nu^\dagger Y_\nu)$ and $ \text{Tr}(Y_\nu^\dagger Y_\nu Y_\nu^\dagger Y_\nu)$. As a consequence, we will find that the requirement of metastability amounts to an upper bound on these quantities of order~$O(0.1-1)$. For such values together with relatively low masses $M_N \lesssim 10^{11}$~GeV, it can be shown that~\cite{Rodejohann:2012px,Lindner:2015qva}
\begin{equation}\label{Traceapprox}
    \text{Tr}(Y_\nu^\dagger Y_\nu)^2\sim \text{Tr}(Y_\nu^\dagger Y_\nu Y_\nu^\dagger Y_\nu).
\end{equation}
Therefore, the metastability bound in this case can be reported directly in terms of $\text{Tr}(Y_\nu^\dagger Y_\nu)$. We outline an analytical proof for this result and show that the bounds are essentially insensitive to the matrix structure of $Y_\nu$ in this high-magnitude limit in Sec.~\ref{sec:traceapprox}. In Sec.~\ref{subsec:Tpar}, we will discuss casees where this condition fails and demonstrate a reliable analysis based on a larger set of parameters. We furthermore find that once the condition~\ref{Traceapprox} is satisfied at the matching scale, it can be recovered for the full range of scales for interest to us. This allows us to directly use $T_2= \text{Tr}(Y_\nu^\dagger Y_\nu)$ in our running rather than the full Yukawa matrix.

At the matching scale $M_N$, we match the SM couplings with the full theory using the threshold corrections given in Sec.~\ref{thresh}. Integrating the beta functions of this model (using $\text{Tr}(Y_\nu^\dagger\, Y_\nu)$ as a parameter), we are able to numerically determine the instanton scale by using relation Eq.~\eqref{eq:saddle}, including the running of the non-minimal coupling $\xi$ at one-loop level. The decay rate can then be evaluated by inserting this scale and the RG trajectories of the relevant parameters into Eq.~\eqref{ratefull}. We furthermore keep track of the convergence of our perturbative treatments of gravity, characterized by the parameter $\epsilon_{\text{grav}}$ defined in Eq.~\eqref{epsgrav}, and higher-order loop effects, controlled by the couplings. We find that the validity of our expansions is guaranteed until deep inside the unstable regime. Following our previous discussion, we scan the degenerate RHN mass $M_N$ and the trace parameter $\text{Tr}(Y_\nu^\dagger\, Y_\nu)$ (evaluated at the RG scale $\mu=M_N$) for different values of the non-minimal coupling parameter $\xi$ (evaluated at the Planck scale). 

The results of our parameter scan (in the regime where Eq.~\eqref{Traceapprox} is satisfied) of the low-scale type-I seesaw are shown in Fig.~\ref{fig:lowSapprox}. In the left panel we display our results for the upper bound on $\text{Tr}(Y_\nu^\dagger\, Y_\nu)$ imposed by metastability of the electroweak vacuum as a function of the RHN mass scale for different values of the non-minimal coupling $\xi$ at the Planck scale. For any given $\xi$, the bound on $\text{Tr}(Y_\nu^\dagger\, Y_\nu)$ becomes weaker with increasing $M_N$. Another noticeable feature is the relaxation of the bounds on $\text{Tr}(Y_\nu^\dagger\, Y_\nu)$ as $\xi$ increases from $\xi=-\frac{1}{6}$ to $\xi=10$. As expected from our previous discussions of the impact of gravitational corrections, the positive contribution to the Euclidean action stabilizes the electroweak vacuum, thus allowing for higher values of $\text{Tr}(Y_\nu^\dagger\, Y_\nu)$ to be compatible with metastability. For $\xi=-\frac{1}{6}$ and $M_N$ at the TeV scale, we find that $\text{Tr}(Y_\nu^\dagger\, Y_\nu) \lesssim 0.22$ is required for metastability, while $\text{Tr}(Y_\nu^\dagger\, Y_\nu) \lesssim 0.32$ becomes viable at $M_N = 10^{10}$~GeV. These bounds get relaxed for higher $\xi$'s, e.g. for $\xi=10$, the allowed value for $M_N =1$~TeV gets relaxed to $0.26$ and to $0.45$ for $M_N = 10^{10}$~GeV. Note that while the bounds are quite independent of the neutrino mass ordering, the choice of lightest neutrino mass scale $m_0$ affects the bound. We have chosen $m_0=0$ eV in our work and the bounds get stronger for non-zero $m_0$. 

An interesting feature is the approximately linear form of our bounds. This can be easily understood through Eq.~\eqref{dl}, as the overall impact of the RHNs' effect on the running of $\lambda$ is given by~$\Delta \lambda \propto \text{Tr}(Y_\nu^\dagger\, Y_\nu) \cdot \ln\left( \mu_S/M_N \right)$, assuming validity of the high-magnitude approximation Eq.~\eqref{Traceapprox}. Thus, to leading order the constrained quantity becomes the product of the trace parameter and $\ln (M_N)$, which is roughly constant along our bounds in Fig.~\ref{fig:lowSapprox}. Another important consequence of this relation is the independence of our results from the number of RHNs, as none of these parameters explicitly depends on it. 

Apart from kinematical factors, the cross section for RHN production and decay along with some lepton flavor $\alpha$ is directly proportional to the mixing angle $U_\alpha$. Since our treatment for vacuum metastability is flavor-blind, we choose to also report our results in terms of total mixing angle $|U|^2$ to allow for better comparison with future experimental searches. In general, the mixing angle satisfies
\begin{equation}
    |U|^2 = \sum |U_\alpha|^2 = \frac{v^2}{2} \text{Tr}\left(M_N^{-1}\,Y_\nu^\dagger \,Y_\nu \,M_N^{-1}\,\right). 
\end{equation}
For degenerate RHN mass scales this reduces to
\begin{equation}
    |U|^2 = \frac{v^2}{2\,M_N^2} \text{Tr}\left(Y_\nu^\dagger\,Y_\nu \,\right),
    \label{eq:mixf}
\end{equation}
allowing for a direct translation of theoretical bounds from the metastability criterion to experimentally interesting bounds on mixing. 

In the right panel of Fig.~\ref{fig:lowSapprox}, we display our bounds from metastability (as already shown in the left panel) in terms of the total mixing angle $|U|^2$ as a function of RHN mass scale. The grey region, labeled \textit{Perturbativity}, can be ruled out by demanding that $\text{Tr}(Y_\nu^\dagger\, Y_\nu)\,\leq 4 \pi$ below the Planck scale. The light green region labeled \textit{Seesaw} in the lower left is incompatible with the light neutrino masses being entirely generated through the type-I seesaw mechanism. The only major experimental constraint above $M_N\,>\, 1$ TeV arises from electroweak precision measurements. The mixing between the active and the heavy state can induce corrections to Electroweak Precision Data (EWPD) observables. These precision observables include various leptonic and hadronic measurements such as the weak mixing angle ($\theta_W$), the mass of the $W$ boson, ratios of certain leptonic weak decays testing EW universality, various ratios of $Z$ boson decay rates as well as its decay width $\Gamma_Z$. Such deviations from new physics are tightly constrained. Assuming the strongest bounds on the mixing parameter (arising from coupling to the first generation), the constraint from EWPD is $|U|^2 \, < \, 0.0025, \text{ for } M_N \gtrsim 1 \text{ GeV}$ \cite{delAguila:2008pw,Akhmedov:2013hec,deBlas:2013gla,Antusch:2014woa,Blennow:2016jkn,Flieger:2019eor}.  For $M_N\,>\,1$ TeV, the metastability of the electroweak vacuum sets the strongest bounds on $|U|^2$.

It is noticeable that the dependence on the non-minimal coupling $\xi$ is barely visible in this plot due to the logarithmic scale of the plot. Therefore, in the right panel the various bounds for different $\xi$ merge into a single curve, labeled \textit{Metastability}. We notice that the metastability bound performs far better than the naive bound arising from requiring perturbativity. It rules out a major portion of the parameter space in the $|U|^2-M_N$ plane. The current constraints from EWPD are stronger only for higher mixing ($|U|^2 \, < \, 10^{-3}$) and lower $M_N\,<\,10$ TeV.  Proposed future direct searches at HL-LHC, FCC-hh, ILC, LHeC, CLIC and $e^- \gamma$ colliders~\cite{Banerjee:2015gca,Mondal:2016kof,Antusch:2016ejd,Pascoli:2018heg,Hernandez:2018cgc,Das:2018usr,FCC:2018vvp,Chakraborty:2018khw, Antusch:2019eiz,Bolton:2019pcu,sterileorg,Das:2023tna} will be able to set competitive bounds\footnote{We again assume strongest mixing with the first generation to get most stringent future exclusions.} for a mixing as low as $|U|^2 \, \sim \, 10^{-6}$.

\subsection{Importance of $T_4$ and efficient RG evolution}\label{subsec:Tpar}

\begin{figure}[t!]
    \centering
    \includegraphics[width=0.49\textwidth]{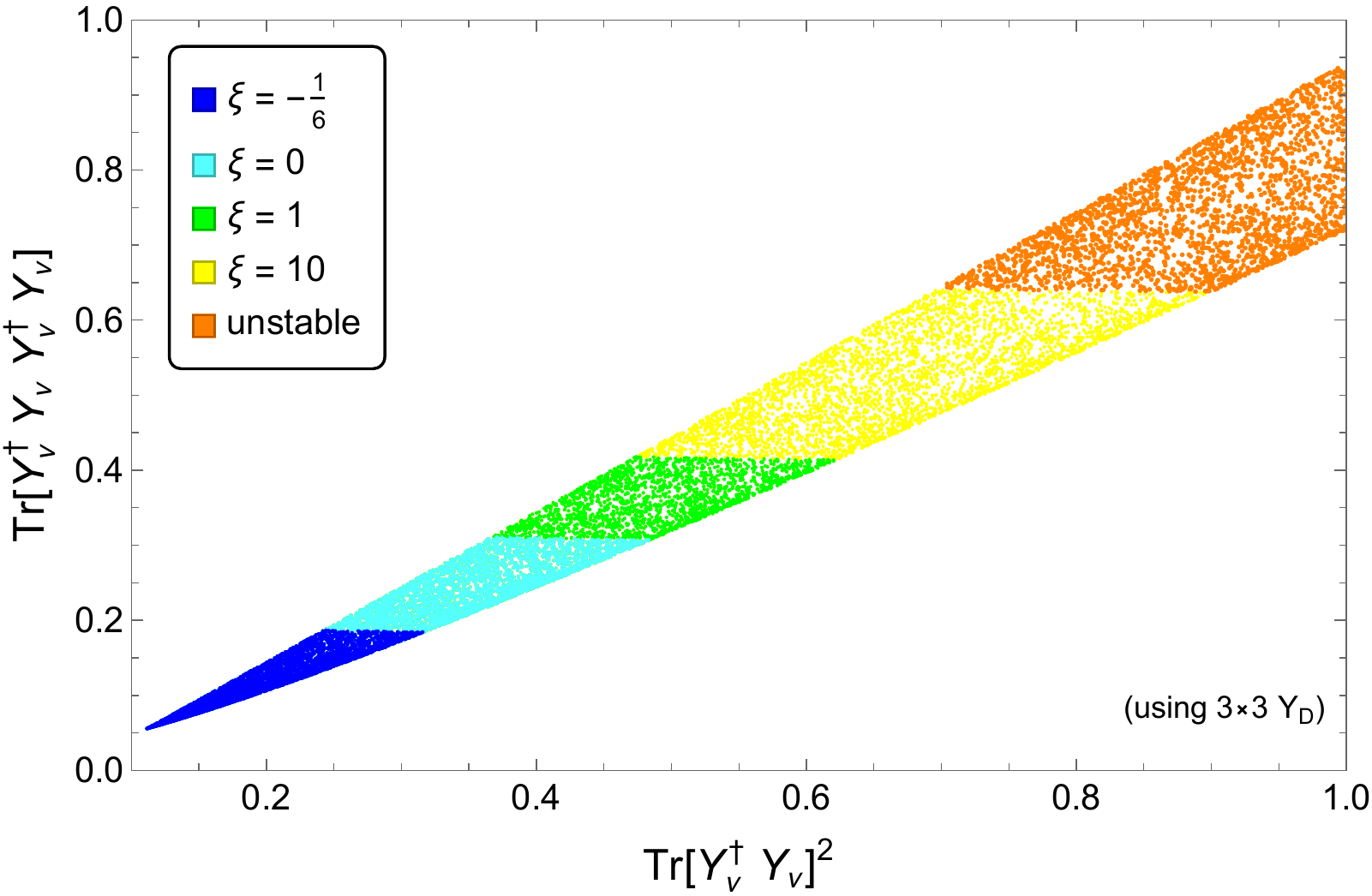}
     \includegraphics[width=0.49\textwidth]{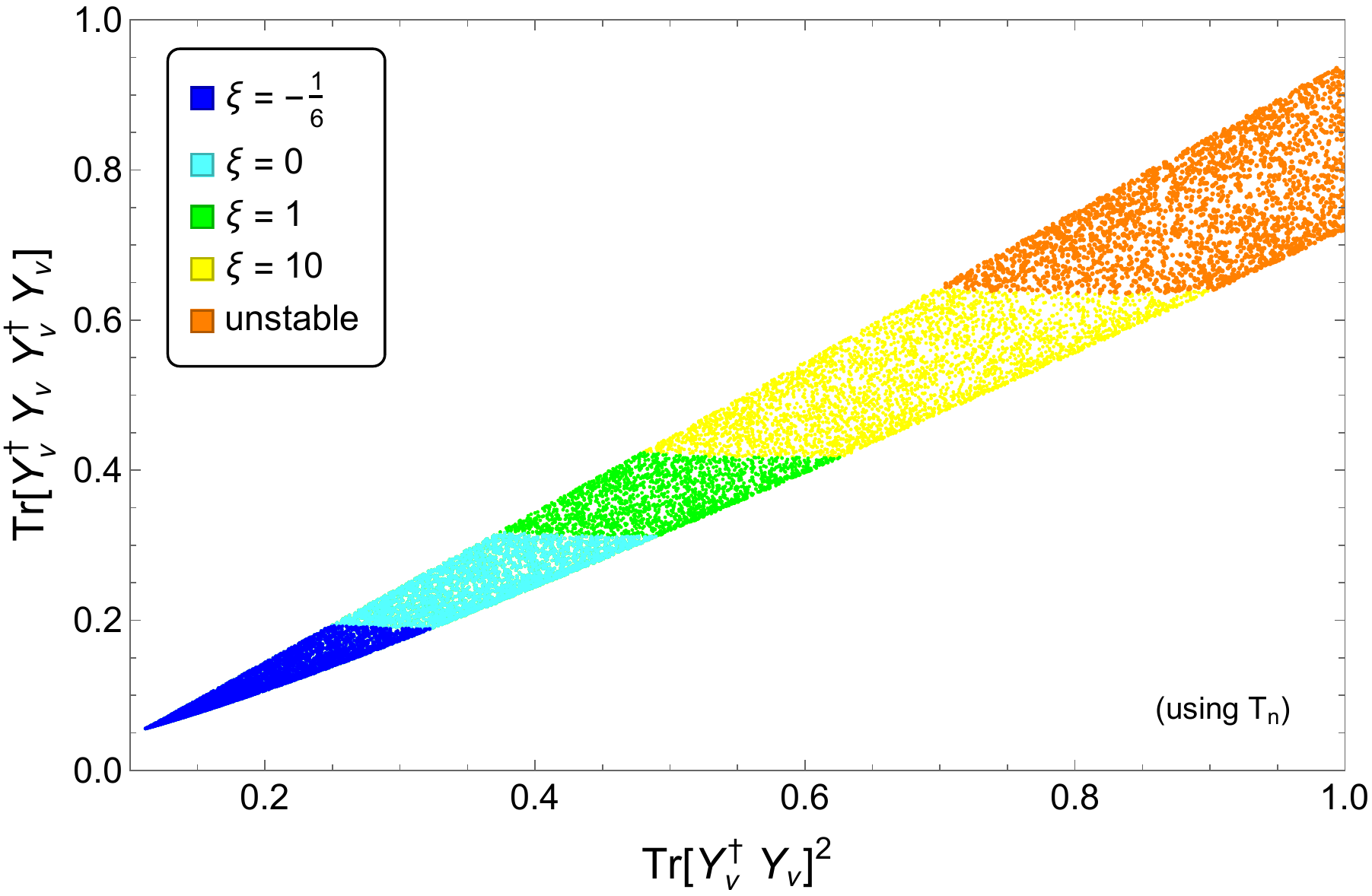}
     \caption{Scatter plot of $\text{Tr}(Y_\nu^\dagger\, Y_\nu)^2$ vs $\text{Tr}(Y_\nu^\dagger\, Y_\nu Y_\nu^\dagger\, Y_\nu)$ for $M_N=10^{14}$ GeV, Normal Hierarchy and non-minimal coupling $\xi = \left(-\frac{1}{6},0,1,10\right)$. Colored points (except orange) correspond to metastable electroweak vacuum for corresponding $\xi$ (in addition to the metastable region for lower value of $\xi$). Orange points signify unstable electroweak vacuum. (Left) RGE evolution has been performed using entire $3 \times 3$ matrix form of $Y_\nu$ and (Right) using $T_n$-parametrization involving only $T_2,T_4,T_6$.}
     \label{fig:lowSmatrix}
\end{figure}

As discussed earlier, previous results were obtained by assuming the high-magnitude limit of $\text{Tr}(Y_\nu^\dagger\, Y_\nu)$, which justifies the approximation $\text{Tr}(Y_\nu^\dagger Y_\nu)^2\sim \text{Tr}(Y_\nu^\dagger Y_\nu Y_\nu^\dagger Y_\nu)$. Thus, the bound on the type-I seesaw mechanism's parameter space is effectively described by the single parameter $\text{Tr}(Y_\nu^\dagger Y_\nu)$ in this limit. The approximation fails when the magnitude of $\text{Tr}(Y_\nu^\dagger\, Y_\nu)$ is in the vicinity of the expected value from the canonical seesaw. This usually happens at high $M_N$, see Fig.~\ref{fig:lowSmatrix} (left panel), which shows the results of our parameter scan in the $\text{Tr}(Y_\nu^\dagger Y_\nu)$- $\text{Tr}(Y_\nu^\dagger Y_\nu Y_\nu^\dagger Y_\nu)$-plane for $M_N=10^{14}$ GeV and taking into account the \textbf{full} set of beta functions for $Y_\nu$ as given in~\ref{beta functions append} and threshold corrections as discussed in Sec.~\ref{thresh}. 

Besides the breakdown of the approximation $T_2^2 \simeq T_4$, we find that these results agree with our previous theoretical expectations. First, the metastability bounds get successively relaxed as we increase the value of $\xi$. Second, the bounds strongly depend on $T_4$ and are almost independent of $T_2$. 

This deviation is a manifestation of the breakdown of the approximation~$T_2^2 \simeq T_4$, which would suggest that a complete RG analysis taking into account the full flavor structure is necessary. While this is, in principle, straightforward to do numerically, there is also a simpler way. Recall that, at the level of accuracy underlying our analysis, the relevant SM parameters depend on the Yukawa couplings only through the trace parameters. We may furthermore observe that it is always possible to express their scale dependence through a set of beta functions which themselves depend only on the SM couplings and trace parameters. Thus, instead of integrating the full RG equations, we may equivalently work with this new set of beta functions as long as we are primarily interested in the decay rate. 

To leading order, this would amount to treating $T_4$ and $T_2$ as constants, which is of course always possible. At next-to-leading order, one would have to incorporate threshold corrections and take into account the running of these two parameters. This requires to specify the value of $T_6$, which is to be treated as a constant, and the combination $\text{Tr}(Y_\nu^\dagger Y_\nu Y_\nu^T Y_\nu^* )$ necessary for the matching at $\mu =M_N$. For non-degenerate RHN masses, the latter takes the more complicated form we present in Sec.~\ref{thresh}. Crucially, this expression cannot be conveniently expressed through the trace parameters, preventing a self-contained analysis. As a consequence, we find that this effective RG running scheme is only reasonably applicable to scenarios with near-degenerate RHN masses.

To illustrate the effectiveness of this procedure, we applied it to perform a next-to-leading order analysis (as defined in the previous paragraph) of the scenario previously discussed in this subsection, i.e., $M_N=10^{14}$~GeV, NH and different values of $\xi$. The left panel of Fig.~\ref{fig:lowSmatrix} shows our results obtained from this complete analysis, while the right panel shows the results obtained through this effective scheme.

\subsection{Comparison with SM uncertainties}\label{subsec:uncertain}

As discussed in Sec.~\ref{sec:mslowscale}, the derivation of metastability bounds involves running the SM couplings up to the instanton scale $\mu_S$. Doing so requires matching these couplings with their initial values at some reference scale, e.g., the top-mass scale $M_t$. Due to the high sensitivity of the decay rate to the quartic coupling and the large hierarchy of scales, it can be expected that the remaining experimental uncertainties in these initial values introduce a significant theoretical error in the metastability bounds. It is well-understood that the uncertainty in~$\lambda$ arises predominantly from that in its initial value, as well as those in the top Yukawa coupling $y_t$ and the strong gauge coupling $g_s$~\cite{Khoury:2021zao,Steingasser:2022yqx,Steingasser:2023ugv}. These have recently been quantified individually by using a combination of linear error propagation and Monte Carlo simulations in Ref.~\cite{Huang:2020hdv}, who found that the $1\sigma$~error ranges for these relevant SM parameters are given by
\begin{gather}
   \Delta \lambda (M_t)=\pm 0.00030\,; \ \ \Delta y_t (M_t)=\pm 0.0022\,; \ \  \Delta g_s (M_t)=\pm 0.0045\,. 
\end{gather}
To obtain the corresponding uncertainty in the decay rate, we first construct the error ellipsoid in the $\left(\lambda (M_t),y_t (M_t), g_s (M_t)\right)$-space as described in Ref.~\cite{Steingasser:2023ugv}.\footnote{As it is sufficient to illustrate our central point, we restrict ourselves to the $68\%$ confidence ellipsoid.} We then repeat the computation described in Sec.~\ref{sec:mslowscale} over a suitable mesh of points on this ellipsoid, calculating the metastability bound for each of the corresponding sets $\left(\lambda (M_t),y_t (M_t), g_s (M_t)\right)$. 

\begin{figure}[t!]
    \centering
    \includegraphics[width=0.49\textwidth]{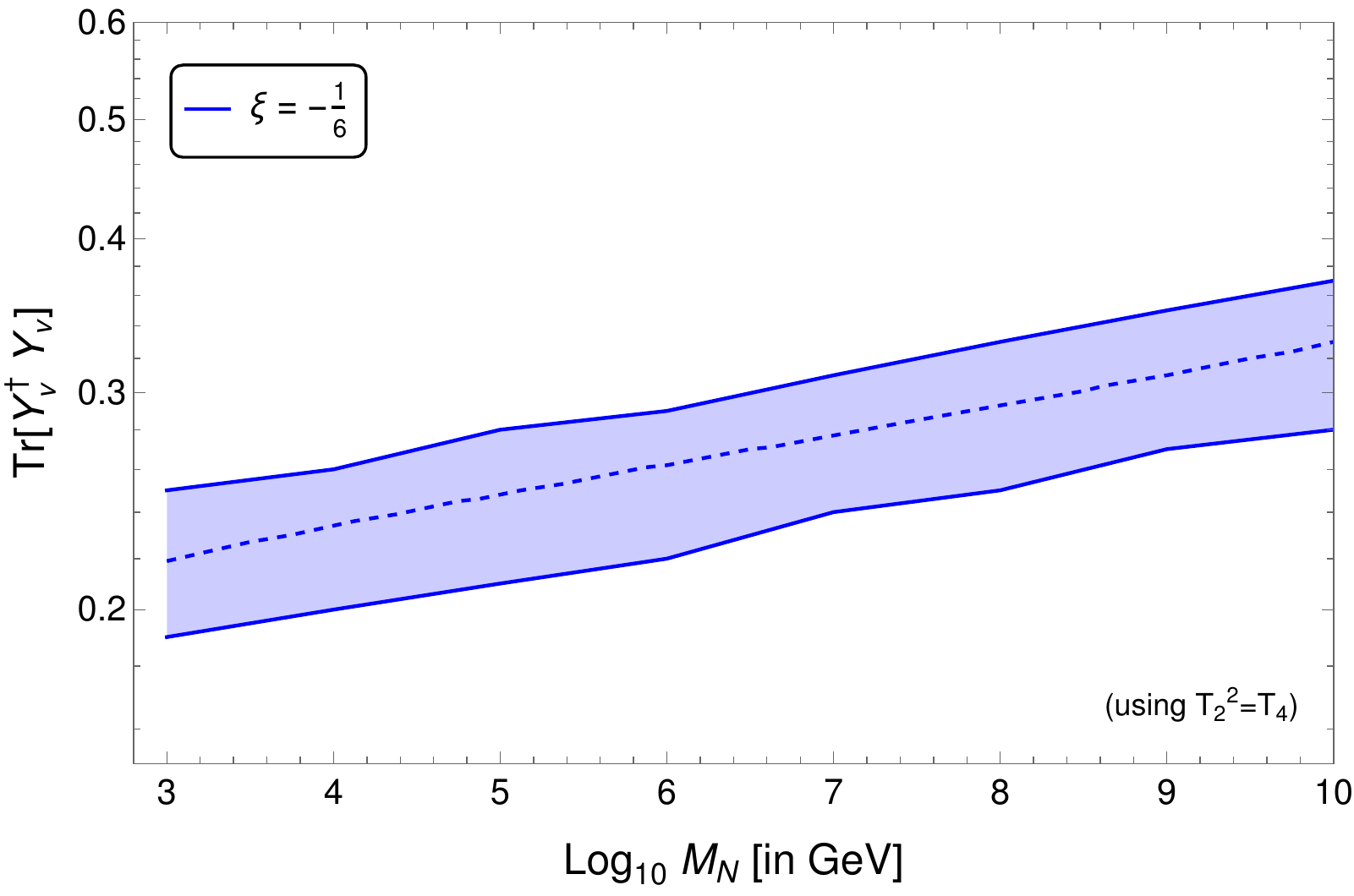}
     \includegraphics[width=0.49\textwidth]{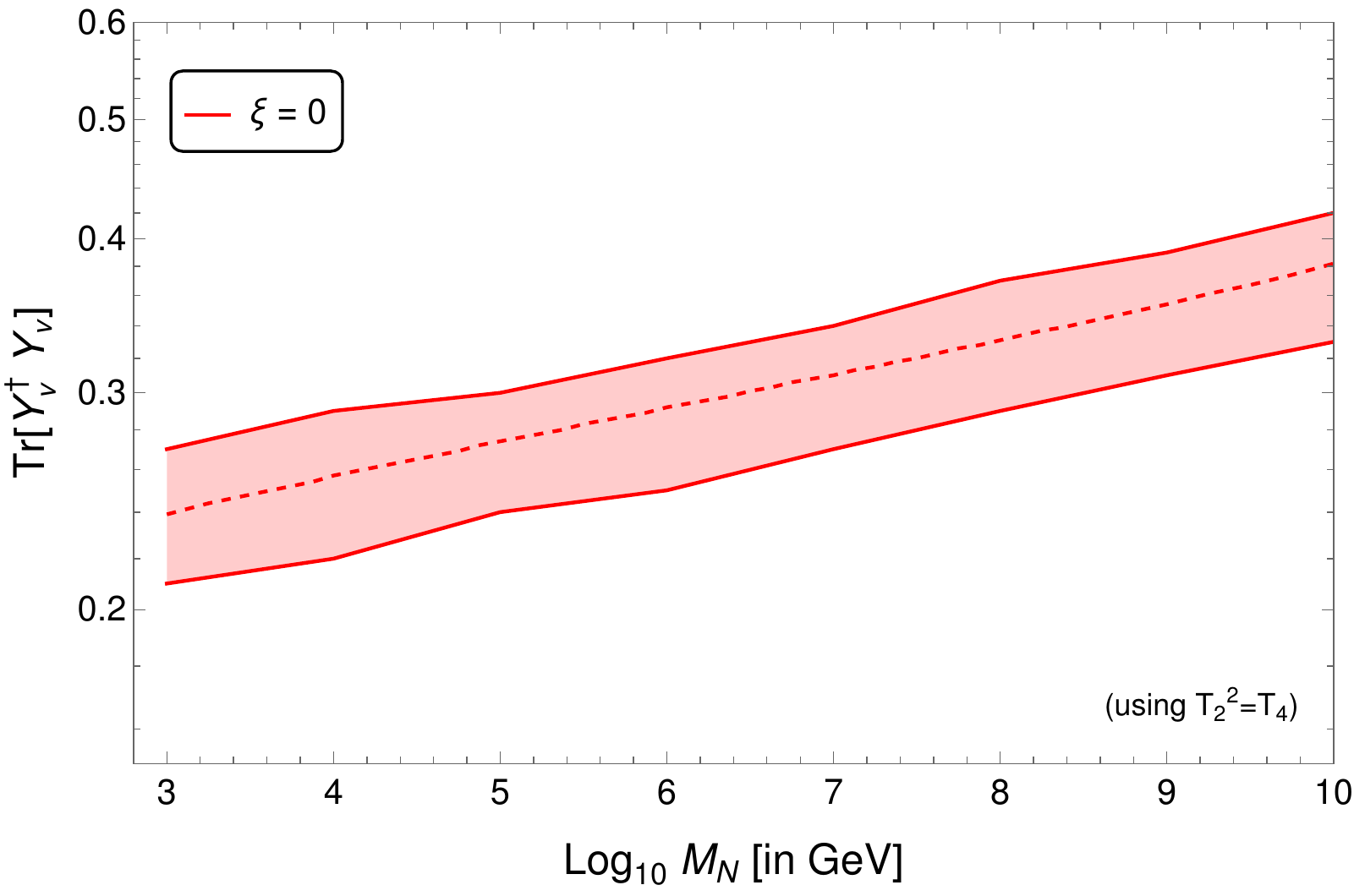} \\
     \includegraphics[width=0.49\textwidth]{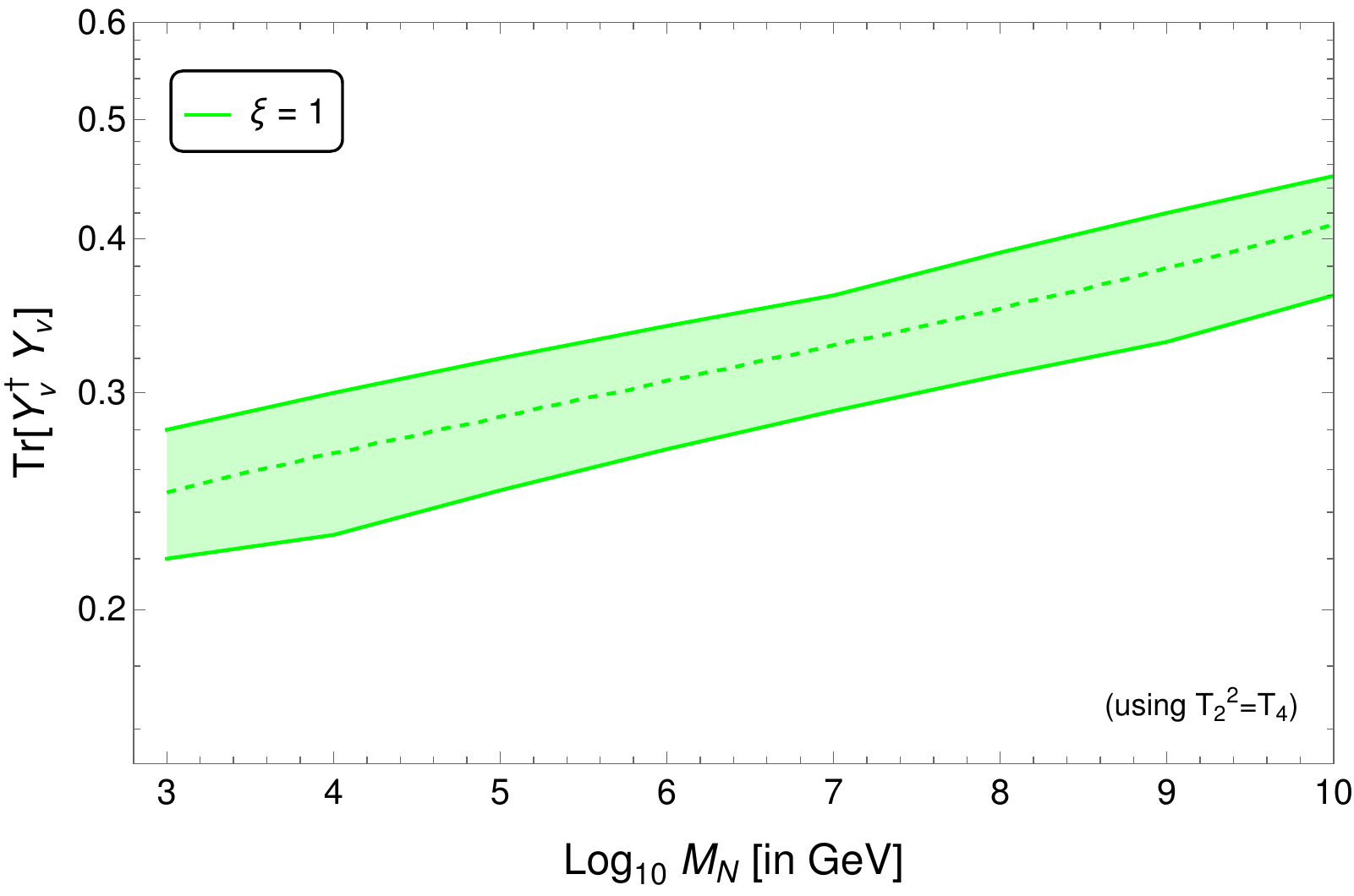}
     \includegraphics[width=0.49\textwidth]{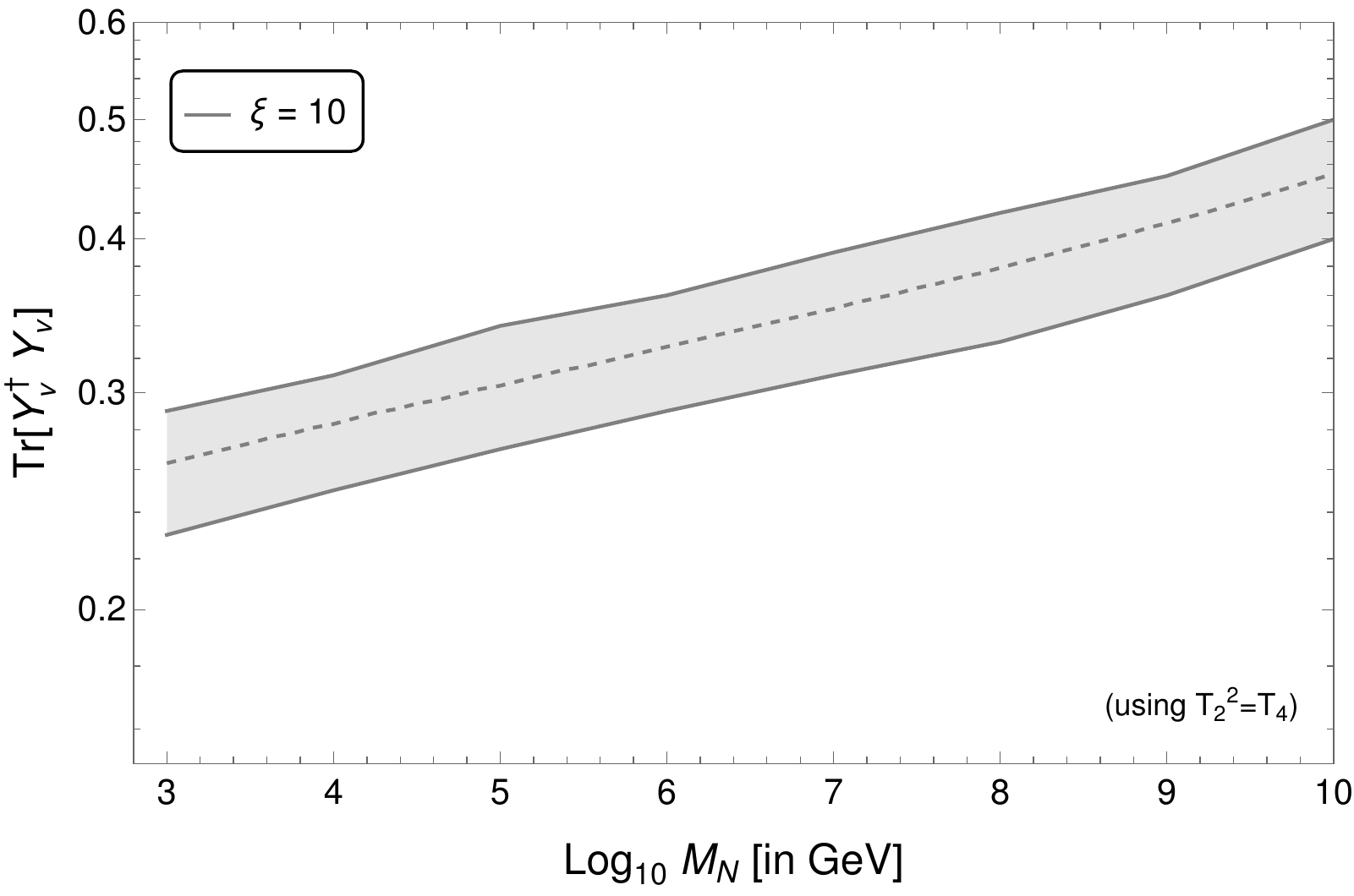} 
     \caption{The $1\sigma$ error band arising from uncertainties in SM parameters for the metastability bound on $\text{Tr}(Y_\nu^\dagger\, Y_\nu)$ as a function of degenerate RHN mass scale $M_N$ for non-minimal couplings $\xi = \left(-\frac{1}{6},0,1,10\right)$. Dashed lines denote the central values also shown in Fig.~\ref{fig:lowSapprox}. }
     \label{fig:errBar}
\end{figure}

We present our results for the $1\sigma$ error band in Fig.~\ref{fig:errBar} for different values of $\xi$, with the dashed line corresponding to the central values of the SM couplings.\footnote{We have also included the error bar in the right panel of Fig.~\ref{fig:lowSapprox}.} Our analysis shows that for any given non-minimal coupling constant $\xi$, the experimental uncertainties imply significant error bars for the metastability bounds already at $68\%$ confidence. Most importantly in the context of our work, we find that these are comparable to the effects of the gravitational corrections we consider. Thus, we find that the existing metastability bounds do not only suffer from a previously ignored theoretical uncertainty arising from the non-minimal coupling to gravity, but also a large experimental error bar. This sensitivity on the couplings' initial values does, on the other hand, also imply that this error bar can be very efficiently reduced through the increased precision of future experiments.

\subsection{Impact on the instability scale}

Following our discussion in Sec.~\ref{NuDecs}, RHNs drive $\lambda$ to smaller values at energies above $M_N$, where they influence its RG running. If this occurs at energies where $\lambda$ is positive, it results in a lower value of the instability scale $\mu_I$ compared to its SM value $\mu_I^{\text{SM}}\sim 10^{11}$~GeV. Although the precise value of this scale is irrelevant for the numerical calculation of the decay rate as long as $\mu_I \ll \mu_S$, it is nevertheless of central importance for related questions.

First, the very idea of metastability bounds depends entirely on the assumption that the vacuum can actually decay. This is, however, a highly non-trivial statement, as it is well-known that there exists a wide class of mechanisms capable of stabilizing the vacuum. The enormous range of scales between $\mu_I^{\text{SM}}$ and the limits of current observations thus allows for numerous scenarios capable of seriously undermining all arguments involving vacuum decay. 

This possible problem could be resolved if the instability scale were lowered to values closer to the electroweak scale, where its consequences could be probed by future detectors. While it has been shown that this would require couplings far exceeding our metastability bounds, it has also been shown that this can be compensated for through the previously mentioned stabilizing effects, e.g., the introduction of any kind of new physics which can be effectively described through a dimension-six operator~$\frac{C_6}{\Lambda_{\text{UV}}^2}$~\cite{Giudice:2021viw,Khoury:2021zao,Steingasser:2022yqx}. In such cases the metastability bounds can be understood as a lower bound on the effectiveness of this additional effect. The generalization of our analysis to such cases is straightforward~\cite{Khoury:2021zao,Steingasser:2022yqx}, and schematically amounts to a replacement $(1+6 \xi)^2 M_{\text{Pl}}^{-2} \to C_6 \Lambda_{\text{UV}}^{-1}$ up to numerical coefficients.

Another important property of the instability scale is its relation to the hierarchy problem. It was first pointed out in~\cite{Buttazzo:2013uya} that the SM symmetry breaking pattern requires the electroweak scale to be smaller than $\mu_I$ by roughly one order of magnitude. This observation has recently proven crucial in attempts to solve the hierarchy problem via independent vacuum selection arguments~\cite{Buttazzo:2013uya,Khoury:2021zao,Denef:2017cxt,Khoury:2019yoo,Khoury:2019ajl,Kartvelishvili:2020thd,Khoury:2021grg,Giudice:2021viw,Steingasser:2022yqx}. Due to the nature of their underlying bound, their effectiveness at actually solving the hierarchy problem hinges on the possibility of a significantly lower instability scale.

While such a lowering can be achieved through the addition of any kind of fermion with a Yukawa coupling to the Higgs (see~\cite{Belfatto:2023tbv} for another recent example), RHNs serve as one of the arguably most natural candidates. Following our discussion in Sec.~\ref{sec:seesawmech}, the neutrinos' effect on the instability scale can be expected to be largest for the kind of scenarios discussed in this section.
\begin{figure}[t!]
    \centering
    \includegraphics[width=0.7\textwidth]{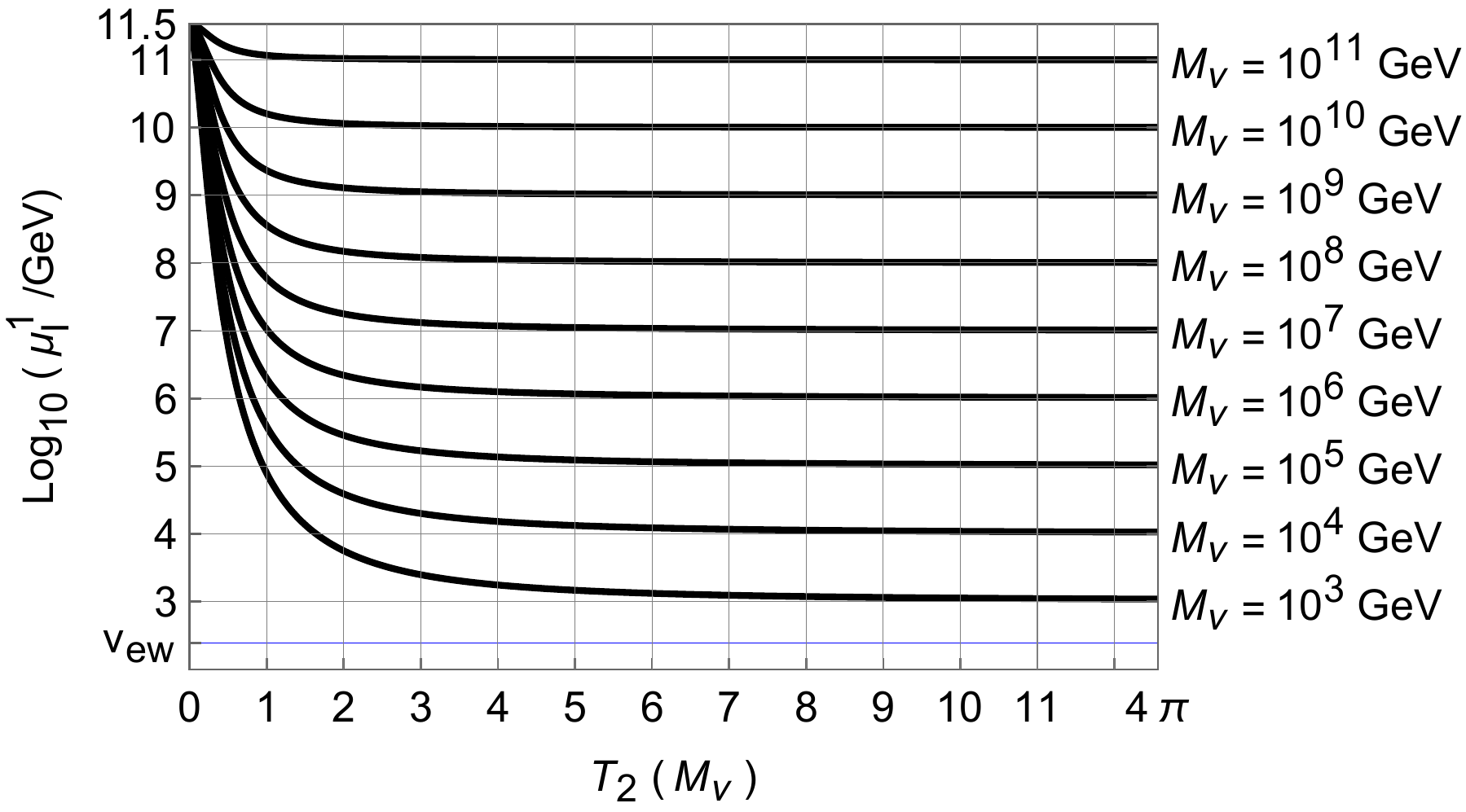}
    \caption{The 1-loop instability scale $\mu_I^{(1)}$ as a function of the parameter $T_2$ at the matching scale for different values of the degenerate RHN mass. Note that we performed this scan over the entire parameter space, including unstable or non-perturbative regions, which could be rendered viable by the effects of other new physics. }
    \label{muiplot}
\end{figure} 
In Fig.~\ref{muiplot}, we present the one-loop instability scale $\mu_I^{1}$ as a function of $T_2\equiv \text{Tr} (Y_{\nu}^\dagger Y_{\nu})$ at the matching scale for different values of $M_\nu$. This scale is defined through the effective quartic coupling including one-loop corrections as
\begin{equation}
    \lambda_{\text{eff}}^{\text{1-loop}}\left(\mu = \mu_I^{1} ,h = \mu_I^{1}\right)=0.
\end{equation}
We find that significantly lowering the instability scale requires $T_2$ to take values of $\mathcal{O}(0.1-1)$, which typically correspond to an unstable vacuum or even a non-perturbative running of the couplings. Thus, values of $\mu_I$ close to current observational bounds are only consistent with the metastability bound Eq.~\eqref{eq:mscond} if the neutrinos' destabilizing effect is at least partially compensated for by additional BSM physics, which could also restore perturbativity.

\section{High-Scale Type-I Seesaw}
\label{sec:highscale}
As discussed in the previous section, the low-scale seesaw mechanism requires degenerate RHN masses to ensure radiative stability of the seesaw relation. Without enforcing the degeneracy in $M_N$, large Yukawas are possible only for the high-scale seesaw in Type-I seesaw mechanism. High-scale seesaw models are a natural alternative to low scale seesaws featuring hierarchical RHN mass scales. A salient feature is the generation of the baryon asymmetry $\eta_B$ without requiring finely tuned $M_N$s (whereas in low-scale leptogenesis, they need to exhibit a quasi-degenerate spectrum for the production of $\eta_B$). Although these heavy RHNs are not directly accessible in current experiments due to the high mass scales involved, these scenarios can be ruled out through the observation of lepton number violating (LNV) signals at the LHC~\cite{Deppisch:2013jxa}. 

To explain the observed $\eta_B$ in the type-I seesaw, it has been shown that thermal leptogenesis provides a lower bound on the mass of the lightest RHN, $M_1 \gtrsim 10^9$ GeV (assuming an $\mathcal{O}(1)$ efficiency factor), known as Davidson-Ibarra bound\footnote{This bound can be relaxed to some degree by $\mathcal{O}(10)$ fine-tuned cancellations between the tree-level and loop-level light neutrino masses. This would however still imply that even the lightest RHN would still lie far beyond the range of near-future experiments~\cite{Moffat:2018wke}.}~\cite{Davidson:2002qv}. It has also been shown that high-scale leptogenesis is viable only for $M_1$ up to $10^{15}$ GeV~\cite{Buchmuller:2002rq,Buchmuller:2004nz}. Therefore, in our parameter scan, we will consider the range $M_1= (10^9-10^{15})$~GeV for the scenarios of 2 as well as 3 RHNs.  

In this section, we will first discuss the RG evolution in the case of non-degenerate RHN masses. We then discuss the results of the parameter scan for the metastable electroweak vacuum in presence of gravity for the high-scale type-I seesaw with both 2 and 3 RHNs. Finally we discuss the implications of these results for high-scale leptogenesis scenarios.

\subsection{RG evolution}
Non-degenerate RHN masses also require a modification of the RG running of the relevant couplings. It is well-understood that the influence of each of these particles manifests only at RG scales above the matching scales $\mu_{\text{match},i}\sim M_{N_i}$. This can be embedded into the standard perturbative framework by using multiple \textit{effective theories} at different scales. The simplest example for this is the SM extended by the dimension-five Weinberg mass term~\cite{Casas:1999tg,Antusch:2003kp} for the light neutrinos, which can be understood as an effective theory of the full seesaw model obtained by integrating out all RHNs\footnote{We will throughout our analysis consistently neglect the effect of this operator on the SM couplings as it is suppressed by a factor of $\mu^2/M_{N}^{-2}$, and hence, small in the relevant regime $\mu \ll M_N$.}. The running of $\kappa$ could, in principle, also be of importance for our analysis, especially for high-scale seesaw models: While the observational input data we use to reduce the number of relevant parameters in our analysis is usually obtained at the mass scale of the light neutrinos, we require its value at the matching scales to determine the corresponding RHN parameters. To understand the effect of the running between these two scales, it is beneficial to consider the beta function for $\kappa$,
\begin{equation}
    \beta_\kappa=\frac{1}{(4\pi)^2} \left( 6y_t^2 \kappa -3 g_2^2\,\kappa + \lambda\, \kappa  \right)
\end{equation}
This implies that, to leading order, the running of the mass eigenvalues is independent of the mixing parameters and only amounts to a rescaling of $\kappa$, and thus, the light neutrino masses. For our purposes, this scaling can simply be compensated for by rescaling the complex orthogonal matrix $R$. While this simplifies the presentation of our results significantly, it also makes an additional step necessary if one wanted to express them in terms of internal parameters.

As can be expected due to presence of threshold corrections and modification of $Y_\nu$ at each RHN mass threshold, it is not possible to implement the trace approximation or $T_n$-parametrization approach to simplify the RG running analysis. Therefore, we use the full matrix form of $Y_\nu$ for the RG evolution in this section.

\subsection{Metastability bounds - 2 RHNs}

\begin{figure}[t!]
    \centering
    \includegraphics[width=0.49\textwidth]{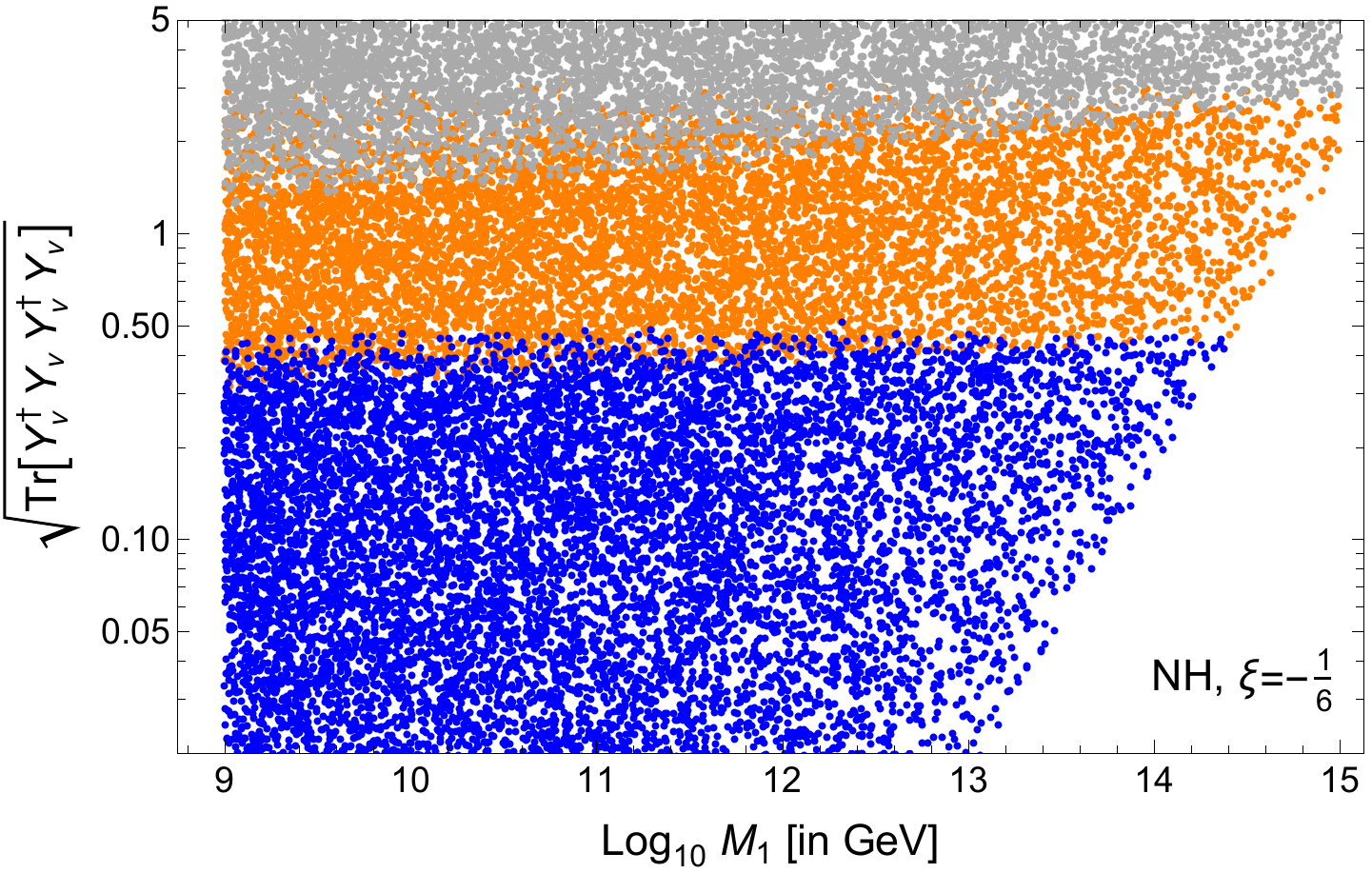}
    \includegraphics[width=0.49\textwidth]{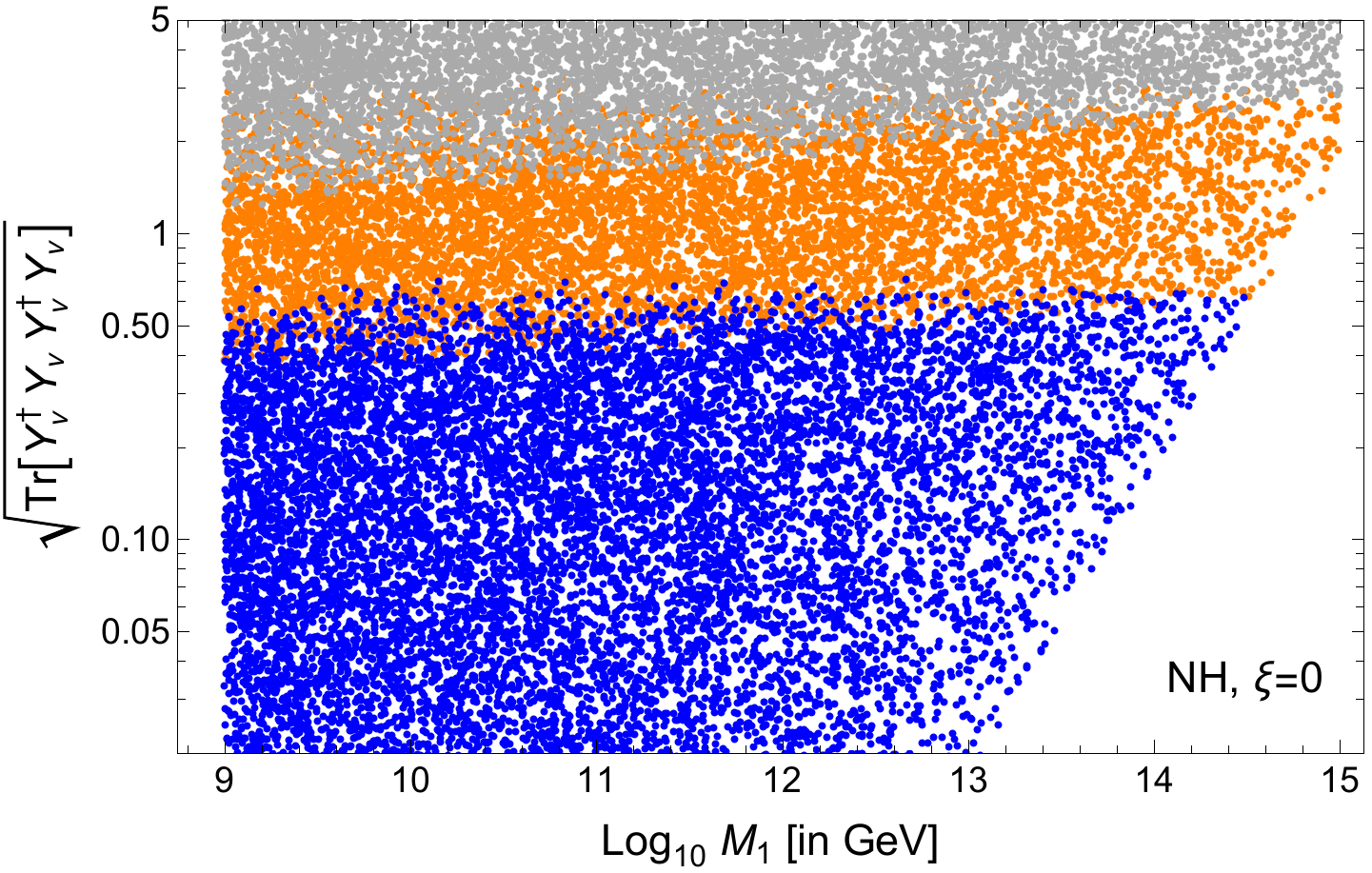}\\
    \includegraphics[width=0.49\textwidth]{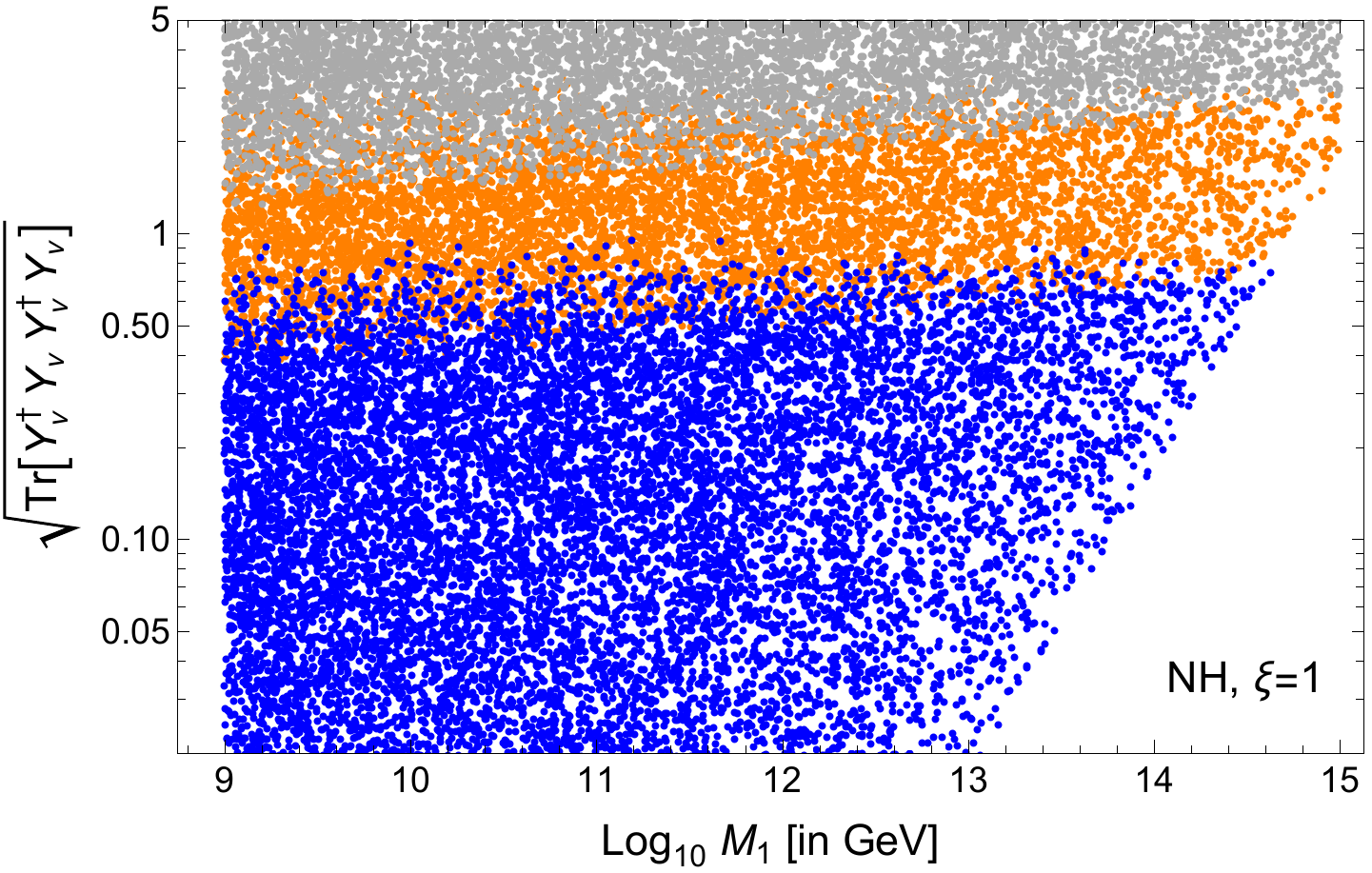}
    \includegraphics[width=0.49\textwidth]{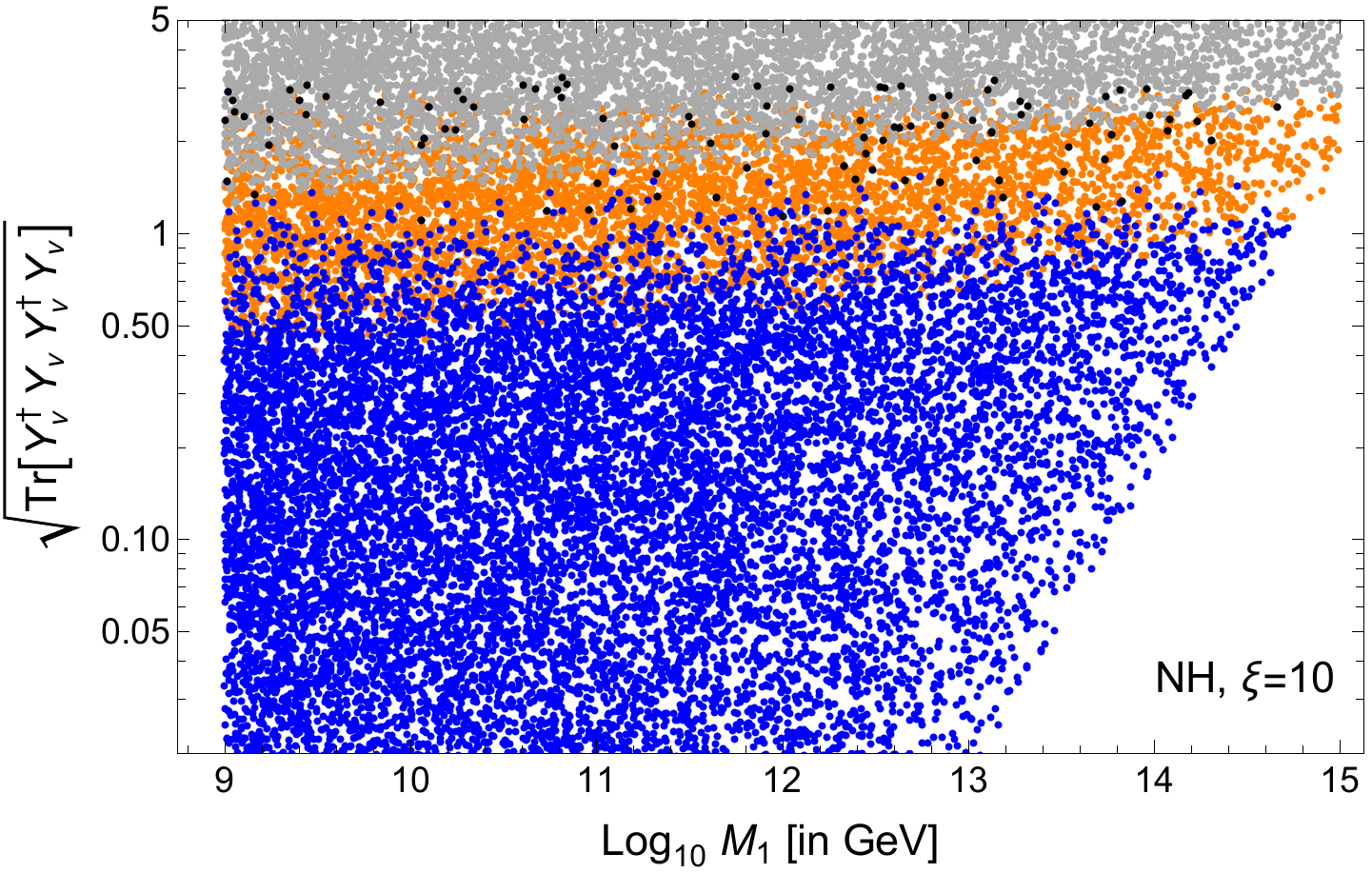}
    \caption{(2 RHNs) Scatter plot of $\sqrt{T_4}$ as a function of lightest RHN mass scale $M_1$ for Normal Hierarchy (NH) and non-minimal coupling $\xi = \left(-\frac{1}{6},0,1,10\right)$. Blue points correspond to a metastable electroweak vacuum while the orange points signify an unstable electroweak vacuum. Gray and black points mark non-perturbative behavior in the loop-expansion and gravitational corrections respectively.}
    \label{fig:NHT2m1}
\end{figure}

\begin{figure}[t!]
    \centering
    \includegraphics[width=0.49\textwidth]{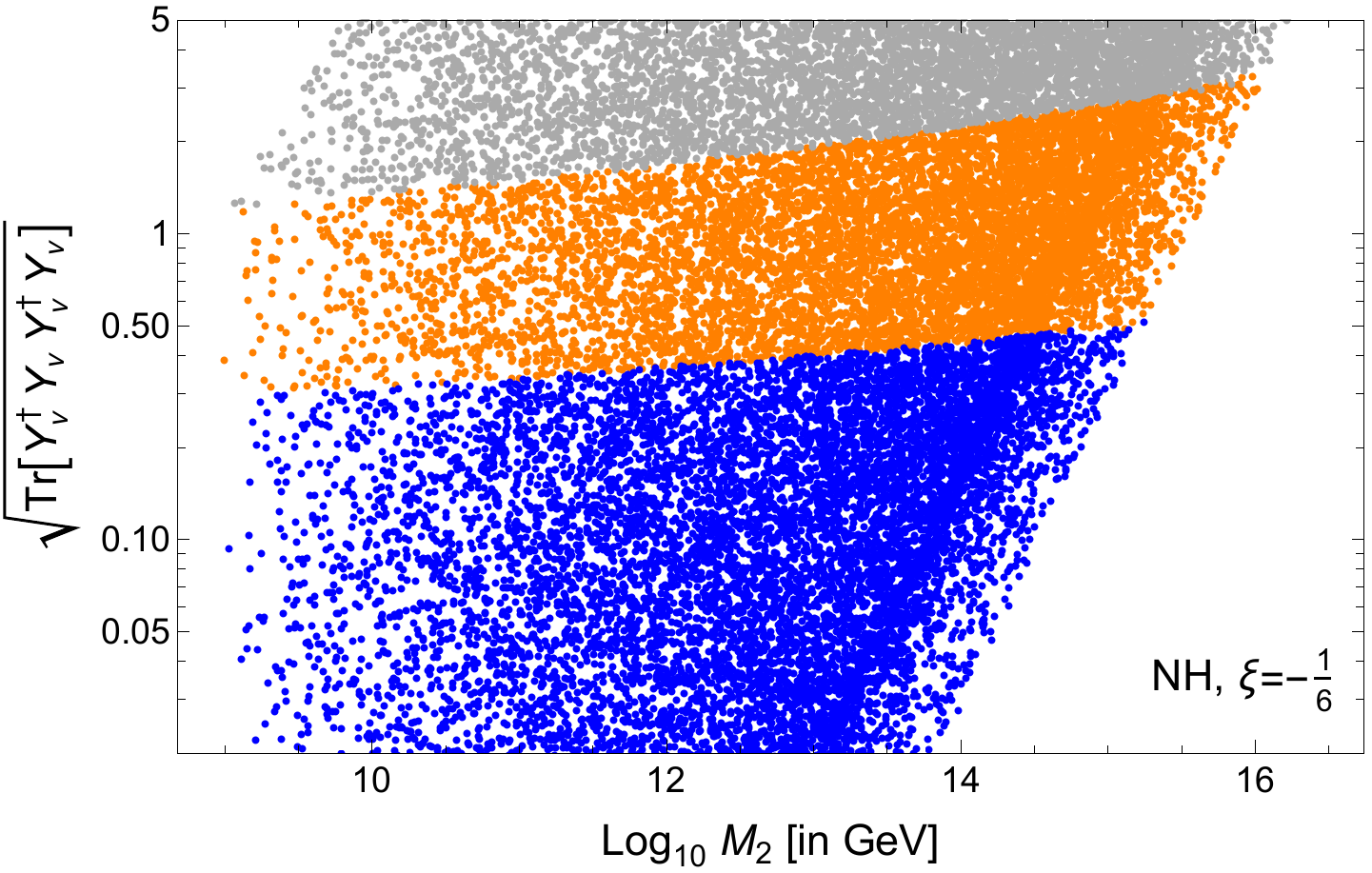}
    \includegraphics[width=0.49\textwidth]{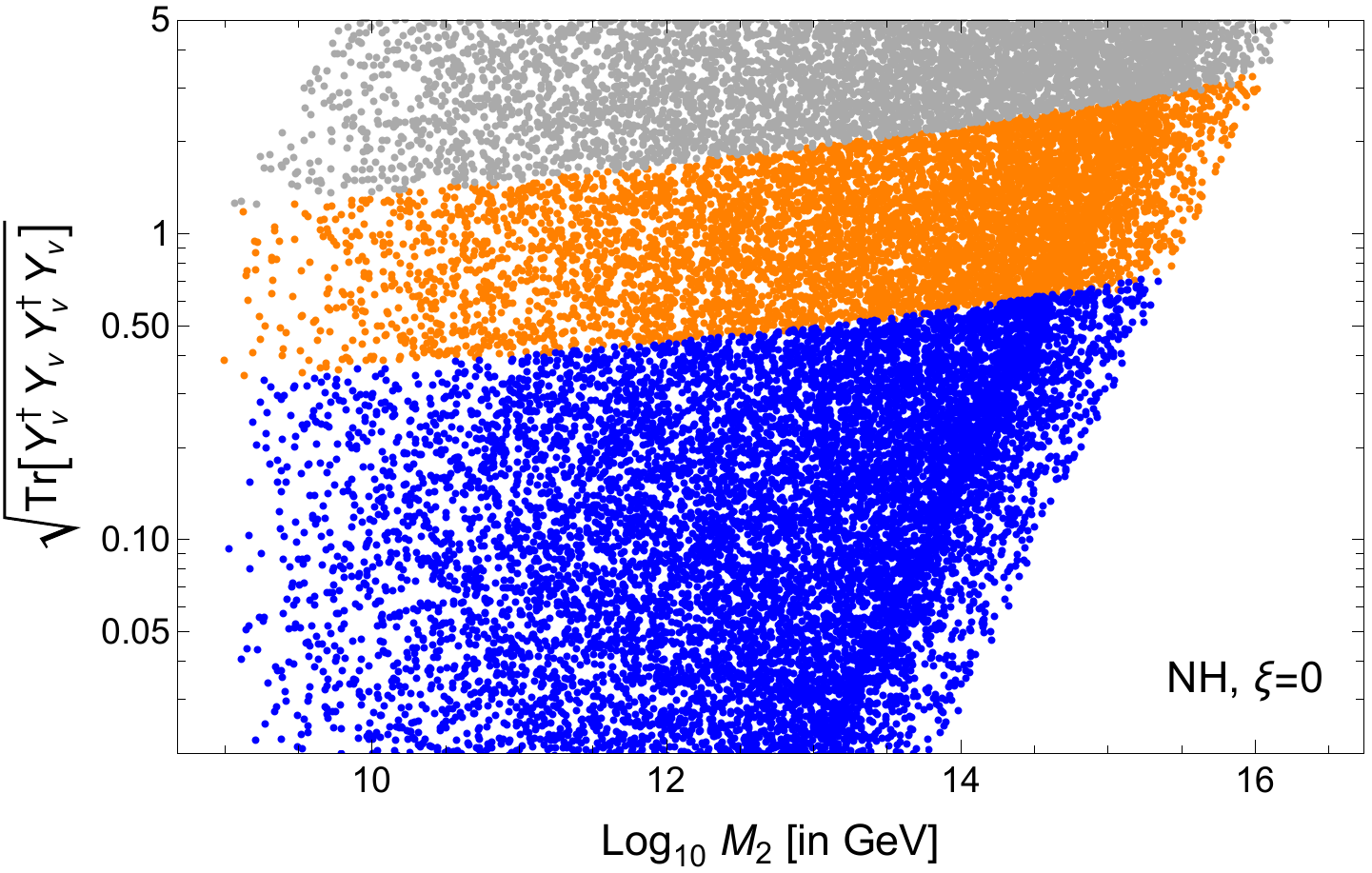}\\
    \includegraphics[width=0.49\textwidth]{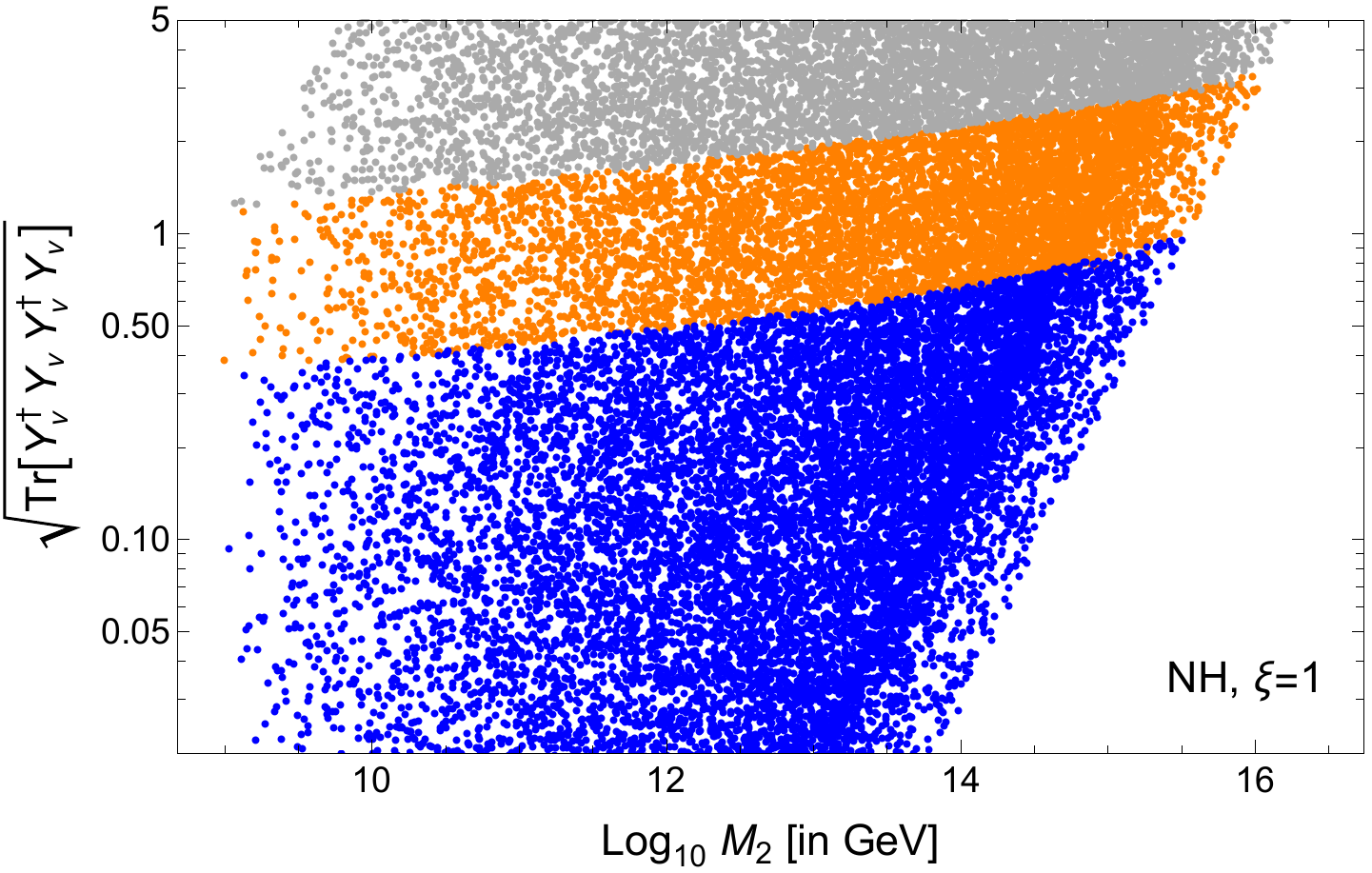}
    \includegraphics[width=0.49\textwidth]{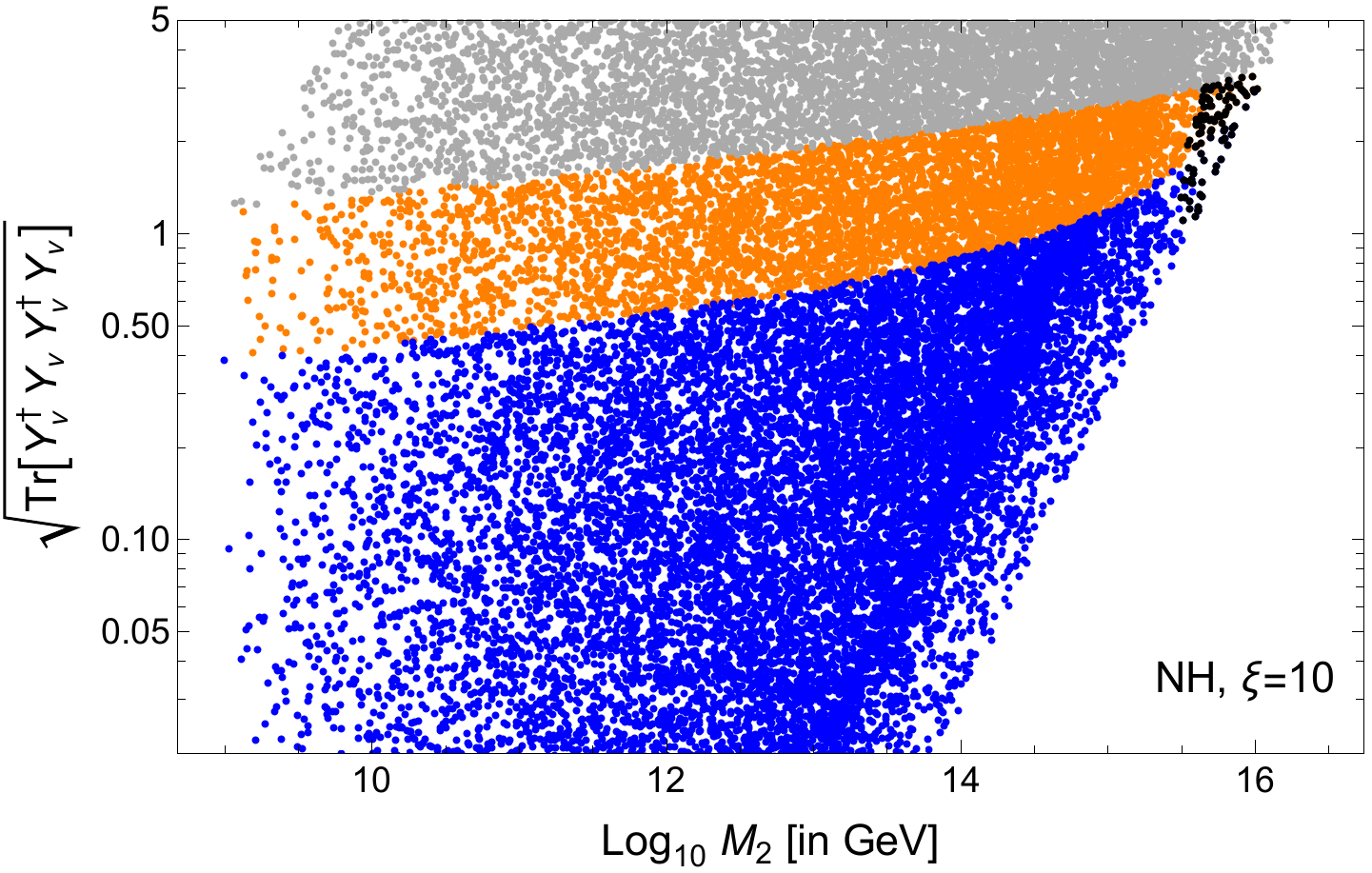}
    \caption{(2 RHNs) Scatter plot of $\sqrt{T_4}$ as a function of heaviest RHN mass scale $M_2$ for Normal Hierarchy (NH) and non-minimal coupling $\xi = \left(-\frac{1}{6},0,1,10\right)$. Blue points correspond to a metastable electroweak vacuum while the orange points signify an unstable electroweak vacuum. Gray and black points mark non-perturbative behavior in the loop-expansion and gravitational corrections respectively.}
    \label{fig:NHT2m2}
\end{figure}

We first apply the procedure described in the previous subsection to scenarios with two RHNs with relevant couplings at the scales of interest. As for the low-scale seesaw models, we ensure the applicability of the perturbative expansions underlying our analysis by keeping track of the couplings and $\epsilon_{\text{grav}}$. Restricting ourselves to scenarios with successful thermal leptogenesis, we scan the RHN masses over $M_1 \in [10^9\text{GeV},10^{15}\text{GeV}]$ and $M_2 \in [M_1,10^{19}\text{GeV}]$. We parametrize the $3\times 2$ complex orthogonal matrix $R$ as
\begin{equation}
R_{\text{NH}} = \left( \begin{array}{ccc}
 0 & \cos{z} & \sin{z}\\
 0 & -\sin{z} & \cos{z}\\
\end{array}
\right) \, ,\\ \,
R_{\text{IH}} = \left( \begin{array}{ccc}
 \cos{z} & \sin{z} & 0\\
 -\sin{z} & \cos{z} & 0\\
\end{array}
\right) \, ,
\end{equation}
where $z$ is a complex parameter with range $\text{Re}(z) \in \left[ -\pi, \pi\right]$ and $\text{Im}(z) \in \left[ -\pi, \pi\right]$.

In Fig.~\ref{fig:NHT2m1}, we display the results of our parameter scan for $T_4$ as a function of the lightest mass scale $M_1$ for NH and different values of the non-minimal coupling $\xi$. We find a clear separation of the metastable, unstable and non-perturbative regions in terms of $T_4$, in agreement with our discussion in Sec.~\ref{relpar}. The fuzziness of the boundary regions is a consequence of the projection onto the $T_4$-$M_1$ plane, and we will find a clear separating line when moving to the $T_4$-$M_2$ plane. Our scan reveals that the metastability bound does not explicitly depend $M_1$. As, however, $M_1$ is bounded from above by phenomenological constraints for any given $T_4$ and the latter \textbf{is} restricted by metastability, our results also imply an upper bound on $M_1$, in agreement with previous works~\cite{Casas:1999cd,Elias-Miro:2011sqh, Ipek:2018sai}. 

\begin{figure}[t!]
    \centering
    \includegraphics[scale=0.7]{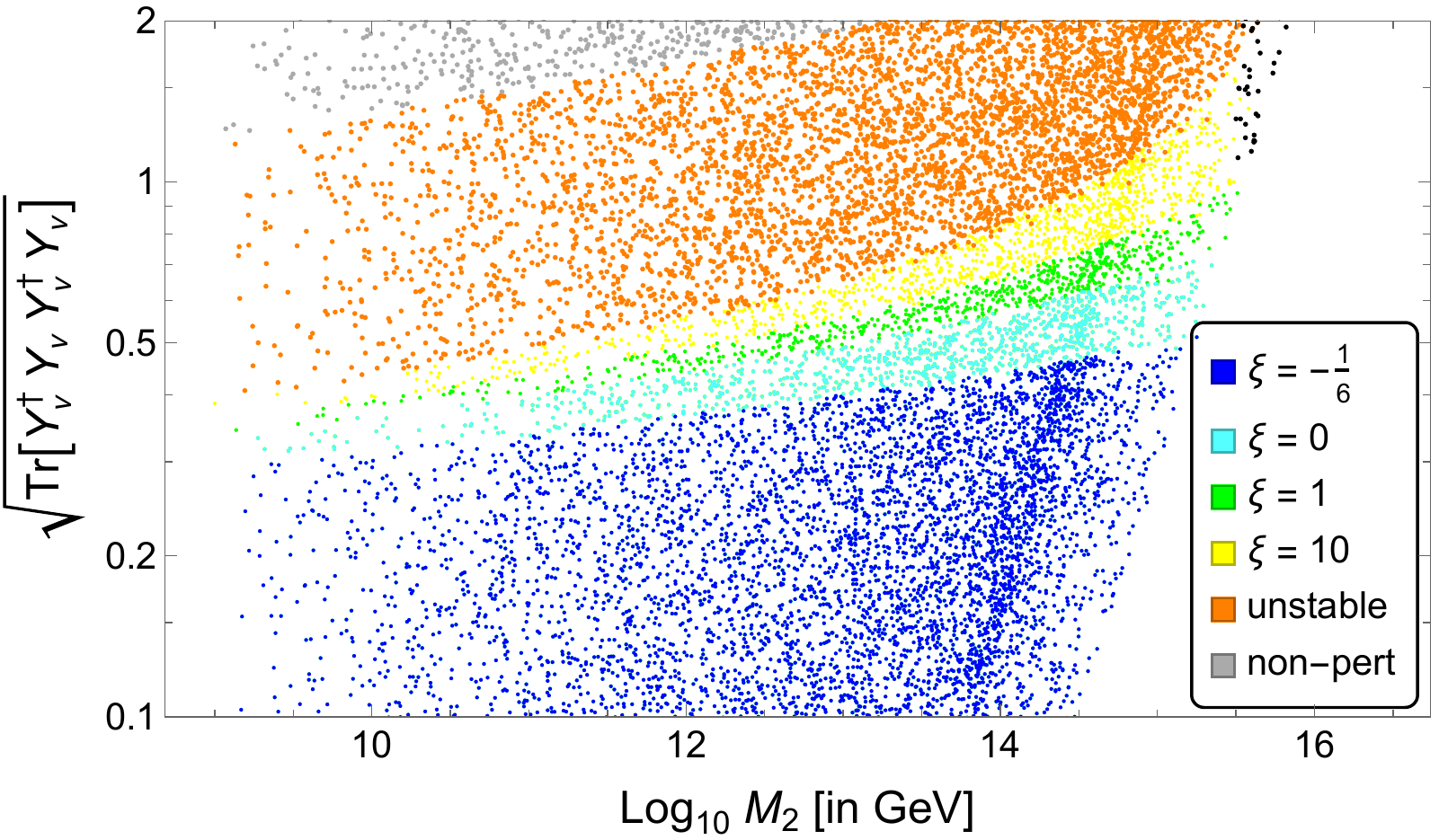}
    \caption{(2 RHNs) Scatter plot of $\sqrt{T_4}$ against the heaviest RHN mass scale $M_2$, Normal Hierarchy and non-minimal coupling $\xi = \left(-\frac{1}{6},0,1,10\right)$. Colored points (except orange) correspond to an electroweak vacuum which is stable for their corresponding $\xi$ (and all larger values). Orange points signify an unstable electroweak vacuum for all considered values of $\xi$. Gray and black points mark non-perturbative behavior in the loop-expansion and gravitational corrections respectively.}
    \label{fig:NHAllscatter}
\end{figure}

We once again notice that gravitational effects loosen the bound on $T_4$, and therefore, also on $M_1$. This effect is illustrated in Fig.~\ref{fig:NHAllscatter}, which presents an overlay of our upper bounds on $T_4$ as a function of $M_3$. For the NH (IH) case without gravity ($\xi=-\frac{1}{6}$), the upper bound on $\sqrt{T_4}$ is around 0.51 (0.46) and $M_1$ is restricted to values below $10^{14.4}$~GeV ($10^{14.2}$~GeV). Meanwhile, for the NH (IH) case with gravity and $\xi=10$, the bound on $\sqrt{T_4}$ gets relaxed to 1.64 (1.09), while that on $M_1$ is loosened to $10^{14.7}$~GeV ($10^{14.5}$~GeV). 

In addition, we observe that gravitational effects seem to amplify the fuzziness of the transition between the different regions, especially for larger $\xi$s. This can also be best understood by keeping in mind that Fig.~\ref{fig:NHT2m1} only represents a projection from the higher-dimensional parameter space. Comparing these results with Fig.~\ref{fig:NHT2m2}, it is easy to see that this behavior is just a manifestation of a distinct feature in the $T_4-M_2$-plane.

The dependence of $T_4$ on $M_2$ is shown in Fig.~\ref{fig:NHT2m2}. Unlike for $M_1$, we find a clear dependence on $M_2$. This is due to the fact that heavier RHNs have a larger contribution to $Y_\nu$ and $T_4$, and thus also the running of $\lambda$. This dependence also explains the amplification of the gravitational stabilization for large $M_2$ and $\xi$, most clearly visible in the plot for $\xi=10$. This feature can be understood from the determinant equation for the instanton scale $\mu_S$, Eq.~\eqref{eq:saddle}, where $\mu_S$ appears either multiplied with $(1 + 6 \xi)^2$ or as the RG scale of running parameters. As the latter dependence is only logarithmic, we find that the instanton scale roughly scales as $\mu_S^2 \propto (1+6 \xi)^{-2}M_{\text{Pl}}^2$. Our numerical analysis shows that, near the transition from metastable to unstable region, the coefficient is such that roughly $\mu_S \sim 10^{16}$~GeV. Since the dominant contribution to $T_4$ is linked to the heaviest RHN, increasing $M_2$ to values close to $\mu_S$ drastically reduces the RHNs total impact on the running of $\lambda$ up to $\mu_S$, which scales with $\ln \left( \frac{\mu_S}{M_2}\right)$. This behavior is also responsible for the breakdown of our pertubative treatment of the gravitational effects for $\xi=10$ in this region.

We provide a summary of our results in Table~\ref{table:T2M2x2}. It is worth recalling that we assumed the lightest neutrino mass scale to be $m_0=0$. The bounds in each case will become stronger for non-zero $m_0$. 

\begin{table}
\centering
\begin{tabular}{ |c|c|c|c| } 
\hline
Case & {Max. $\sqrt{T_4}$ }  & {Max. $M_1$ (in GeV)} & {Max. $M_2$ (in GeV)} \\
\hline
\textbf{$\xi=-\frac{1}{6}$} & & & \\
NH & $0.51 $ & $ 10^{14.4}$& $ 10^{15.3}$ \\
IH & $0.46 $ & $ 10^{14.2}$ & $ 10^{14.4}$ \\
\hline
\textbf{$\xi=0$} & & & \\
NH & $0.71 $ & $ 10^{14.5}$& $ 10^{15.3}$ \\
IH & $0.63 $ & $ 10^{14.3}$ & $ 10^{14.6}$ \\
\hline
\textbf{$\xi=1$} & & & \\
NH & $0.95 $ & $ 10^{14.6}$& $ 10^{15.5}$ \\
IH & $0.76 $ & $ 10^{14.3}$ & $ 10^{14.7}$ \\
\hline
\textbf{$\xi=10$} & & & \\
NH & $1.64 $ & $ 10^{14.7}$& $ 10^{16.0}$ \\
IH & $1.09 $ & $ 10^{14.5}$ & $ 10^{14.8}$ \\
\hline
\end{tabular}
\caption{Summary of metastability bounds on $T_4$, $M_1$ and $M_2$ in the case of 2 RHNs and Normal Hierarchy (NH) as well as Inverse Hierarchy (IH).}
\label{table:T2M2x2}
\end{table}

\subsection{Metastability bounds - 3 RHNs}

Once again imposing the condition of successful thermal leptogenesis, we scan the masses over the range $M_1 \in [10^9\text{GeV},10^{15}\text{GeV}]$, $M_2 \in [M_1,10^{19}\text{GeV}]$ and $M_3 \in [M_2,10^{19}\text{GeV}]$. The $3\times 3$ complex orthogonal matrix $R$ can be parametrized as
\begin{equation}
R = \left( \begin{array}{ccc}
 1 & 0  & 0\\
 0 & \cos{z_1} & \sin{z_1}\\
 0 & -\sin{z_1} & \cos{z_1}\\
\end{array}
\right) \,
\left( \begin{array}{ccc}
\cos{z_2} & 0 & \sin{z_2}\\
 0 & 1  & 0\\
-\sin{z_2} & 0 & \cos{z_2}\\
\end{array}
\right) \,
\left( \begin{array}{ccc}
 \cos{z_3} & \sin{z_3} & 0  \\
-\sin{z_3} & \cos{z_3} & 0 \\
 0 & 0 & 1\\
\end{array}
\right) \,
\end{equation}
where $z_i\, (i=1,2,3)$ are complex parameters with ranges, $\text{Re}(z_i) \in \left[ -\pi, \pi\right]$ and $\text{Im}(z_i) \in \left[ -\pi, \pi\right]$.

\begin{figure}[t!]
    \centering
    \includegraphics[width=0.49\textwidth]{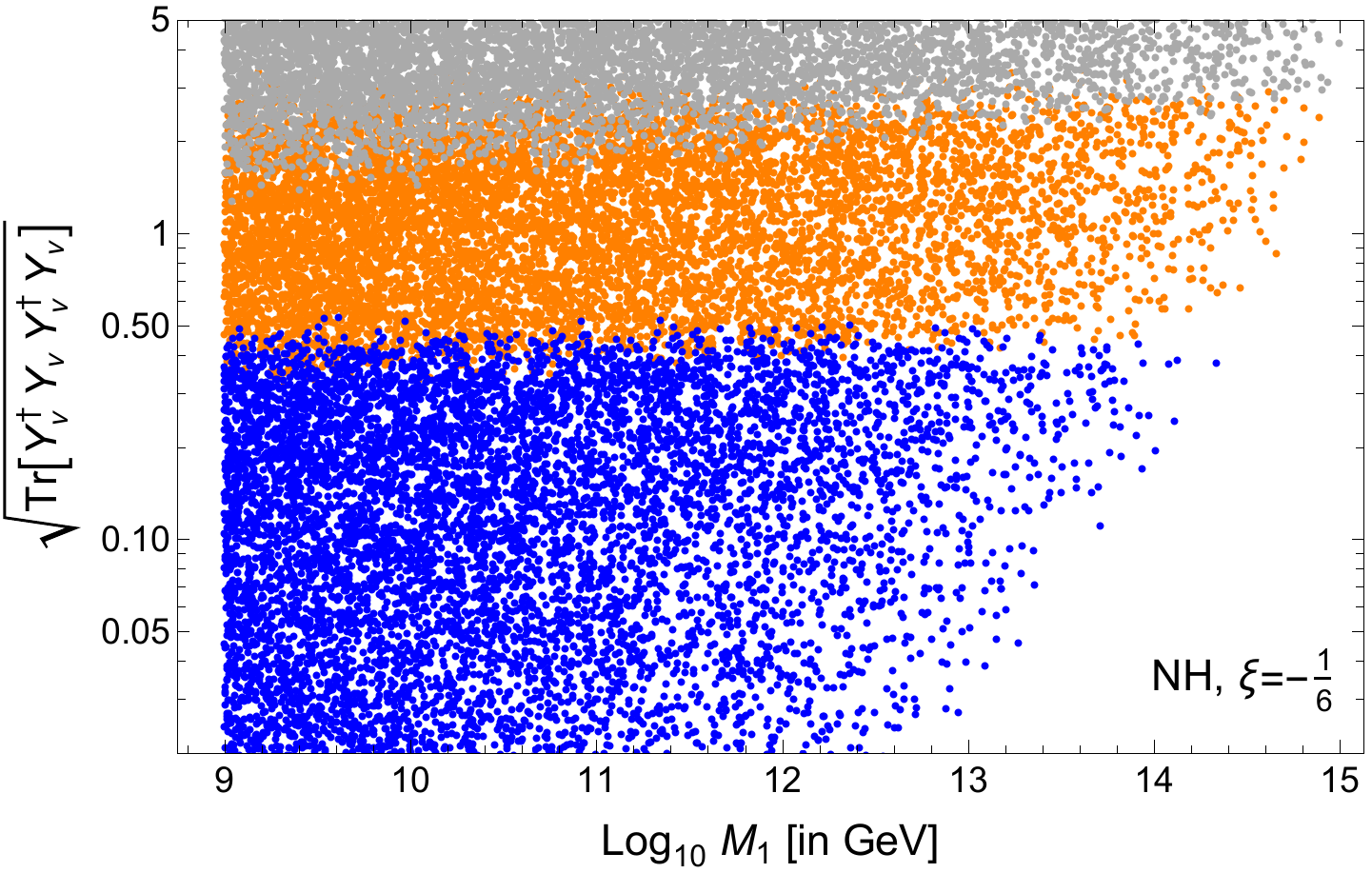}
    \includegraphics[width=0.49\textwidth]{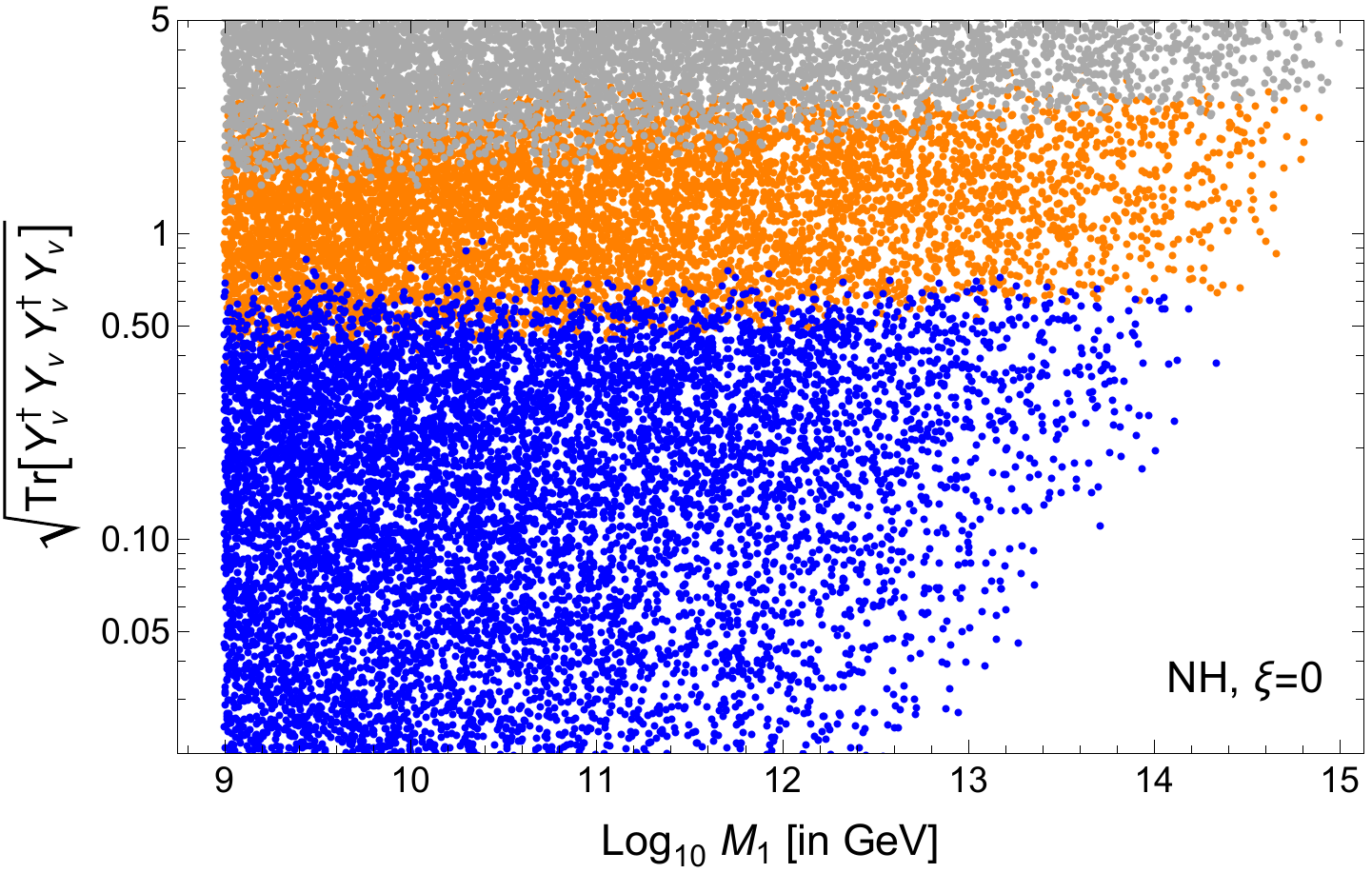}
    \\
    \includegraphics[width=0.49\textwidth]{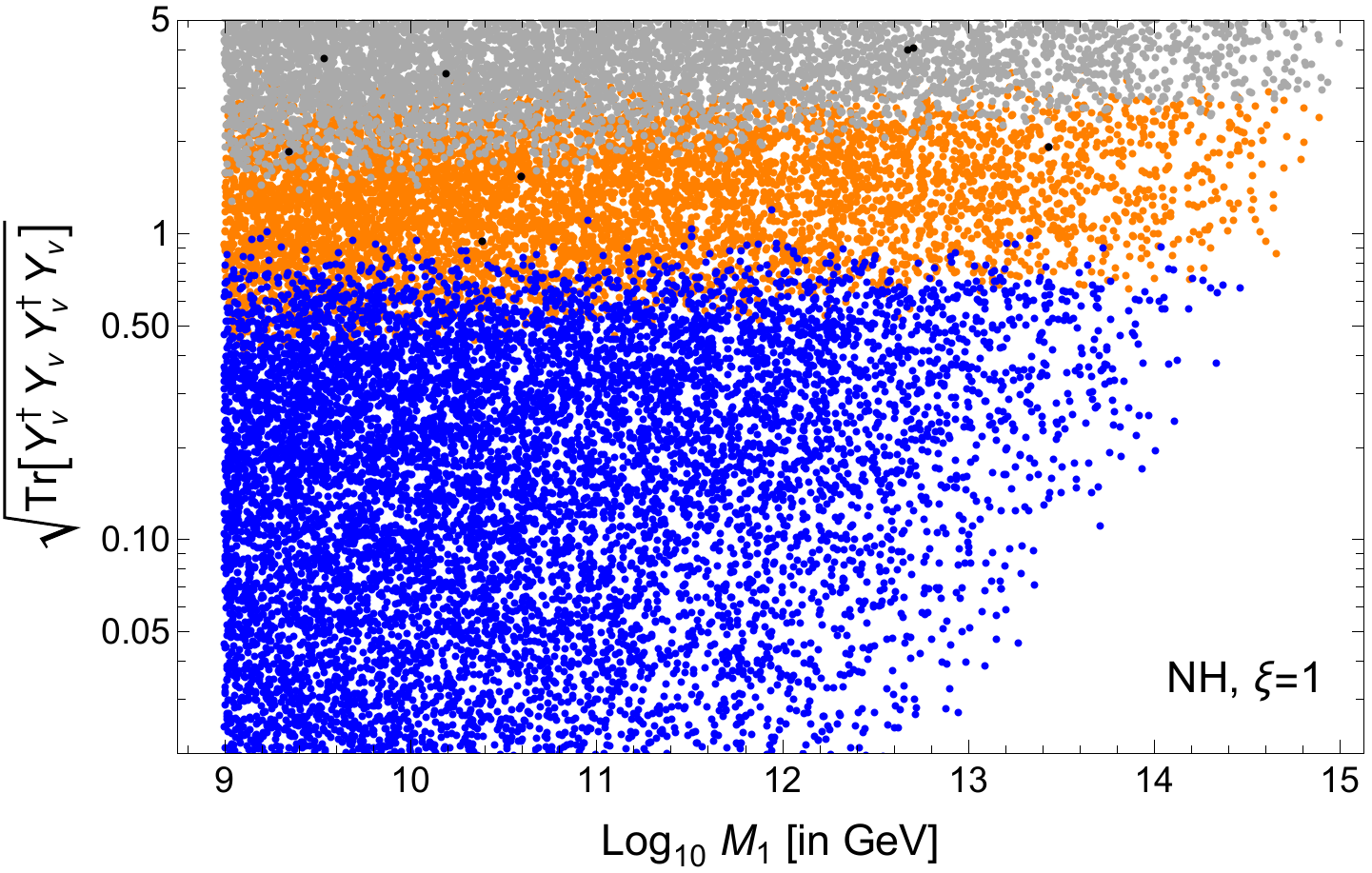}
    \includegraphics[width=0.49\textwidth]{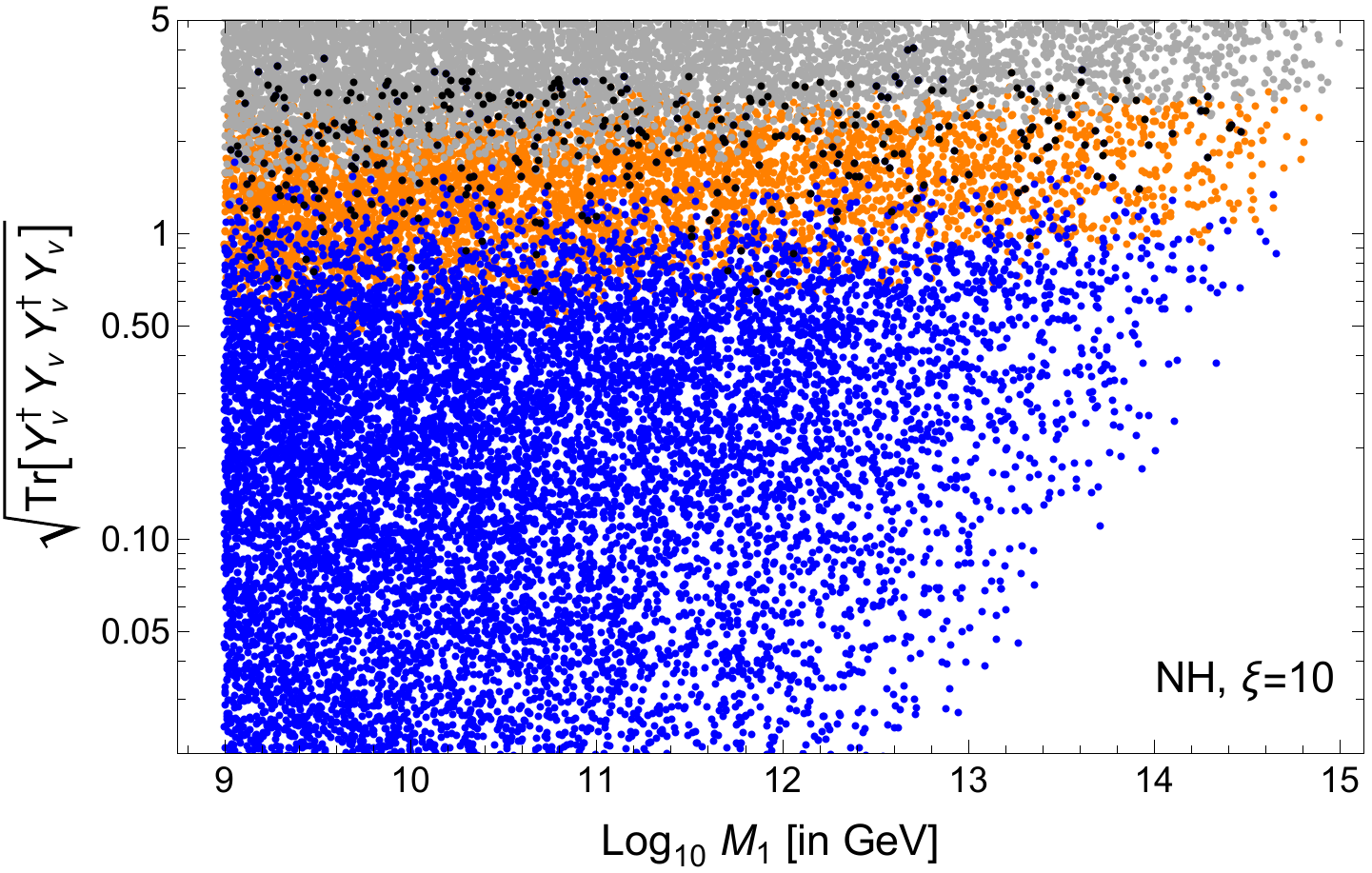}
    \caption{(3 Heavy $N$) Scatter plot of $\sqrt{T_4}$ as a function of lightest RHN mass scale $M_1$ (3 RHN case) for NH and non-minimal coupling $\xi = \left(-\frac{1}{6},0,1,10\right)$. Blue points correspond to metastable electroweak vacuum while the orange points signify unstable electroweak vacuum. Gray and black points mark non-perturbative behavior in the loop-expansion and gravitational corrections respectively.}
    \label{fig:NHT2m13x3}
\end{figure}

\begin{figure}[t!]
    \centering
    \includegraphics[width=0.49\textwidth]{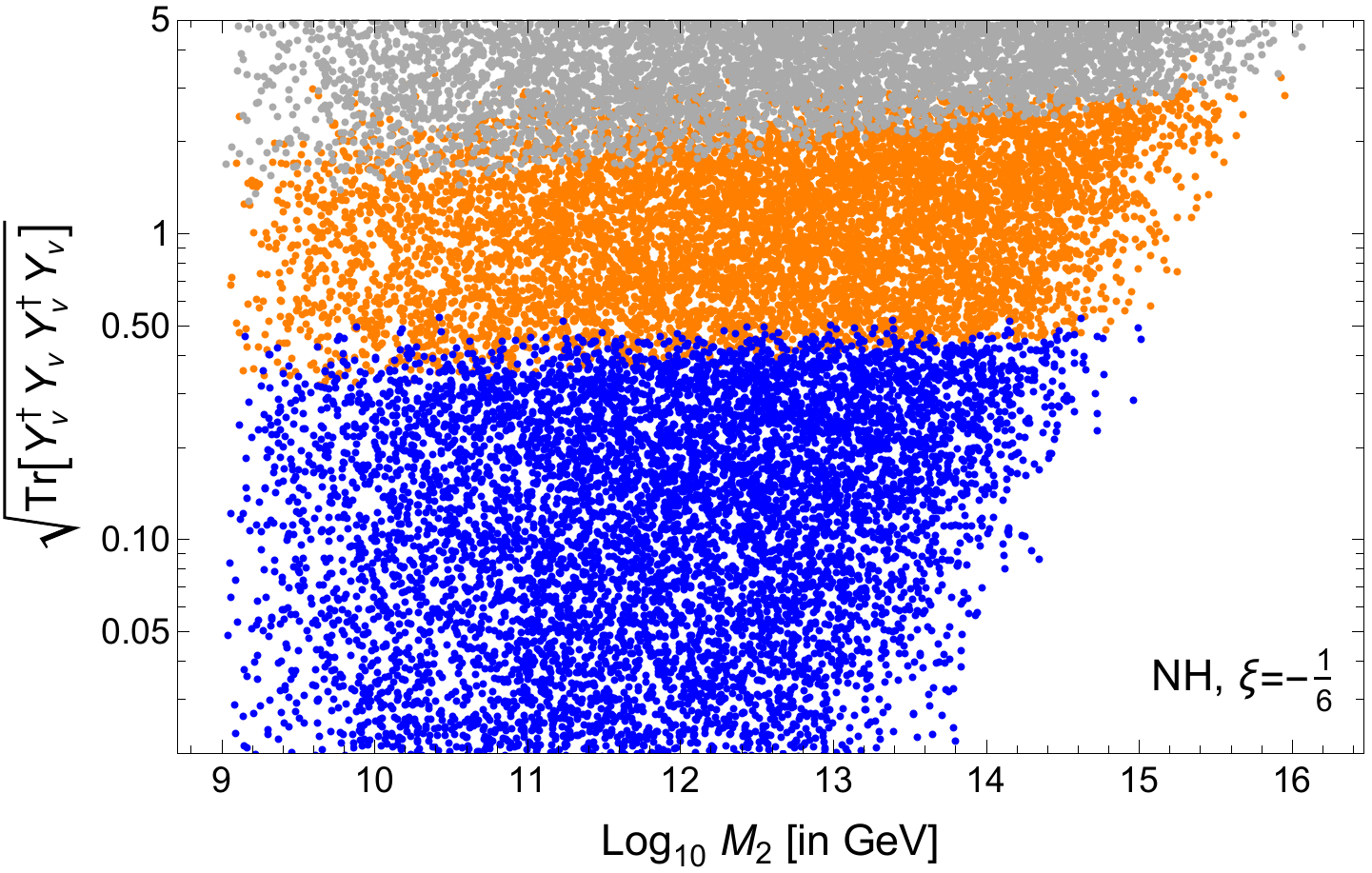}
    \includegraphics[width=0.49\textwidth]{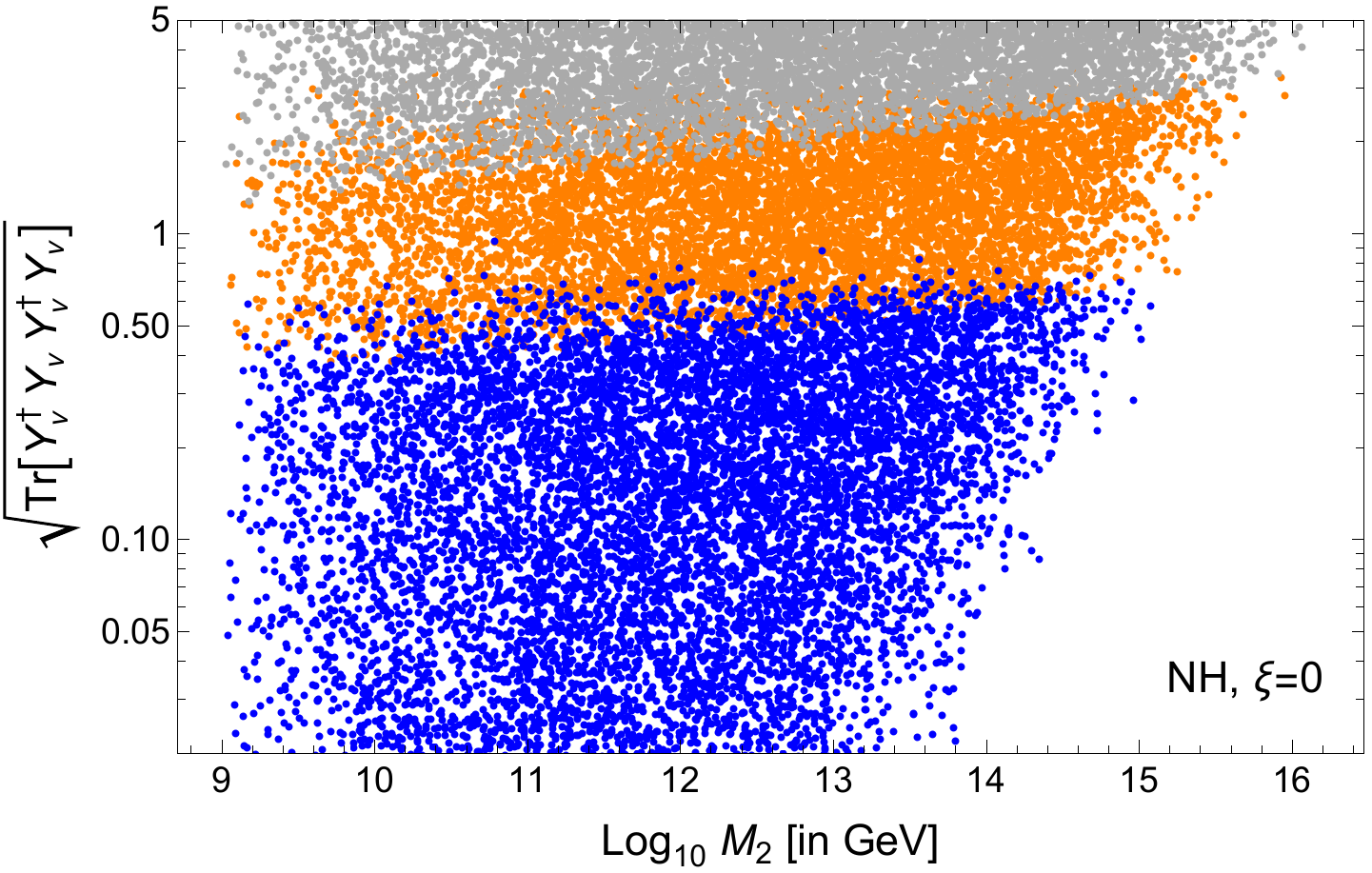}
    \\
    \includegraphics[width=0.49\textwidth]{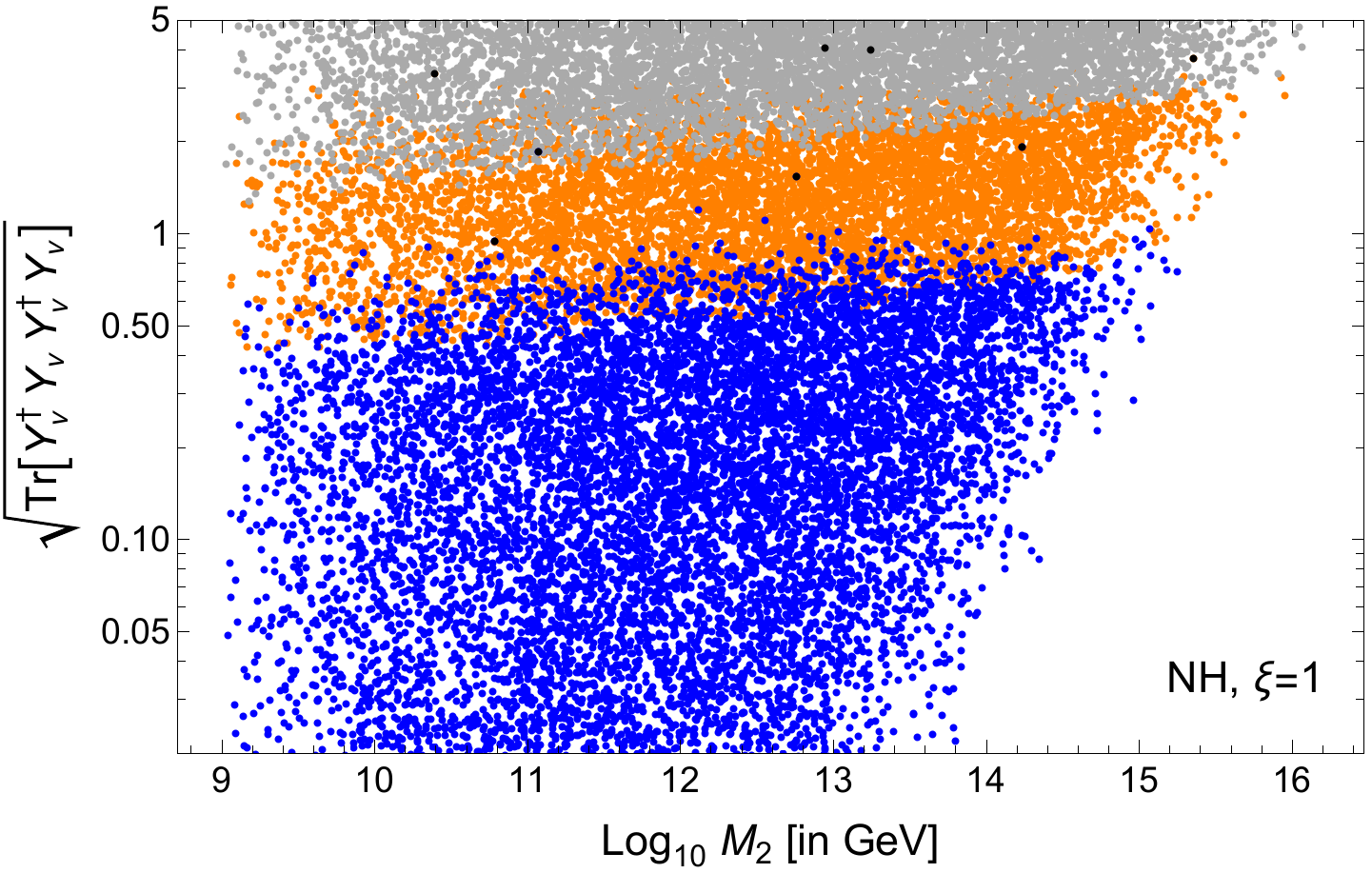}
    \includegraphics[width=0.49\textwidth]{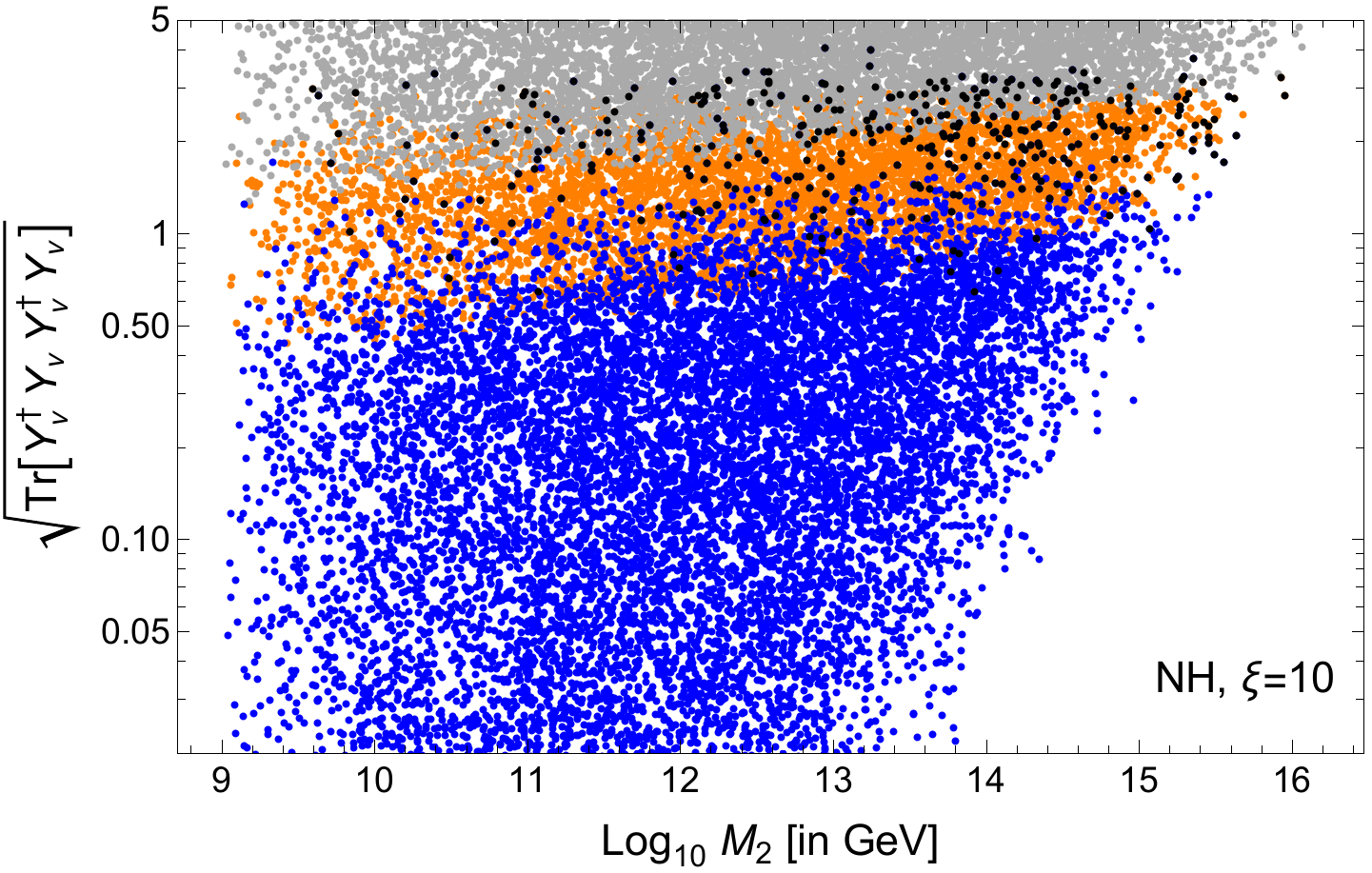}
    \caption{(3 Heavy $N$) Scatter plot of $\sqrt{T_4}$ as a function of RHN mass scale $M_1$ (3 RHN case) for NH and non-minimal coupling $\xi = \left(-\frac{1}{6},0,1,10\right)$. Blue points correspond to metastable electroweak vacuum while the orange points signify unstable electroweak vacuum. Gray and black points mark non-perturbative behavior in the loop-expansion and gravitational corrections respectively.}
    \label{fig:NHT2m23x3}
\end{figure}

\begin{figure}[t!]
    \centering
    \includegraphics[width=0.49\textwidth]{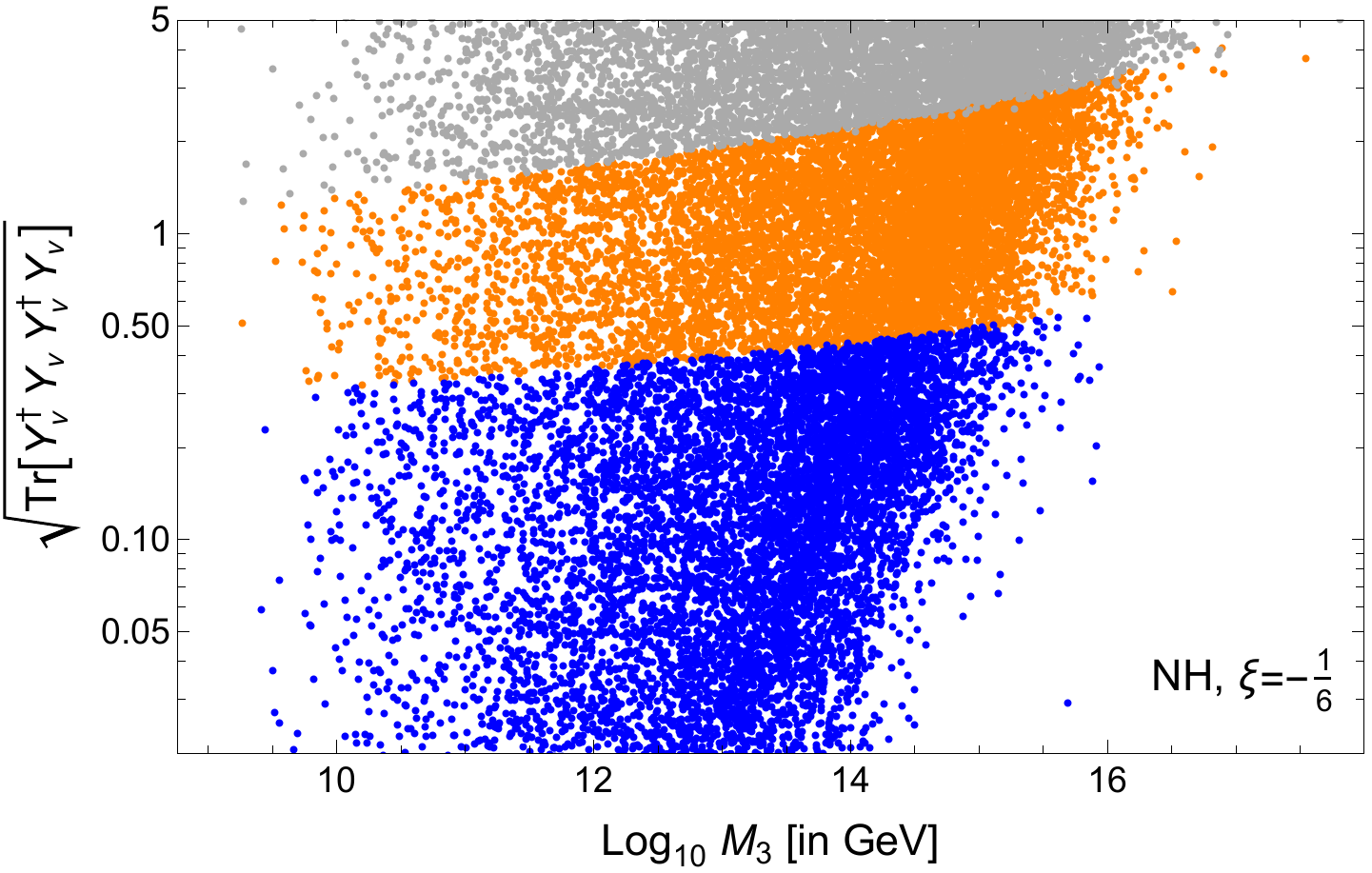}
    \includegraphics[width=0.49\textwidth]{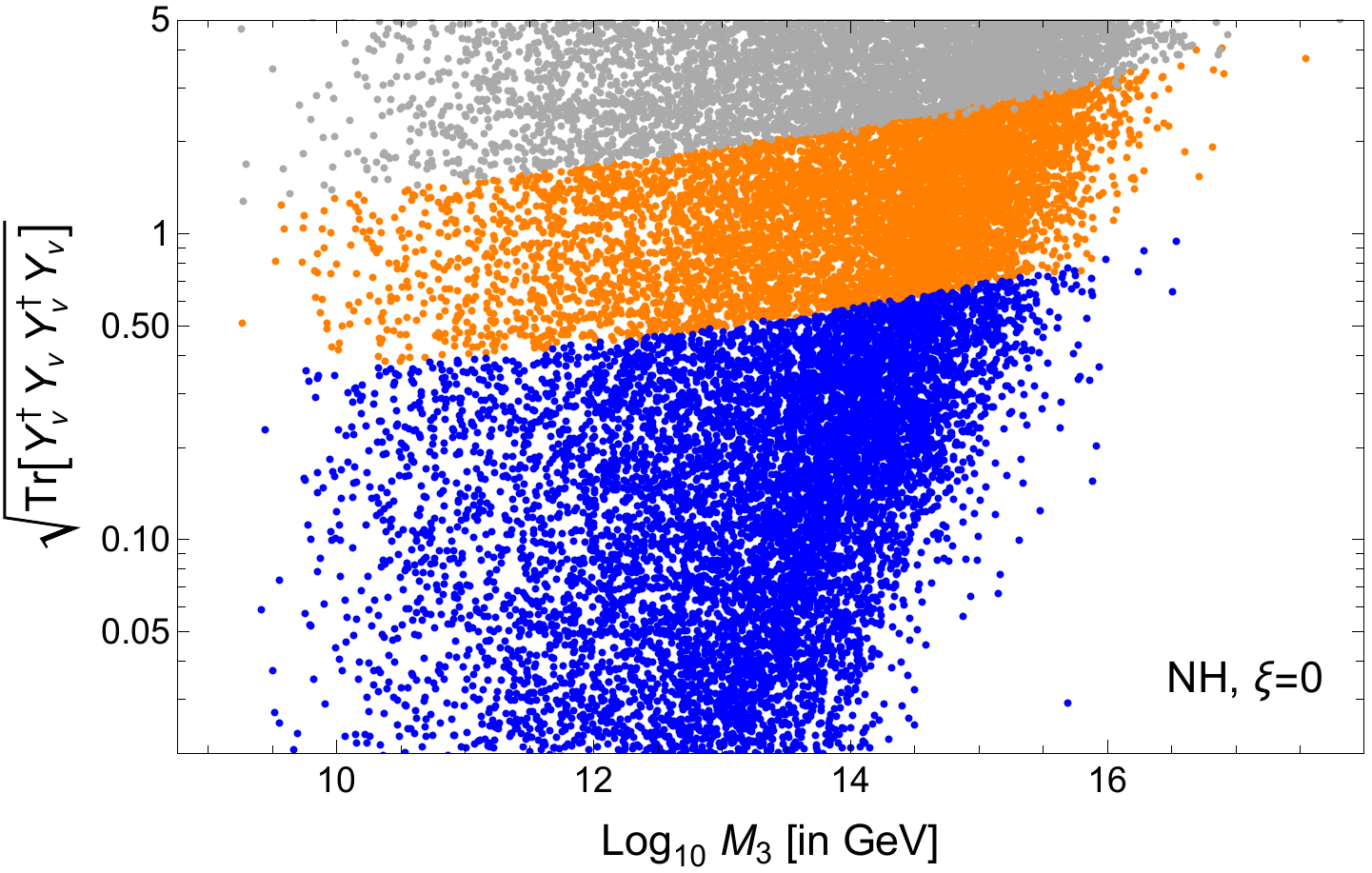}
    \\
    \includegraphics[width=0.49\textwidth]{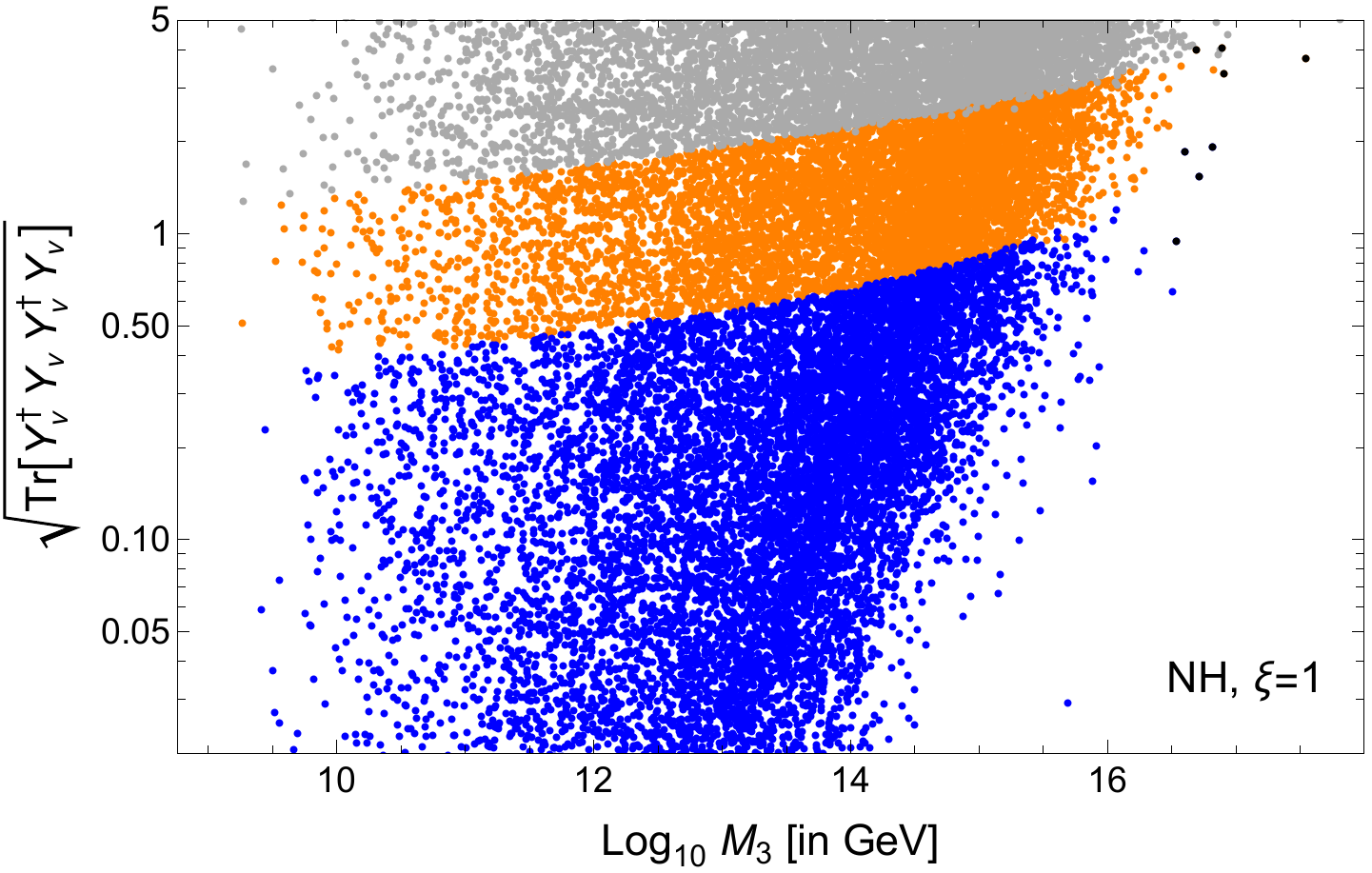}
    \includegraphics[width=0.49\textwidth]{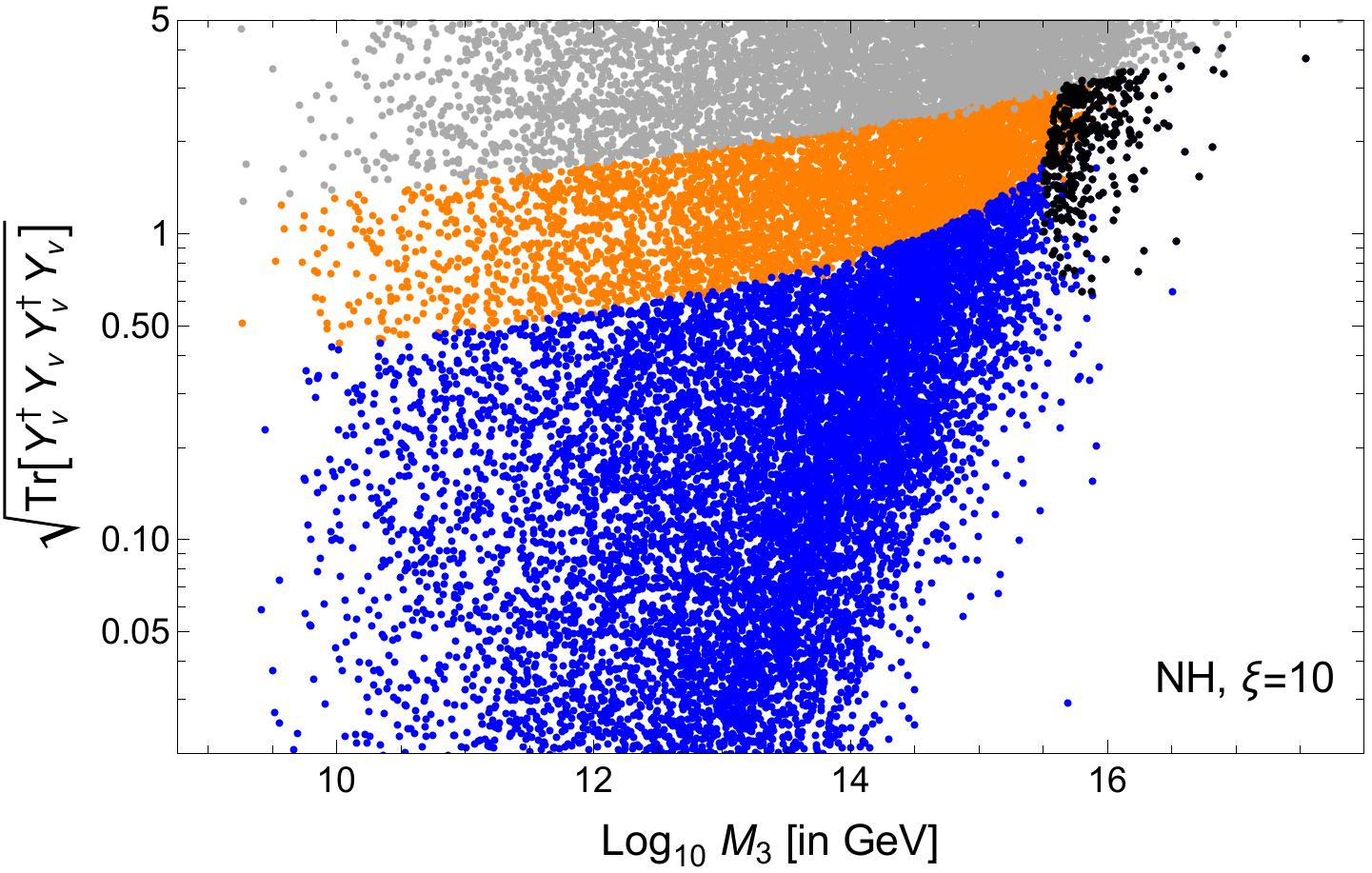}
    \caption{(3 Heavy $N$) Scatter plot of $\sqrt{T_4}$ as a function of heaviest RHN mass scale $M_3$ (3 RHN case) for NH and non-minimal coupling $\xi = \left(-\frac{1}{6},0,1,10\right)$. Blue points correspond to metastable electroweak vacuum while the orange points signify unstable electroweak vacuum. Gray and black points mark non-perturbative behavior in the loop-expansion and gravitational corrections respectively.}
    \label{fig:NHT2m33x3}
\end{figure}

In Fig.~\ref{fig:NHT2m13x3}, we display the results of our scan for $T_4$ as a function of the lightest mass scale $M_1$ for NH and different non-minimal couplings $\xi$. Similar to the case of two heavy RHNs, we again notice the clear separation of metastable and unstable/non-perturbative regions with almost no dependence on the lightest RHN mass scale $M_1$. In addition to these regions, we again find that our perturbative treatment of the gravitational effects breaks down for $M_3\gtrsim 10^{15.5}$~GeV and $\xi =10$.

For the NH (IH) case with conformal coupling ($\xi=-\frac{1}{6}$), which nullifies the gravitational contributions to leading order, the upper bound on $\sqrt{T_4}$ is around 0.53 (0.49) and $M_1$ is restricted to be below $10^{14.3}$~GeV ($10^{13.9}$~GeV). Meanwhile, for the NH (IH) case with gravity and $\xi=10$, the upper bound $T_4$ gets relaxed to 1.77 (1.59) and the bound on $M_1$ is loosened to $10^{14.7}$ GeV ($10^{14.3}$ GeV). However, as these latter bounds were obtained near the breakdown of our perturbative treatment of the gravitational effects (see Fig.~\ref{fig:NHT2m33x3}), it is likely that these values could be modified by a higher-order analysis.

In Fig.~\ref{fig:NHT2m23x3}, we display the results of our scan for $T_4$ as a function of the mass scale $M_2$ for NH and different non-minimal couplings $\xi$. We notice that although the parameter scan range for $M_2$ was up to $10^{19}$ GeV, th mass range allowed by the metastability bounds on $T_4$ restricts $M_2$ to values below $10^{16}$ GeV. Unlike in the case of two heavy RHNs, we find no dependence of the bounds on $T_4$ on $M_2$.

In Fig.~\ref{fig:NHT2m33x3}, we display the results of our scan for $T_4$ as a function of the heaviest RHN mass scale $M_3$ for NH and different non-minimal couplings $\xi$. We again find a dependence of the $T_4$ metastability bound on $M_3$. Just as the analogous behavior in our analysis involving two RHNs, this is because the largest contribution to $T_4$ arises from the heaviest RHN. We also recover the amplification of the gravitational stabilization for large $M_3$ and $\xi$, which can, in the same way as we discussed for the case of two RHNs, be understood through a lowering of the instanton scale $\mu_S$. In general, we again find that gravitational effects significantly stabilize the electroweak vacuum, corresponding to weaker bounds on $T_4$.

\begin{table}
\centering
\begin{tabular}{ |c|c|c|c|c| } 
\hline
Case & {Max. $\sqrt{T_4}$ }  & {Max. $M_1$ (in GeV)} & {Max. $M_2$ (in GeV)} & {Max. $M_3$ (in GeV)} \\
\hline
\textbf{$\xi=-\frac{1}{6}$} & & & &\\
NH & $0.53$ & $ 10^{14.3}$& $ 10^{15.0}$ & $ 10^{15.9}$ \\
IH & $0.49$ & $ 10^{13.9}$ & $ 10^{14.2}$ & $ 10^{15.5}$ \\
\hline
\textbf{$\xi=0$} & & & &\\
NH & $0.94$ & $ 10^{14.3}$& $ 10^{15.1}$ & $ 10^{16.5}$ \\
IH & $0.71$ & $ 10^{14.1}$ & $ 10^{14.4}$ & $ 10^{16.0}$ \\
\hline
\textbf{$\xi=1$} & & & &\\
NH & $1.93 $ & $ 10^{14.5}$& $ 10^{15.3}$ & $ 10^{16.8}$ \\
IH & $0.92 $ & $ 10^{14.3}$ & $ 10^{14.6}$ & $ 10^{16.0}$ \\
\hline
\textbf{$\xi=10$} & & & &\\
NH & $1.77(2.91) $ & $ 10^{14.7}$& $ 10^{15.6}$ & $ 10^{17.5}$ \\
IH & $1.59(3.14) $ & $ 10^{14.3}$ & $ 10^{14.7}$ & $ 10^{16.4}$ \\
\hline
\end{tabular}
\caption{Summary of metastability bounds on $T_4$, $M_1$, $M_2$ and $M_3$ in the case of 3 RHNs and Normal Hierarchy (NH) as well as Inverse Hierarchy (IH).}
\label{table:T2M3x3}
\end{table}

We provide a summary of our results in Table~\ref{table:T2M3x3}, assuming a light neutrino mass scale $m_0=0$. As before, we expect stronger bounds for a non-zero $m_0$. 

\subsection{Implications for High-scale Leptogenesis}

For the case of two RHNs, metastability sets an upper bound on $M_1$ and $M_2$, see Figs.~\ref{fig:NHT2m1} and~\ref{fig:NHT2m2}, as summarized in Table~\ref{table:T2M2x2} for different $\xi$s. Comparing all these values, we find a hierarchy between these bounds, $M_2^{\text{max}}$ and $M_1^{\text{max}}$. For a NH, we find $M_2^{\text{max}} \sim 10 M_1^{\text{max}}$, while for an IH $M_2^{\text{max}} \sim 2 M_1^{\text{max}}$ holds loosely.

These bounds have important consequences for high-scale leptogenesis. For the case of a conformal coupling ($\xi=-\frac{1}{6}$), $M_{1,2}$ is restricted to values below $10^{15.3}$~GeV. In conjunction with previous discussions, the range of RHN masses for high-scale leptogenesis gets restricted to $10^{9}-10^{16}$ GeV. This bound is only valid for RHNs which contribute to light neutrino mass generation through the seesaw relation and if no other new degrees of freedom are introduced below the RHN mass scales. If the RHNs are decoupled, their masses can easily be set to higher scales (usually set to $M_{\text{Planck}}$ for RGE studies). New degrees of freedom like scalars can usually relax the stringent bounds on $Y_\nu$ and $M_N$.

Similar to the 2 RHNs case, we notice that there are upper bounds on $M_1$, $M_2$ and $M_3$. We find that the bounds on $M_1$ and $M_2$ are slightly stronger in the 3 RHN case and again exhibit a hierarchy. For a NH, we find $M_2^{\text{max}} \sim 6.5 M_1^{\text{max}}$, while for an IH $M_2^{\text{max}} \sim 2 M_1^{\text{max}}$ holds loosely. The upper bound on $M_3$ with a metastable vacuum tends to be an order of magnitude higher than the bounds on $M_1$ and $M_2$. For comparison, in the case of an IH without gravity ($\xi=-\frac{1}{6}$), $M_1$ and $M_2$ are restricted to be below $10^{13.9}$ GeV and $10^{14.2}$ GeV respectively, while the bound on $M_3$ lies around $10^{15.5}$ GeV. As pointed out earlier, the bounds on $M_3$ can be relaxed if it decouples from the seesaw mechanism, which is usually the case for high-scale leptogenesis with two heavy RHNs. A complete summary of the bounds on $M_{1,2,3}$ is given in Table~\ref{table:T2M3x3}.

\section{Conclusions}
\label{sec:concl}
Introducing RHNs to the SM allows to simultaneously explain the generation of light neutrino masses and the observed matter-antimatter asymmetry of the universe. Despite a plethora of experimental searches, the parameter space of the type-I seesaw mechanism remains restricted only for $M_N \lesssim 1$~TeV. These limitations of current experimental accessibility motivate the search for theoretical constraints on the theory. In this article, we investigated the bounds arising from the stability of the electroweak vacuum. The introduction of additional fermions, such as RHNs, has the potential to lower the vacuum's expected lifetime below the current age of the universe. This allows for the derivation of stringent bounds on these Yukawa couplings to the Higgs in terms of their corresponding RHNs' masses. 

We performed a complete systematic study of the parameter space for the type-I seesaw mechanism (for $M_N>1$ TeV) from the perspective of vacuum stability, covering both low- and high-scale type-I seesaws including both two and three RHNs. In most cases, non-zero $Y_\nu$s push the scale where $\lambda$ takes its minimum, $\mu_*$, beyond the Planck scale. This motivated us to discuss the crucial conceptual and numerical importance of gravitational corrections to the decay rate. We find that doing so keeps all relevant scales below $M_{\text{Pl}}$ while simultaneously stabilizing the vacuum. We show that the non-minimal coupling $\xi$ of the Higgs with gravity can enhance this stabilizing effect and help relieve stringent bounds on $Y_\nu$ found in previous works. This effect is naturally strongest for large values of $\xi$ (usually of interest in the context of Higgs inflation). However, even small values, which are almost inevitably generated radiatively, lead to a notable effect.

For a low-scale seesaw, we have shown that the bounds are almost independent of the matrix structure of $Y_\nu$ and the number of RHNs in the high-magnitude limit. We presented our bounds on $Y_\nu$ in terms of $\text{Tr}(Y_\nu^\dagger Y_\nu)$ and the total mixing parameter $|\text{U}|^2$ in fig.~\ref{fig:lowSapprox}. These metastability bounds perform far better than the perturbativity bound as well as all proposed future searches for RHN masses above~$1$~TeV. When the approximation breaks down, we find that due to the strong dependence of $\beta_\lambda$ on $\text{Tr}(Y_\nu^\dagger Y_\nu)$ and $\text{Tr}(Y_\nu^\dagger Y_\nu Y_\nu^\dagger Y_\nu)$ (more prominently on the latter), the RG evolution can be carried out with good accuracy using only the trace parameters. Lastly, we also provide extensive results for the the instability scale, which has recently gained some interest in the context of cosmology and BSM Higgs physics. 

For the case of a high-scale seesaw, our results are summarized in Table~\ref{table:T2M2x2} and~\ref{table:T2M3x3}. As expected, gravitational corrections ensure consistency of our computations and significantly stabilize the vacuum, leading to looser bounds. Our results also have strong implications for high-scale leptogenesis. We find that metastability imposes clear upper bounds on the allowed values of the smallest ($M_1$) and largest RHN masses ($M_2$/$M_3$).

\section*{Acknowledgments}

We thank David Kaiser, Juraj Klaric, Yannis Georis and Sudip Jana for many useful discussions and comments on the manuscript. The work of GC is supported by the U.S. Department of Energy under the award number DE-SC0020250. TS's contributions to this work were made possible by the Walter Benjamin Programme of the Deutsche Forschungsgemeinschaft (DFG, German Research Foundation) – 512630918. Portions of this work were conducted in MIT’s Center for Theoretical Physics and partially supported by the U.S. Department of Energy under Contract No. DE-SC0012567. This project/publication is funded in part by the Gordon and Betty Moore Foundation. It was also made possible through the support of a grant from the John Templeton Foundation. The opinions expressed in this publication are those of the author(s) and do not necessarily reflect the views of these Foundations.

\appendix
\section{Neutrinos and RG running}
\label{RGappendix}

\subsection{Trace approximation}
\label{sec:traceapprox}
The quantities that affect the RGE running for Higgs quartic coupling at leading order are the traces $\text{Tr}(Y_\nu^\dagger Y_\nu)$ and $ \text{Tr}(Y_\nu^\dagger Y_\nu Y_\nu^\dagger Y_\nu)$. The running is significantly affected only when the magnitudes of these traces are of order $O(0.1-1)$. Under this assumption, we outline an analytical proof below to show that $\text{Tr}(Y_\nu^\dagger Y_\nu)^2\sim \text{Tr}(Y_\nu^\dagger Y_\nu Y_\nu^\dagger Y_\nu)$ in this high-magnitude limit, s.t. the bounds are almost insensitive to the matrix structure of $Y_\nu$.

The Casas-Ibarra parametrization of $Y_\nu$ is given by
\begin{equation}
    Y_\nu= -\frac{i}{v_{\text{ew}}} \,U^*_{\text{PMNS}} \:\sqrt{m_{\nu}}\, R\, \sqrt{M_{N}}
\end{equation}
where $v_{\text{ew}}$ is the vacuum expectation value of SM the Higgs field, $U_{\text{PMNS}}$ is the PMNS matrix, $m_{\nu}(M_{N})$ is the diagonalized light (heavy right-handed) neutrino mass matrix and $R$ is a general complex orthogonal matrix.

The complex orthogonal matrix $R$ can be decomposed in several ways \cite{Rodejohann:2012px,Lindner:2015qva,Drewes:2021nqr}. For transparency and easier physical understanding, we use a parametrization recently proposed in~\cite{Drewes:2021nqr}. In the usual Euler decomposition of $R$, we have $N$ real and $N$ complex angles. The real angles correspond to rotations in the $N\times N$ complex matrix space, while the complex angles correspond to scalings. In~\cite{Drewes:2021nqr}, the orthogonal matrix is decomposed as 5 real angles and 1 complex angle. This implies that the magnitude of any quantity which is a function of $Y_\nu$ will depend primarily on the complex angle/scaling:
\begin{equation}
    R = O_{13}(\theta_{13})\:O_{23}(\theta_{23})\:R_{12}(z_{12})\:O_{23}(\theta'_{23})\:O_{13}(\theta'_{13}),
\end{equation}
where $O_{ij}(R_{ij})$ is the real(complex) orthogonal rotation matrix in the $ij$-th plane. Note that the case of $2$ RHNs can be recovered by setting all angles except the complex angle $z_{12}$ to zero. The complex rotation matrix $R_{12}$ can be written as
\begin{equation}
   R_{12}= \begin{pmatrix}
\cos{z_{12}} & \sin{z_{12}} & 0\\
-\sin{z_{12}} & \cos{z_{12}} & 0\\
0 & 0 & 1
\end{pmatrix} = O_{12}\begin{pmatrix}
\cosh{y} & -i\sinh{y} & 0\\
i\sinh{y} & \cosh{y} & 0\\
0 & 0 & 1
\end{pmatrix},
\label{eq:Rdec}
\end{equation}
where $z_{12}= x+ i\ y$. The high-magnitude limit for $\text{Tr}(Y^\dagger Y)$ corresponds to a large $y$. In this limit
\begin{equation}
    \cosh{y}\sim\sinh{y}\sim \frac{e^y}{2} .
\end{equation}
This implies that all possible trace combinations of $Y_\nu(Y_\nu^\dagger)$ can be decomposed as a polynomial series in $e^y$. For $\text{Tr}(Y_\nu^\dagger Y_\nu)$, the contribution in the high-magnitude limit is dominated\footnote{The coefficients of different powers of $e^y$ cannot affect the conclusion since the rest of the entries in the decomposition of $R$ are only around $O(1)$.} by the term proportional to $e^{2y}$,
\begin{equation}
    \text{Tr}(Y_\nu^\dagger Y_\nu) \sim \frac{\mathcal{A}\times \mathcal{B}}{2} e^{2y},
    \label{eq:trY2}
\end{equation}
where 
\begin{align}
    \mathcal{A} = &  m_2\cos^2{\theta_{23}}\,+\, \cos^2{\theta_{13}}(m_1+m_3\sin^2{\theta_{23}})\,+\, \sin^2{\theta_{13}}(m_3+m_1\cos^2{\theta_{23}})  , \nonumber \\
    \mathcal{B} = &  M_2\cos^2{\theta'_{23}}\,+\, \cos^2{\theta'_{13}}(M_1+M_3\sin^2{\theta'_{23}})\,+\, \sin^2{\theta'_{13}}(M_3+M_1\cos^2{\theta'_{23}}).
\end{align}
Similarly applying Eq.~\eqref{eq:Rdec}, $ \text{Tr}(Y_\nu^\dagger Y_\nu Y_\nu^\dagger Y_\nu)$ can be written as a polynomial series in $e^y$ dominated by the contribution of order $e^{4y}$ if $\text{Tr}(Y^\dagger Y)$ is large. On retaining the dominant terms, 
\begin{equation}
    \text{Tr}(Y_\nu^\dagger Y_\nu Y_\nu^\dagger Y_\nu) \sim \left( \frac{\mathcal{A} \times \mathcal{B}}{2}\right)^2 e^{4y}
    \label{eq:trY4}
\end{equation}
Finally on comparing Eqs.~\eqref{eq:trY2} and \eqref{eq:trY4}, we have our final result
\begin{equation}\label{Tappr}
    \text{Tr}(Y_\nu^\dagger Y_\nu)^2\sim \text{Tr}(Y_\nu^\dagger Y_\nu Y_\nu^\dagger Y_\nu)
\end{equation}
in the high-magnitude $\text{Tr}(Y_\nu^\dagger Y_\nu)$ limit. Therefore, we can use the above result for our RGE analysis and report the results as bounds on $\text{Tr}(Y_\nu^\dagger Y_\nu)$ irrespective of the matrix structure of $Y$. These bounds on $\text{Tr}(Y^\dagger Y)$ can be transformed back to study the dependence on internal parameters such as $\theta_{ij}$, $m_i$ or $M_{N_i}$ by using eq.~\ref{eq:trY2}.

\subsection{Threshold corrections}
\label{thresh}
The introduction of RHNs affects the Higgs' dynamics not only through the running of its parameters, but also by inducing direct corrections to its parameters as
\begin{equation}
    \lambda_{\rm eff}(\mu,h) = \lambda(\mu) + \delta \lambda_{\rm SM} (\mu, h)+ \delta \lambda_{\nu} (\mu, h)
\end{equation}
This simple observation affects our discussion in two ways. First, taking seriously our EFT-approach we describe the running of our couplings using different theories at different energies. The matching of these theories is usually performed through the requirement that all physical quantities, such as scattering amplitudes, be continuous at the matching scale. This typically implies that the running parameter needs to be corrected through a \textit{threshold correction}, which compensates for the absence of loops involving the heavy particle in the theory below the matching scale. Taking as an example the quartic coupling, this implies that
\begin{gather}
    \lambda^{\rm UV}(M_i) + \delta \lambda^{\rm UV} (M_i) \overset{!}{=}  \lambda^{\rm IR}(M_i) +\delta \lambda^{\rm IR} ( M_i) \\ 
    \Leftrightarrow  \lambda^{\rm UV}(M_i)  \overset{!}{=}  \lambda^{\rm IR}(M_i) - \big( \delta \lambda^{\rm UV} (M_i) -\delta \lambda^{\rm IR} (M_i) \big).
\end{gather}
In other words, at the matching scale $\mu=M_{N_i}$ one has to subtract from the running coupling the loop-corrections induced by the particles that enter the spectrum.

The neutrinos' contributions to these threshold corrections for all SM couplings of interest has recently been derived, e.g., in~\cite{Brdar:2019iem}. For the readers' convenience, we repeat their results for the Higgs quartic and top Yukawa coupling here in the limit $m^2 \ll M_i^2$ for all $i$:
\begin{align}
    (4 \pi)^2 \delta \lambda^{UV \rm}=& - \lambda^{\rm IR} \left(Y_\nu^\dagger Y_\nu \right)_{ii} \left[1+ \ln \left( \frac{\mu^2}{M_i^2} \right) \right] + \\
    &+\left(Y_\nu^\dagger Y_\nu \right)_{ik}\left(Y_\nu^\dagger Y_\nu \right)_{ki}\frac{M_i^2 \left(1+ \ln \left( \frac{\mu^2}{M_i^2} \right)\right)-M_k^2 \left(1+ \ln \left( \frac{\mu^2}{M_k^2} \right)\right)}{M_i^2 - M_k^2} - \\ 
    &- \left(Y_\nu^\dagger Y_\nu \right)_{ik}\left(Y_\nu^\dagger Y_\nu \right)_{ik} \frac{M_i M_k \ln \left( \frac{M_i^2}{M_k^2} \right)}{M_i^2 - M_k^2} \\ 
    (4 \pi)^2 \delta y_t^{UV \rm}=& - y_t^{IR \rm} \cdot \frac{1}{4} \left(Y_\nu^\dagger Y_\nu \right)_{ii} \left[1+ \ln \left( \frac{\mu^2}{M_i^2} \right) \right] ,
\end{align}
where all indices are summed over all RHNs in the spectrum at the scale $\mu$. All implicit contractions, such as in $Y_\nu^\dagger Y_\nu$, are performed using the indices representing left-handed neutrinos. For degenerate masses, these expressions simplify to
\begin{align}
    (4 \pi)^2 \delta \lambda^{UV \rm}=& - \lambda^{\rm IR} \text{Tr}\left(Y_\nu^\dagger Y_\nu \right) \left[1+ \ln \left( \frac{\mu^2}{M^2} \right) \right] + \\
    &+\text{Tr}\left(Y_\nu^\dagger Y_\nu Y_\nu^\dagger Y_\nu \right)\ln \left( \frac{\mu^2}{M} \right)- \text{Tr}\left(Y_\nu^\dagger Y_\nu Y_\nu^T Y_\nu^* \right) \\ 
    (4 \pi)^2 \delta y_t^{UV \rm}=& - y_t^{IR \rm} \cdot \frac{1}{4} \text{Tr}\left(Y_\nu^\dagger Y_\nu \right) \left[1+ \ln \left( \frac{\mu^2}{M^2} \right) \right] .
\end{align}

\section{$\beta$-functions}
\label{beta functions append}
We provide the beta functions used in our calculations, as originally derived in~\cite{Buttazzo:2013uya,Pirogov:1998tj}. For the quartic coupling~$\lambda$, we have
\begin{align}
\beta_\lambda = \frac{1}{(4 \pi)^2}\bigg[ & 24 \lambda^2 - 6 y_{\rm t}^4 + \frac{3}{8} \left(2 g^4 + \left(g^2+{g^{\prime}}^2\right)^2 \right) - \lambda\left(9 g^2+ 3 {g^{\prime}}^2 - 12 y_{\rm t}^2\right) + \label{eq:betalambdafull} \\[-3pt]
+ & 4 \lambda \text{Tr} \left(Y_\nu^{\dagger} Y_\nu \right)  - 2  \text{Tr}\left(Y_\nu^{\dagger} Y_\nu Y_\nu^{\dagger} Y_\nu  \right)  \bigg] + \nonumber \\[-3pt]
+\frac{1}{(4 \pi)^4}\bigg[&\frac{1}{48} \left(915 g^6 - 289 g^4 {g^{\prime}}^2 -559 g^2 {g^{\prime}}^4 -379 {g^{\prime}}^6 \right) +30 y_{\rm t}^6 -y_{\rm t}^4 \left( \frac{8}{3}{g^{\prime}}^2 +32 g_s^2 +3 \lambda \right) + \nonumber \\[-3pt]
\nonumber
 +  &  \lambda\left(- \frac{73}{8} g^4 +\frac{39}{4} g^2 {g^{\prime}}^2 + \frac{629}{24} {g^{\prime}}^4 +108 g^2 \lambda +36 {g^{\prime}}^2 \lambda -312 \lambda^2 \right) + \\[-3pt]
\nonumber
 +  & y_{\rm t}^2 \left(- \frac{9}{4}g^4 + \frac{21}{2} g^2 {g^{\prime}}^2 - \frac{19}{4}{g^{\prime}}^4 + \lambda \left(\frac{45}{2} g^2 +\frac{85}{6} {g^{\prime}}^2 +80 g_s^2 -144 \lambda \right)\right) \nonumber + \\[-3pt]
\nonumber
  + &   \text{Tr} \left(Y_\nu^{\dagger} Y_\nu \right)\left( - \frac{3}{4} g^4 - \frac{1}{4} {g^{\prime}}^4 - \frac{1}{4} g^2 {g^{\prime}}^2 + \frac{5}{2} \left( 3 g^2 + {g^{\prime}}^2 \right) - 48 \lambda^2 \right) - \nonumber \\[-3pt] 
 - & \lambda  \text{Tr}\left(Y_\nu^{\dagger} Y_\nu Y_\nu^{\dagger} Y_\nu  \right) + 10  \text{Tr}\left(Y_\nu^{\dagger} Y_\nu Y_\nu^{\dagger} Y_\nu  Y_\nu^{\dagger} Y_\nu  \right) \bigg] + \nonumber \\[-3pt] 
\nonumber
+\frac{1}{(4 \pi)^6}\bigg[ & \lambda^3 \left(12022.7 \lambda + 1746 y_{\rm t}^2 - 774.904 g^2 - 258.3 {g^\prime}^2 \right) + \\[-3pt]
\nonumber
  + & \lambda^2 y_{\rm t}^2 \left(3536.52 y_{\rm t}^2 + 321.54 g_s^2 - 719.078 g^2 - 212.896 {g^\prime}^2\right) + \\[-3pt] 
\nonumber
+  & \lambda^2 \left(-1580.56 g^4 - 1030.734 {g^\prime}^4 - 1055.466 g^2 {g^\prime}^2\right) + \\[-3pt] 
\nonumber
 +  &  \lambda y_{\rm t}^4 \left(-446.764 y_{\rm t}^2 - 1325.732 g_s^2 - 10.94 g^2 - 70.05 {g^\prime}^2\right) +  \\[-3pt] 
\nonumber
 +  &  \lambda y_{\rm t}^2 \left(713.936 g_s^4 - 639.328 g^4 - 415.888 {g^\prime}^4 + \right.\\ 
 & ~~~~~~ \left. +30.288 g_s^2 g^2 + 58.18 g_s^2 {g^\prime}^2 + 18.716 g^2 {g^\prime}^2\right) + \nonumber  \\[-3pt]
\nonumber
 +  &  \lambda g^4 \left(-114.288 g_s^2 + 1730.966 g^2 + 265.46 {g^\prime}^2\right) + \\[-3pt] \nonumber
+  & \lambda {g^\prime}^4 \left(-46.562 g_s^2 + 343.072 g^2 + 260.814 {g^\prime}^2\right) + \\[-3pt] 
\nonumber
+  &  y_{\rm t}^6 \left(-486.298 y_{\rm t}^2 + 500.988 g_s^2 + 146.276 g^2 + 113.1 {g^\prime}^2\right) + \\[-3pt]
\nonumber
 +  & y_{\rm t}^4 \left(-100.402 g_s^4 + 31.768 g^4 + 88.6 {g^\prime}^4 + 26.698 g_s^2 g^2 +\right.  \nonumber \\[-3pt] 
  & ~~~~~~ \left. + 58.566 g_s^2 {g^\prime}^2 - 234.52 g^2 {g^\prime}^2\right) + \nonumber \\[-3pt] 
\nonumber
 +  &  y_{\rm t}^2 g_s^2 \left(32.928 g^4 + 3.644 {g^\prime}^4 + 37.954 g^2 {g^\prime}^2\right) + y_{\rm t}^2 g^4 \left(125 g^2 + 43.470 {g^\prime}^2\right) + \\[-3pt] 
\nonumber
 +  &  y_{\rm t}^2 {g^\prime}^4 \left(58.318 g^2 + 102.936 {g^\prime}^2\right)+ \\[-3pt]
 + & g_s^2 \left(15.072 g^6 + 7.138 {g^\prime}^6 + 5.024 g^4 {g^\prime}^2 + 6.138 g^2 {g^\prime}^4\right) \nonumber - \\[-3pt]
  -  & 228.182 g^8 -    23.272 {g^\prime}^8 -  126.296 g^6 {g^\prime}^2 + 36.112 g^4 {g^\prime}^4 - 14.288 g^2 {g^\prime}^6 \bigg]   \nonumber.
\end{align}
For the top Yukawa coupling~$y_{\rm t}$, we have, at the same order,
\begin{align}
\beta_{y_{\rm t}} = \frac{y_{\rm t}}{(4 \pi)^2}\bigg[ &- \frac{9}{4} g^2 - \frac{17}{12}{g^{\prime}}^2 -8 g_s^2 + \frac{9}{2} y_{\rm t}^2 +  \text{Tr}\left(Y_\nu^{\dagger} Y_\nu \right) \bigg] + \\ 
\nonumber
+\frac{y_{\rm t}}{(4 \pi)^4}\bigg[& - \frac{23}{4} g^4 - \frac{3}{4}g^2 {g^{\prime}}^2 + \frac{1187}{216} {g^{\prime}}^4 + 9 g^2 g_s^2 + \frac{19}{9} {g^{\prime}}^2 g_s^2 -108 g_s^4 + \\ 
\nonumber
 +& y_{\rm t}^2 \left( \frac{225}{16} g^2 +\frac{131}{16} +36 g_s^2 \right) + 6 \left(\lambda^2 - 2 y_{\rm t}^4 -2 \lambda y_{\rm t}^2\right)  + \nonumber \\ 
 +&  \text{Tr}\left(Y_\nu^{\dagger} Y_\nu  \right)\left(-\frac{9}{8} y_{\rm t}^2 + \frac{5}{8} {g^\prime}^2 +  \frac{15}{8} g^2 \right) - \frac{9}{4}  \text{Tr}\left( Y_\nu^{\dagger} Y_\nu Y_\nu^{\dagger} Y_\nu  \right)\bigg] +
\nonumber \\
+\frac{y_{\rm t}}{(4 \pi)^6} \bigg[& y_{\rm t}^4 \left(58.6028 y_{\rm t}^2 + 198 \lambda - 157 g_s^2 -  \frac{1593}{16} g^2 - \frac{2437}{48} {g^\prime}^2\right) + \nonumber \\
\nonumber
 +& \lambda y_{\rm t}^2 \left( \frac{15}{4} \lambda + 16 g_s^2 - \frac{135}{2} g^2 - \frac{127}{6} {g^\prime}^2\right) + \\
\nonumber
 +& y_{\rm t}^2 \left(363.764 g_s^4 + 16.990 g^4 - 67.839 {g^\prime}^4 + 48.370 g_s^2 g^2 + \right. \nonumber \\ 
&~~~~~ + \left.30.123 g_s^2 {g^\prime}^2 + 58.048 g^2 {g^\prime}^2 \right) \nonumber +\\
\nonumber
 +&\lambda^2 \left(-36 \lambda + 45 g^2 + 15 {g^\prime}^2\right) + \lambda \left(- \frac{171}{16} g^4 - \frac{1089}{144} {g^\prime}^4 + \frac{39}{8} g^2 {g^\prime}^2\right)+ \\
\nonumber  
 +& 619.35 g_s^6 + 169.829 g^6 + 74.074 {g^\prime}^6 +  73.654 g_s^4 g^2 - 25.16 g_s^4 {g^\prime}^2-  \\
\nonumber
 -&  21.072 g_s^2 g^4 - 61.997 g_s^2 {g^\prime}^4 -\frac{107}{4} g_s^2 g^2 {g^\prime}^2 - 7.905 g^4 {g^\prime}^2 - 12.339 g^2 {g^\prime}^4 \bigg] .
\end{align}
The beta functions of the gauge couplings~$g^\prime$,~$g$ and~$g_s$ are respectively given by
\begin{align}
\beta_{g^\prime} =& \frac{{g^{\prime}}^3}{(4 \pi)^2} \cdot  \frac{41}{6} + \frac{{g^{\prime}}^3}{(4 \pi)^4}\bigg[ \frac{199}{18}{g^{\prime}}^2 + \frac{9}{2}g^2 + \frac{44}{3} g_s^2 - \frac{17}{6} y_{\rm t}^2 - \frac{1}{2}  \text{Tr}\left(Y_\nu^{\dagger} Y_\nu \right) \bigg] + \\ 
\nonumber
+& \frac{{g^{\prime}}^3}{(4 \pi)^6}\bigg[ y_{\rm t}^2 \bigg( \frac{315}{16} y_{\rm t}^2 -\frac{29}{5} g_s^2 - \frac{785}{32} g^2 - \frac{2827}{288} {g^\prime}^2 \bigg) + \lambda \bigg(-3 \lambda + \frac{3}{2} g^2 + \frac{3}{2} {g^\prime }^2 \bigg) +\nonumber \\ 
 & ~~~~~~~~ +\, 99 g_s^4 + \frac{1315}{64} g^4 - \frac{388613}{5184} {g^\prime}^4 - \frac{25}{9} g_s^2 g^2 - \frac{137}{27} g_s^2 {g^\prime}^2 +\frac{205}{96} g^2 {g^\prime}^2 \bigg] , \nonumber \\
\beta_{g} =& - \frac{g^3}{(4 \pi)^2} \cdot  \frac{19}{6}  + \frac{g^3}{(4 \pi)^4}\bigg[ \frac{3}{2}{g^{\prime}}^2 + \frac{35}{6}g^2 + 12 g_s^2 - \frac{3}{2} y_{\rm t}^2 - \frac{1}{2}  \text{Tr}\left(Y_\nu^{\dagger} Y_\nu \right)  \bigg]+  \\ 
\nonumber
+& \frac{g^3}{(4 \pi)^6}\bigg[ y_{\rm t}^2 \left( \frac{147}{16} y_{\rm t}^2 - 7 g_s^2 - \frac{729}{32} g^2 - \frac{593}{96} {g^\prime}^2 \right) + \lambda \left( - 3 \lambda + \frac{3}{2} g^2 + \frac{1}{2} {g^\prime}^2 \right) +\\ 
 & ~~~~~~~~ +\, 81 g_s^4 + \frac{324953}{1728} g^4 - \frac{5597}{576} {g^\prime}^4 + 39 g_s^2 g^2 - \frac{1}{3} g_s^2 {g^\prime}^2 + \frac{291}{32} g^2 {g^\prime}^2 \bigg],\nonumber
\end{align}
and
\begin{align}
\beta_{g_s} =&- \frac{g_s^3}{(4 \pi)^2} \cdot 7  + \frac{g_s^3}{(4 \pi)^4}\bigg[ \frac{11}{6}g_s^2 + \frac{9}{2}g^2 - 26 g_s^2 - 2 y_{\rm t}^2 \bigg] +\\ 
& + \frac{g_s^3}{(4 \pi)^6} \bigg[y_{\rm t}^2 \left(15 y_{\rm t}^2 - 40 g_s^2 - 93/8 g^2 - 101/24 {g^\prime}^2\right) +\nonumber \\ 
 & ~~~~~~~~ +\, \frac{65}{2} g_s^4 + \frac{109}{8} g^4 - \frac{2615}{216} {g^\prime}^4 +  21 g_s^2 g^2 + \frac{77}{9} g_s^2 {g^\prime}^2 - \frac{1}{8} g^2 {g^\prime}^2 \bigg].\nonumber
\end{align}
For the neutrinos' Yukawa couplings, the beta function at 2-loop order is 
\begin{align}
\beta_{Y_{\nu}} =\frac{Y_\nu}{(4 \pi)^2} \bigg[&\frac{3}{2} Y_\nu^\dagger Y_\nu+3 y_{\rm t}^2 - \frac{3}{4} {g^\prime}^2 - \frac{9}{4} g^2 +  \text{Tr}\left(Y_\nu^{\dagger} Y_\nu \right) \bigg]+ \\
\nonumber
+ \frac{Y_\nu}{(4 \pi)^4} \bigg[ &\frac{3}{2} Y_\nu^\dagger Y_\nu Y_\nu^\dagger Y_\nu - \frac{27}{8} y_{\rm t}^2 Y_\nu^\dagger Y_\nu - \frac{9}{4} Y_\nu^\dagger Y_\nu \text{Tr}\left( Y_\nu^\dagger Y_\nu \right) - \frac{27}{16} y_{\rm t}^4 - \frac{9}{4} \text{Tr}\left( Y_\nu^\dagger Y_\nu Y_\nu^\dagger Y_\nu \right) + \nonumber \\ 
+&6 \lambda^2 -12 \lambda Y_\nu^\dagger Y_\nu + \frac{1}{16} \left(93 {g^{\prime}}^2 + 135 g^2 \right) Y_\nu^\dagger Y_\nu + \frac{5}{4} y_{\rm t}^2 \left( \frac{17}{12} {g^{\prime}}^2 + \frac{9}{4} g^2 + 8 g_s^2 \right)+ \nonumber \\ 
+&\frac{5}{8}   \text{Tr}\left(Y_\nu^{\dagger} Y_\nu \right) \left(  {g^\prime}^2 + 3 g^2 \right) + \frac{5}{8} {g^{\prime}}^4 - \frac{45}{8}  {g^{\prime}}^2 g^2 - \frac{115}{8} g^4 \bigg]
\nonumber
\end{align}
The running of the two most important trace parameters, as defined in Eqs.~\eqref{dl}-\eqref{dt4}, is to leading order given by
\begin{align}
    \beta_{T_2}= 2 \frac{T_2}{(4\pi)^2} \bigg[ T_2 + \frac{3}{2} y_t^2 - \frac{3}{4} {g^{\prime}}^2 - \frac{9}{4} g^2 + \frac{3}{2} T_4  \bigg] \\ 
    \beta_{T_6}=4 \frac{T_4}{(4 \pi)^2}\bigg[ T_2 + \frac{3}{2} y_t^2 - \frac{3}{4} {g^{\prime}}^2 - \frac{9}{4} g^2  + \frac{3}{2} T_6\bigg]
\end{align}

\section{NLO tunneling rate for the SM with RHNs}\label{Instantonloops}
For completeness, we include here the NLO formula for the vacuum decay rate derived in~\cite{Andreassen:2016cvx}. Before performing the integral over the dilatation modes, the decay rate per unit volume is given by
\begin{equation}
\frac{\Gamma}{V}=  \int \frac{\text{d}R}{R^5} \ {\rm e}^{- S_{\text{E}} \left(\lambda \left(R^{-1}\right),R\right)} D \big(R^{-1}\big).
\end{equation}
The factor~$D \big(R^{-1}\big)$ is defined as
\begin{align}
D \big(R^{-1}\big)  \equiv & \frac{72}{\sqrt{6}\pi^2} S_{\text{E}}^4 \left(\lambda \left(R^{-1}\right),R\right) \cdot \\
\nonumber
&\cdot  \exp\left[12 \zeta^\prime (-1) - \frac{25}{3} + \pi^2 - \gamma_{\rm E} - \frac{3}{2} \ln 2 - \frac{3}{2} S_{\text{fin}}^+ \left(X\right) - 3 S_{\text{fin}}^+ \left(Y\right)  +  \frac{3}{2} S_{\text{fin}}^{\bar{\psi}\psi} \left(\sqrt{Z} \right)  \right. \\
& ~~~~~\left.  -3\,S_{\text{loops}}^{\bar{\psi}\psi} \left(Z\right) -\frac{1}{2}S_{\text{diff}}^{\text{AG}}\left(X \right) - S_{\text{diff}}^{\text{AG}}\left(Y \right) - S_{\text{loops}}^{\text{AG}}\left(X \right) - 2 S_{\text{loops}}^{\text{AG}}\left(Y \right) + \Delta S_\nu \right] , \nonumber
\label{fullformula}
\end{align}
where~$X\equiv - \frac{g^2+{g^{\prime}}^2}{12 \lambda}$,~$Y \equiv -\frac{g^2}{12 \lambda}$, and~$Z\equiv \frac{y_{\rm t}^2}{\lambda}$ and $\Delta S_\nu$ parametrizes the neutrinos' contributions. The correction~$S_{\text{fin}}^+ (x)$ appearing in the exponent is given by
\begin{align}
& S_{\text{fin}}^+ (x) = x^2 \left(6 \gamma_{\rm E} +51-6\pi^2\right) + 6 x + \frac{11}{36} + \ln 2 \pi + \frac{3 \zeta (3)}{4 \pi^2} - 4 \zeta^\prime (-1)  - \ln \left( \frac{\cos \left(\frac{\pi}{2} \kappa_x\right)}{6 \pi x} \right) - \nonumber \\
\nonumber
 - & x \kappa_x \left[ \psi^{(-1)} \left( \frac{3+\kappa_x}{2}\right)  - \psi^{(-1)} \left( \frac{3-\kappa_x}{2}   \right) \right] + \left( 6x - \frac{1}{6}\right) \left[ \psi^{(-2)} \left( \frac{3+\kappa_x}{2}\right) + \psi^{(-2)} \left( \frac{3-\kappa_x}{2}   \right) \right] \\ 
+ & \kappa_x \left[ \psi^{(-3)} \left( \frac{3+\kappa_x}{2}\right) - \psi^{(-3)} \left( \frac{3-\kappa_x}{2}\right) \right]  - 2\left[ \psi^{(-4)} \left( \frac{3+\kappa_x}{2}\right) + \psi^{(-4)} \left( \frac{3-\kappa_x}{2}\right) \right],
\end{align}
where~$\kappa_x \equiv \sqrt{1-24 x}$ and~$\psi^{n}$ is the Polygamma function. The other corrections to the action are 
\begin{align}
S_{\text{diff}}^{\text{AG}} (x) =& x^2 \left(121 - 12 \pi^2 \right) - \frac{45}{2} x^2, \\
S_{\text{loops}}^{\text{AG}} (x)  =& - \frac{5}{18}  - \frac{1}{3} \left(\gamma_{\rm E}  - \ln 2\right) -   x \Big(7 + 6 (\gamma_{\rm E} - \ln 2) \Big) -   9 x^2 \bigg( \frac{1}{2} + \gamma_{\rm E} - \ln 2 \bigg), \\ 
S_{\text{loops}}^{\bar{\psi}\psi} (x) =& - x \bigg( \frac{13}{8} + \frac{2}{3} (\gamma_{\rm E} -  \ln 2) \bigg) + x^2 \bigg( \frac{5}{18} + \frac{1}{3}(\gamma_{\rm E} - \ln 2) \bigg) ,
\end{align}
as well as 
\begin{align}
S_{\text{fin}}^{\bar{\psi}\psi}& (x) =  16 \psi^{(-1)} (2) - \frac{8}{3} \psi^{(-2)} (2) + \frac{4}{3} x^2 (1-\gamma_{\rm E}) - \frac{x^4}{3} (1- 2 \gamma_{\rm E}) - \\
\nonumber
-& \frac{4}{3}x (1-x^2) \bigg[ \psi^{(-1)}(2+x)- \psi^{(-1)}(2-x) \bigg] + \frac{4}{3}x (1-3 x^2) \bigg[ \psi^{(-2)}(2+x) +  \psi^{(-2)}(2-x) \bigg] + \\ 
+&8x \bigg[ \psi^{(-3)}(2+x)- \psi^{(-3)}(2-x) \bigg]-8 \bigg[ \psi^{(-4)}(2+x) +  \psi^{(-4)}(2-x) \bigg] . \nonumber
\end{align}
A full derivation of these corrections can be found in~\cite{Andreassen:2017rzq}. 

We consistently find that the scale of interest for our analysis, $\mu_S$, is larger than the RHNs explicit masses. As a consequence, we can safely neglect them during our calculation of the vacuum decay rate and move to a flavor space basis of interaction eigenstates in which the Yukawa matrix becomes diagonal. After doing so, the neutrinos' combine into three Dirac fermions with masses $\propto y_{\nu,i} h_R$, where $y_{\nu,i}^2$ are the eigenvalues of $Y_\nu^\dagger Y_\nu$. Thus, the neutrinos' fluctuation operator can be factorized into three copies of the top quark's operator with the replacement $y_t \to y_{\nu,i}$. This allows a straightforward application of the computation given in~\cite{Andreassen:2017rzq}, ultimately yielding
\begin{equation}
    \Delta_\nu S = \sum_{i} \frac{1}{2} S_{\text{fin}}^{\bar{\psi}\psi} \left(\sqrt{Z_i} \right)  -\,S_{\text{loops}}^{\bar{\psi}\psi} \left(Z_i\right),
\end{equation}
where now $Z_i= \frac{y_{\nu,i}^2}{\lambda}$.

This prescription also preserves the relation $T_2^2 \sim T_4$ we used for low-scale seesaw models. As first pointed out in~\cite{Khoury:2021zao}, the cancellations necessary to achieve this relation while upholding agreement with observations and, crucially, simultaneously allowing for large enough Yukawa couplings to be of interest for our discussion can be used to show that
\begin{equation}\label{nuMSMnuloop}
    \Delta_\nu S_{\text{low-scale}} =  \frac{1}{2} S_{\text{fin}}^{\bar{\psi}\psi} \left(\sqrt{Z_{\text{low-scale}}} \right)  -\,S_{\text{loops}}^{\bar{\psi}\psi} \left(Z_{\text{low-scale}}\right),
\end{equation}
where now $Z_{\text{low-scale}}= \frac{T_2}{\lambda}$. In essence, the criteria above allow to move to a basis in flavor space in which two of the RHNs can be combined into a single Dirac fermion, whose Yukawa coupling is now given by $\sqrt{T_2}$~\cite{Shaposhnikov:2006nn}. 

The latter observation can, however, not be immediately generalized to our effective running in terms of $T$-parameters. As the latter only serves as an approximate scheme we nevertheless use Eq.~\eqref{nuMSMnuloop} for this discussion, as the deviation from our results obtained from a full analysis is within our desired level of accuracy.

\bibliographystyle{JHEP}
\bibliography{ref}

\providecommand{\href}[2]{#2}\begingroup\raggedright\begin{thebibliography}{10}

\bibitem{Minkowski:1977sc}
P.~Minkowski, \emph{{$\mu \to e\gamma$ at a Rate of One Out of $10^{9}$ Muon
  Decays?}}, \href{https://doi.org/10.1016/0370-2693(77)90435-X}{\emph{Phys.
  Lett. B} {\bfseries 67} (1977) 421}.

\bibitem{Mohapatra:1979ia}
R.~N. Mohapatra and G.~Senjanovic, \emph{{Neutrino Mass and Spontaneous Parity
  Nonconservation}},
  \href{https://doi.org/10.1103/PhysRevLett.44.912}{\emph{Phys. Rev. Lett.}
  {\bfseries 44} (1980) 912}.

\bibitem{Yanagida:1979as}
T.~Yanagida, \emph{{Horizontal gauge symmetry and masses of neutrinos}},
  {\emph{Conf. Proc. C} {\bfseries 7902131} (1979) 95}.

\bibitem{GellMann:1980vs}
M.~Gell-Mann, P.~Ramond and R.~Slansky, \emph{{Complex Spinors and Unified
  Theories}}, {\emph{Conf. Proc. C} {\bfseries 790927} (1979) 315}
  [\href{https://arxiv.org/abs/1306.4669}{{\ttfamily 1306.4669}}].

\bibitem{Glashow:1979nm}
S.~Glashow, \emph{{The Future of Elementary Particle Physics}},
  \href{https://doi.org/10.1007/978-1-4684-7197-7\_15}{\emph{NATO Sci. Ser. B}
  {\bfseries 61} (1980) 687}.

\bibitem{Schechter:1980gr}
J.~Schechter and J.~W.~F. Valle, \emph{{Neutrino Masses in SU(2) x U(1)
  Theories}}, \href{https://doi.org/10.1103/PhysRevD.22.2227}{\emph{Phys. Rev.
  D} {\bfseries 22} (1980) 2227}.

\bibitem{Grzadkowski:1987tf}
B.~Grzadkowski and M.~Lindner, \emph{{Nonlinear Evolution of Yukawa
  Couplings}}, \href{https://doi.org/10.1016/0370-2693(87)90458-8}{\emph{Phys.
  Lett. B} {\bfseries 193} (1987) 71}.

\bibitem{Casas:1999tg}
J.~A. Casas, J.~R. Espinosa, A.~Ibarra and I.~Navarro, \emph{{General RG
  equations for physical neutrino parameters and their phenomenological
  implications}},
  \href{https://doi.org/10.1016/S0550-3213(99)00781-6}{\emph{Nucl. Phys. B}
  {\bfseries 573} (2000) 652}
  [\href{https://arxiv.org/abs/hep-ph/9910420}{{\ttfamily hep-ph/9910420}}].

\bibitem{Antusch:2002rr}
S.~Antusch, J.~Kersten, M.~Lindner and M.~Ratz, \emph{{Neutrino mass matrix
  running for nondegenerate seesaw scales}},
  \href{https://doi.org/10.1016/S0370-2693(02)01960-3}{\emph{Phys. Lett. B}
  {\bfseries 538} (2002) 87}
  [\href{https://arxiv.org/abs/hep-ph/0203233}{{\ttfamily hep-ph/0203233}}].

\bibitem{Pirogov:1998tj}
Y.~F. Pirogov and O.~V. Zenin, \emph{{Two loop renormalization group
  restrictions on the standard model and the fourth chiral family}},
  \href{https://doi.org/10.1007/s100520050602}{\emph{Eur. Phys. J. C}
  {\bfseries 10} (1999) 629}
  [\href{https://arxiv.org/abs/hep-ph/9808396}{{\ttfamily hep-ph/9808396}}].

\bibitem{Rodejohann:2012px}
W.~Rodejohann and H.~Zhang, \emph{{Impact of massive neutrinos on the Higgs
  self-coupling and electroweak vacuum stability}},
  \href{https://doi.org/10.1007/JHEP06(2012)022}{\emph{JHEP} {\bfseries 06}
  (2012) 022} [\href{https://arxiv.org/abs/1203.3825}{{\ttfamily 1203.3825}}].

\bibitem{Lindner:2015qva}
M.~Lindner, H.~H. Patel and B.~Radov\v{c}i\'c, \emph{{Electroweak Absolute,
  Meta-, and Thermal Stability in Neutrino Mass Models}},
  \href{https://doi.org/10.1103/PhysRevD.93.073005}{\emph{Phys. Rev. D}
  {\bfseries 93} (2016) 073005}
  [\href{https://arxiv.org/abs/1511.06215}{{\ttfamily 1511.06215}}].

\bibitem{Bambhaniya:2016rbb}
G.~Bambhaniya, P.~S. Bhupal~Dev, S.~Goswami, S.~Khan and W.~Rodejohann,
  \emph{{Naturalness, Vacuum Stability and Leptogenesis in the Minimal Seesaw
  Model}}, \href{https://doi.org/10.1103/PhysRevD.95.095016}{\emph{Phys. Rev.
  D} {\bfseries 95} (2017) 095016}
  [\href{https://arxiv.org/abs/1611.03827}{{\ttfamily 1611.03827}}].

\bibitem{Mandal:2019ndp}
S.~Mandal, R.~Srivastava and J.~W.~F. Valle, \emph{{Consistency of the
  dynamical high-scale type-I seesaw mechanism}},
  \href{https://doi.org/10.1103/PhysRevD.101.115030}{\emph{Phys. Rev. D}
  {\bfseries 101} (2020) 115030}
  [\href{https://arxiv.org/abs/1903.03631}{{\ttfamily 1903.03631}}].

\bibitem{Ipek:2018sai}
S.~Ipek, A.~D. Plascencia and J.~Turner, \emph{{Assessing Perturbativity and
  Vacuum Stability in High-Scale Leptogenesis}},
  \href{https://doi.org/10.1007/JHEP12(2018)111}{\emph{JHEP} {\bfseries 12}
  (2018) 111} [\href{https://arxiv.org/abs/1806.00460}{{\ttfamily
  1806.00460}}].

\bibitem{Khan:2012zw}
S.~Khan, S.~Goswami and S.~Roy, \emph{{Vacuum Stability constraints on the
  minimal singlet TeV Seesaw Model}},
  \href{https://doi.org/10.1103/PhysRevD.89.073021}{\emph{Phys. Rev. D}
  {\bfseries 89} (2014) 073021}
  [\href{https://arxiv.org/abs/1212.3694}{{\ttfamily 1212.3694}}].

\bibitem{DelleRose:2015bms}
L.~Delle~Rose, C.~Marzo and A.~Urbano, \emph{{On the stability of the
  electroweak vacuum in the presence of low-scale seesaw models}},
  \href{https://doi.org/10.1007/JHEP12(2015)050}{\emph{JHEP} {\bfseries 12}
  (2015) 050} [\href{https://arxiv.org/abs/1506.03360}{{\ttfamily
  1506.03360}}].

\bibitem{Andreassen:2017rzq}
A.~Andreassen, W.~Frost and M.~D. Schwartz, \emph{{Scale Invariant Instantons
  and the Complete Lifetime of the Standard Model}},
  \href{https://doi.org/10.1103/PhysRevD.97.056006}{\emph{Phys. Rev. D}
  {\bfseries 97} (2018) 056006}
  [\href{https://arxiv.org/abs/1707.08124}{{\ttfamily 1707.08124}}].

\bibitem{Salvio:2016mvj}
A.~Salvio, A.~Strumia, N.~Tetradis and A.~Urbano, \emph{{On gravitational and
  thermal corrections to vacuum decay}},
  \href{https://doi.org/10.1007/JHEP09(2016)054}{\emph{JHEP} {\bfseries 09}
  (2016) 054} [\href{https://arxiv.org/abs/1608.02555}{{\ttfamily
  1608.02555}}].

\bibitem{Espinosa:2020qtq}
J.~R. Espinosa, \emph{{Vacuum Decay in the Standard Model: Analytical Results
  with Running and Gravity}},
  \href{https://doi.org/10.1088/1475-7516/2020/06/052}{\emph{JCAP} {\bfseries
  06} (2020) 052} [\href{https://arxiv.org/abs/2003.06219}{{\ttfamily
  2003.06219}}].

\bibitem{Khoury:2021zao}
J.~Khoury and T.~Steingasser, \emph{{Gauge hierarchy from electroweak vacuum
  metastability}},
  \href{https://doi.org/10.1103/PhysRevD.105.055031}{\emph{Phys. Rev. D}
  {\bfseries 105} (2022) 055031}
  [\href{https://arxiv.org/abs/2108.09315}{{\ttfamily 2108.09315}}].

\bibitem{Callan:1970ze}
C.~G. Callan, Jr., S.~R. Coleman and R.~Jackiw, \emph{{A New improved energy -
  momentum tensor}},
  \href{https://doi.org/10.1016/0003-4916(70)90394-5}{\emph{Annals Phys.}
  {\bfseries 59} (1970) 42}.

\bibitem{Bunch:1980bs}
T.~S. Bunch and P.~Panangaden, \emph{{On renormalization of lambda phi**4 field
  theory in curved spacetime. II}},
  \href{https://doi.org/10.1088/0305-4470/13/3/023}{\emph{J. Phys. A}
  {\bfseries 13} (1980) 919}.

\bibitem{Birrell:1982ix}
N.~D. Birrell and P.~C.~W. Davies, \emph{{Quantum Fields in Curved Space}},
  Cambridge Monographs on Mathematical Physics. Cambridge Univ. Press,
  Cambridge, UK, 2, 1984,
  \href{https://doi.org/10.1017/CBO9780511622632}{10.1017/CBO9780511622632}.

\bibitem{Odintsov:1990mt}
S.~D. Odintsov, \emph{{Renormalization Group, Effective Action and Grand
  Unification Theories in Curved Space-time}}, {\emph{Fortsch. Phys.}
  {\bfseries 39} (1991) 621}.

\bibitem{Buchbinder:1992rb}
I.~L. Buchbinder, S.~D. Odintsov and I.~L. Shapiro, \emph{{Effective action in
  quantum gravity}}. 1992.

\bibitem{Parker:2009uva}
L.~E. Parker and D.~Toms, \emph{{Quantum Field Theory in Curved Spacetime}:
  {Quantized Field and Gravity}}, Cambridge Monographs on Mathematical Physics.
  Cambridge University Press, 8, 2009,
  \href{https://doi.org/10.1017/CBO9780511813924}{10.1017/CBO9780511813924}.

\bibitem{Markkanen:2013nwa}
T.~Markkanen and A.~Tranberg, \emph{{A Simple Method for One-Loop
  Renormalization in Curved Space-Time}},
  \href{https://doi.org/10.1088/1475-7516/2013/08/045}{\emph{JCAP} {\bfseries
  08} (2013) 045} [\href{https://arxiv.org/abs/1303.0180}{{\ttfamily
  1303.0180}}].

\bibitem{Kaiser:2015usz}
D.~I. Kaiser, \emph{{Nonminimal Couplings in the Early Universe: Multifield
  Models of Inflation and the Latest Observations}},
  \href{https://doi.org/10.1007/978-3-319-31299-6_2}{\emph{Fundam. Theor.
  Phys.} {\bfseries 183} (2016) 41}
  [\href{https://arxiv.org/abs/1511.09148}{{\ttfamily 1511.09148}}].

\bibitem{Markkanen:2018bfx}
T.~Markkanen, S.~Nurmi, A.~Rajantie and S.~Stopyra, \emph{{The 1-loop effective
  potential for the Standard Model in curved spacetime}},
  \href{https://doi.org/10.1007/JHEP06(2018)040}{\emph{JHEP} {\bfseries 06}
  (2018) 040} [\href{https://arxiv.org/abs/1804.02020}{{\ttfamily
  1804.02020}}].

\bibitem{Andreassen:2016cvx}
A.~Andreassen, D.~Farhi, W.~Frost and M.~D. Schwartz, \emph{{Precision decay
  rate calculations in quantum field theory}},
  \href{https://doi.org/10.1103/PhysRevD.95.085011}{\emph{Phys. Rev. D}
  {\bfseries 95} (2017) 085011}
  [\href{https://arxiv.org/abs/1604.06090}{{\ttfamily 1604.06090}}].

\bibitem{Steingasser:2022yqx}
T.~Steingasser, \emph{{New perspectives on solitons and instantons in the
  Standard Model and beyond}}, Ph.D. thesis, Munich U., 2022.
\newblock 10.5282/edoc.30495.

\bibitem{Zhang:2021jdf}
D.~Zhang and S.~Zhou, \emph{{Complete one-loop matching of the type-I seesaw
  model onto the Standard Model effective field theory}},
  \href{https://doi.org/10.1007/JHEP09(2021)163}{\emph{JHEP} {\bfseries 09}
  (2021) 163} [\href{https://arxiv.org/abs/2107.12133}{{\ttfamily
  2107.12133}}].

\bibitem{Huang:2020hdv}
G.-y. Huang and S.~Zhou, \emph{{Precise Values of Running Quark and Lepton
  Masses in the Standard Model}},
  \href{https://doi.org/10.1103/PhysRevD.103.016010}{\emph{Phys. Rev. D}
  {\bfseries 103} (2021) 016010}
  [\href{https://arxiv.org/abs/2009.04851}{{\ttfamily 2009.04851}}].

\bibitem{Bezrukov:2007ep}
F.~L. Bezrukov and M.~Shaposhnikov, \emph{{The Standard Model Higgs boson as
  the inflaton}},
  \href{https://doi.org/10.1016/j.physletb.2007.11.072}{\emph{Phys. Lett. B}
  {\bfseries 659} (2008) 703}
  [\href{https://arxiv.org/abs/0710.3755}{{\ttfamily 0710.3755}}].

\bibitem{Barvinsky:2009fy}
A.~O. Barvinsky, A.~Y. Kamenshchik, C.~Kiefer, A.~A. Starobinsky and
  C.~Steinwachs, \emph{{Asymptotic freedom in inflationary cosmology with a
  non-minimally coupled Higgs field}},
  \href{https://doi.org/10.1088/1475-7516/2009/12/003}{\emph{JCAP} {\bfseries
  12} (2009) 003} [\href{https://arxiv.org/abs/0904.1698}{{\ttfamily
  0904.1698}}].

\bibitem{Barvinsky:2009ii}
A.~O. Barvinsky, A.~Y. Kamenshchik, C.~Kiefer, A.~A. Starobinsky and C.~F.
  Steinwachs, \emph{{Higgs boson, renormalization group, and naturalness in
  cosmology}}, \href{https://doi.org/10.1140/epjc/s10052-012-2219-3}{\emph{Eur.
  Phys. J. C} {\bfseries 72} (2012) 2219}
  [\href{https://arxiv.org/abs/0910.1041}{{\ttfamily 0910.1041}}].

\bibitem{Bezrukov:2012sa}
F.~Bezrukov, M.~Y. Kalmykov, B.~A. Kniehl and M.~Shaposhnikov, \emph{{Higgs
  Boson Mass and New Physics}},
  \href{https://doi.org/10.1007/JHEP10(2012)140}{\emph{JHEP} {\bfseries 10}
  (2012) 140} [\href{https://arxiv.org/abs/1205.2893}{{\ttfamily 1205.2893}}].

\bibitem{Schutz:2013fua}
K.~Schutz, E.~I. Sfakianakis and D.~I. Kaiser, \emph{{Multifield Inflation
  after Planck: Isocurvature Modes from Nonminimal Couplings}},
  \href{https://doi.org/10.1103/PhysRevD.89.064044}{\emph{Phys. Rev. D}
  {\bfseries 89} (2014) 064044}
  [\href{https://arxiv.org/abs/1310.8285}{{\ttfamily 1310.8285}}].

\bibitem{Iacconi:2021ltm}
L.~Iacconi, H.~Assadullahi, M.~Fasiello and D.~Wands, \emph{{Revisiting
  small-scale fluctuations in \ensuremath{\alpha}-attractor models of
  inflation}}, \href{https://doi.org/10.1088/1475-7516/2022/06/007}{\emph{JCAP}
  {\bfseries 06} (2022) 007}
  [\href{https://arxiv.org/abs/2112.05092}{{\ttfamily 2112.05092}}].

\bibitem{Geller:2022nkr}
S.~R. Geller, W.~Qin, E.~McDonough and D.~I. Kaiser, \emph{{Primordial black
  holes from multifield inflation with nonminimal couplings}},
  \href{https://doi.org/10.1103/PhysRevD.106.063535}{\emph{Phys. Rev. D}
  {\bfseries 106} (2022) 063535}
  [\href{https://arxiv.org/abs/2205.04471}{{\ttfamily 2205.04471}}].

\bibitem{Braglia:2022phb}
M.~Braglia, A.~Linde, R.~Kallosh and F.~Finelli, \emph{{Hybrid
  \ensuremath{\alpha}-attractors, primordial black holes and gravitational wave
  backgrounds}},
  \href{https://doi.org/10.1088/1475-7516/2023/04/033}{\emph{JCAP} {\bfseries
  04} (2023) 033} [\href{https://arxiv.org/abs/2211.14262}{{\ttfamily
  2211.14262}}].

\bibitem{Qin:2023lgo}
W.~Qin, S.~R. Geller, S.~Balaji, E.~McDonough and D.~I. Kaiser, \emph{{Planck
  Constraints and Gravitational Wave Forecasts for Primordial Black Hole Dark
  Matter Seeded by Multifield Inflation}},
  \href{https://arxiv.org/abs/2303.02168}{{\ttfamily 2303.02168}}.

\bibitem{Lebedev:2022vwf}
O.~Lebedev, T.~Solomko and J.-H. Yoon, \emph{{Dark matter production via a
  non-minimal coupling to gravity}},
  \href{https://arxiv.org/abs/2211.11773}{{\ttfamily 2211.11773}}.

\bibitem{Joti:2017fwe}
A.~Joti, A.~Katsis, D.~Loupas, A.~Salvio, A.~Strumia, N.~Tetradis et~al.,
  \emph{{(Higgs) vacuum decay during inflation}},
  \href{https://doi.org/10.1007/JHEP07(2017)058}{\emph{JHEP} {\bfseries 07}
  (2017) 058} [\href{https://arxiv.org/abs/1706.00792}{{\ttfamily
  1706.00792}}].

\bibitem{Shkerin:2015exa}
A.~Shkerin and S.~Sibiryakov, \emph{{On stability of electroweak vacuum during
  inflation}},
  \href{https://doi.org/10.1016/j.physletb.2015.05.012}{\emph{Phys. Lett. B}
  {\bfseries 746} (2015) 257}
  [\href{https://arxiv.org/abs/1503.02586}{{\ttfamily 1503.02586}}].

\bibitem{Kobakhidze:2013tn}
A.~Kobakhidze and A.~Spencer-Smith, \emph{{Electroweak Vacuum (In)Stability in
  an Inflationary Universe}},
  \href{https://doi.org/10.1016/j.physletb.2013.04.013}{\emph{Phys. Lett. B}
  {\bfseries 722} (2013) 130}
  [\href{https://arxiv.org/abs/1301.2846}{{\ttfamily 1301.2846}}].

\bibitem{Mantziris:2020rzh}
A.~Mantziris, T.~Markkanen and A.~Rajantie, \emph{{Vacuum decay constraints on
  the Higgs curvature coupling from inflation}},
  \href{https://doi.org/10.1088/1475-7516/2021/03/077}{\emph{JCAP} {\bfseries
  03} (2021) 077} [\href{https://arxiv.org/abs/2011.03763}{{\ttfamily
  2011.03763}}].

\bibitem{Isidori:2007vm}
G.~Isidori, V.~S. Rychkov, A.~Strumia and N.~Tetradis, \emph{{Gravitational
  corrections to standard model vacuum decay}},
  \href{https://doi.org/10.1103/PhysRevD.77.025034}{\emph{Phys. Rev. D}
  {\bfseries 77} (2008) 025034}
  [\href{https://arxiv.org/abs/0712.0242}{{\ttfamily 0712.0242}}].

\bibitem{Gialamas:2022gxv}
I.~D. Gialamas, A.~Karam and T.~D. Pappas, \emph{{Gravitational corrections to
  electroweak vacuum decay: metric vs. Palatini}},
  \href{https://doi.org/10.1016/j.physletb.2023.137885}{\emph{Phys. Lett. B}
  {\bfseries 840} (2023) 137885}
  [\href{https://arxiv.org/abs/2212.03052}{{\ttfamily 2212.03052}}].

\bibitem{Buttazzo:2013uya}
D.~Buttazzo, G.~Degrassi, P.~P. Giardino, G.~F. Giudice, F.~Sala, A.~Salvio
  et~al., \emph{{Investigating the near-criticality of the Higgs boson}},
  \href{https://doi.org/10.1007/JHEP12(2013)089}{\emph{JHEP} {\bfseries 12}
  (2013) 089} [\href{https://arxiv.org/abs/1307.3536}{{\ttfamily 1307.3536}}].

\bibitem{Esteban:2020cvm}
I.~Esteban, M.~C. Gonzalez-Garcia, M.~Maltoni, T.~Schwetz and A.~Zhou,
  \emph{{The fate of hints: updated global analysis of three-flavor neutrino
  oscillations}}, \href{https://doi.org/10.1007/JHEP09(2020)178}{\emph{JHEP}
  {\bfseries 09} (2020) 178}
  [\href{https://arxiv.org/abs/2007.14792}{{\ttfamily 2007.14792}}].

\bibitem{Pilaftsis:1991ug}
A.~Pilaftsis, \emph{{Radiatively induced neutrino masses and large Higgs
  neutrino couplings in the standard model with Majorana fields}},
  \href{https://doi.org/10.1007/BF01482590}{\emph{Z. Phys. C} {\bfseries 55}
  (1992) 275} [\href{https://arxiv.org/abs/hep-ph/9901206}{{\ttfamily
  hep-ph/9901206}}].

\bibitem{Gluza:2002vs}
J.~Gluza, \emph{{On teraelectronvolt Majorana neutrinos}}, {\emph{Acta Phys.
  Polon. B} {\bfseries 33} (2002) 1735}
  [\href{https://arxiv.org/abs/hep-ph/0201002}{{\ttfamily hep-ph/0201002}}].

\bibitem{Xing:2009in}
Z.-z. Xing, \emph{{Naturalness and Testability of TeV Seesaw Mechanisms}},
  \href{https://doi.org/10.1143/PTPS.180.112}{\emph{Prog. Theor. Phys. Suppl.}
  {\bfseries 180} (2009) 112}
  [\href{https://arxiv.org/abs/0905.3903}{{\ttfamily 0905.3903}}].

\bibitem{He:2009ua}
X.-G. He, S.~Oh, J.~Tandean and C.-C. Wen, \emph{{Large Mixing of Light and
  Heavy Neutrinos in Seesaw Models and the LHC}},
  \href{https://doi.org/10.1103/PhysRevD.80.073012}{\emph{Phys. Rev. D}
  {\bfseries 80} (2009) 073012}
  [\href{https://arxiv.org/abs/0907.1607}{{\ttfamily 0907.1607}}].

\bibitem{Ibarra:2010xw}
A.~Ibarra, E.~Molinaro and S.~T. Petcov, \emph{{TeV Scale See-Saw Mechanisms of
  Neutrino Mass Generation, the Majorana Nature of the Heavy Singlet Neutrinos
  and $(\beta\beta)_{0\nu}$-Decay}},
  \href{https://doi.org/10.1007/JHEP09(2010)108}{\emph{JHEP} {\bfseries 09}
  (2010) 108} [\href{https://arxiv.org/abs/1007.2378}{{\ttfamily 1007.2378}}].

\bibitem{Mitra:2011qr}
M.~Mitra, G.~Senjanovic and F.~Vissani, \emph{{Neutrinoless Double Beta Decay
  and Heavy Sterile Neutrinos}},
  \href{https://doi.org/10.1016/j.nuclphysb.2011.10.035}{\emph{Nucl. Phys. B}
  {\bfseries 856} (2012) 26} [\href{https://arxiv.org/abs/1108.0004}{{\ttfamily
  1108.0004}}].

\bibitem{Lee:2013htl}
C.-H. Lee, P.~S. Bhupal~Dev and R.~N. Mohapatra, \emph{{Natural TeV-scale
  left-right seesaw mechanism for neutrinos and experimental tests}},
  \href{https://doi.org/10.1103/PhysRevD.88.093010}{\emph{Phys. Rev. D}
  {\bfseries 88} (2013) 093010}
  [\href{https://arxiv.org/abs/1309.0774}{{\ttfamily 1309.0774}}].

\bibitem{Kersten:2007vk}
J.~Kersten and A.~Y. Smirnov, \emph{{Right-Handed Neutrinos at CERN LHC and the
  Mechanism of Neutrino Mass Generation}},
  \href{https://doi.org/10.1103/PhysRevD.76.073005}{\emph{Phys. Rev. D}
  {\bfseries 76} (2007) 073005}
  [\href{https://arxiv.org/abs/0705.3221}{{\ttfamily 0705.3221}}].

\bibitem{Moffat:2017feq}
K.~Moffat, S.~Pascoli and C.~Weiland, \emph{{Equivalence between massless
  neutrinos and lepton number conservation in fermionic singlet extensions of
  the Standard Model}},  \href{https://arxiv.org/abs/1712.07611}{{\ttfamily
  1712.07611}}.

\bibitem{Shaposhnikov:2006nn}
M.~Shaposhnikov, \emph{{A Possible symmetry of the nuMSM}},
  \href{https://doi.org/10.1016/j.nuclphysb.2006.11.003}{\emph{Nucl. Phys. B}
  {\bfseries 763} (2007) 49}
  [\href{https://arxiv.org/abs/hep-ph/0605047}{{\ttfamily hep-ph/0605047}}].

\bibitem{delAguila:2008pw}
F.~del Aguila, J.~de~Blas and M.~Perez-Victoria, \emph{{Effects of new leptons
  in Electroweak Precision Data}},
  \href{https://doi.org/10.1103/PhysRevD.78.013010}{\emph{Phys. Rev. D}
  {\bfseries 78} (2008) 013010}
  [\href{https://arxiv.org/abs/0803.4008}{{\ttfamily 0803.4008}}].

\bibitem{Akhmedov:2013hec}
E.~Akhmedov, A.~Kartavtsev, M.~Lindner, L.~Michaels and J.~Smirnov,
  \emph{{Improving Electro-Weak Fits with TeV-scale Sterile Neutrinos}},
  \href{https://doi.org/10.1007/JHEP05(2013)081}{\emph{JHEP} {\bfseries 05}
  (2013) 081} [\href{https://arxiv.org/abs/1302.1872}{{\ttfamily 1302.1872}}].

\bibitem{deBlas:2013gla}
J.~de~Blas, \emph{{Electroweak limits on physics beyond the Standard Model}},
  \href{https://doi.org/10.1051/epjconf/20136019008}{\emph{EPJ Web Conf.}
  {\bfseries 60} (2013) 19008}
  [\href{https://arxiv.org/abs/1307.6173}{{\ttfamily 1307.6173}}].

\bibitem{Antusch:2014woa}
S.~Antusch and O.~Fischer, \emph{{Non-unitarity of the leptonic mixing matrix:
  Present bounds and future sensitivities}},
  \href{https://doi.org/10.1007/JHEP10(2014)094}{\emph{JHEP} {\bfseries 10}
  (2014) 094} [\href{https://arxiv.org/abs/1407.6607}{{\ttfamily 1407.6607}}].

\bibitem{Blennow:2016jkn}
M.~Blennow, P.~Coloma, E.~Fernandez-Martinez, J.~Hernandez-Garcia and
  J.~Lopez-Pavon, \emph{{Non-Unitarity, sterile neutrinos, and Non-Standard
  neutrino Interactions}},
  \href{https://doi.org/10.1007/JHEP04(2017)153}{\emph{JHEP} {\bfseries 04}
  (2017) 153} [\href{https://arxiv.org/abs/1609.08637}{{\ttfamily
  1609.08637}}].

\bibitem{Flieger:2019eor}
W.~Flieger, J.~Gluza and K.~Porwit, \emph{{New limits on neutrino non-unitary
  mixings based on prescribed singular values}},
  \href{https://doi.org/10.1007/JHEP03(2020)169}{\emph{JHEP} {\bfseries 03}
  (2020) 169} [\href{https://arxiv.org/abs/1910.01233}{{\ttfamily
  1910.01233}}].

\bibitem{Banerjee:2015gca}
S.~Banerjee, P.~S.~B. Dev, A.~Ibarra, T.~Mandal and M.~Mitra, \emph{{Prospects
  of Heavy Neutrino Searches at Future Lepton Colliders}},
  \href{https://doi.org/10.1103/PhysRevD.92.075002}{\emph{Phys. Rev. D}
  {\bfseries 92} (2015) 075002}
  [\href{https://arxiv.org/abs/1503.05491}{{\ttfamily 1503.05491}}].

\bibitem{Mondal:2016kof}
S.~Mondal and S.~K. Rai, \emph{{Probing the Heavy Neutrinos of Inverse Seesaw
  Model at the LHeC}},
  \href{https://doi.org/10.1103/PhysRevD.94.033008}{\emph{Phys. Rev. D}
  {\bfseries 94} (2016) 033008}
  [\href{https://arxiv.org/abs/1605.04508}{{\ttfamily 1605.04508}}].

\bibitem{Antusch:2016ejd}
S.~Antusch, E.~Cazzato and O.~Fischer, \emph{{Sterile neutrino searches at
  future $e^-e^+$, $pp$, and $e^-p$ colliders}},
  \href{https://doi.org/10.1142/S0217751X17500786}{\emph{Int. J. Mod. Phys. A}
  {\bfseries 32} (2017) 1750078}
  [\href{https://arxiv.org/abs/1612.02728}{{\ttfamily 1612.02728}}].

\bibitem{Pascoli:2018heg}
S.~Pascoli, R.~Ruiz and C.~Weiland, \emph{{Heavy neutrinos with dynamic jet
  vetoes: multilepton searches at $ \sqrt{s}=14 $ , 27, and 100 TeV}},
  \href{https://doi.org/10.1007/JHEP06(2019)049}{\emph{JHEP} {\bfseries 06}
  (2019) 049} [\href{https://arxiv.org/abs/1812.08750}{{\ttfamily
  1812.08750}}].

\bibitem{Hernandez:2018cgc}
P.~Hern\'andez, J.~Jones-P\'erez and O.~Suarez-Navarro, \emph{{Majorana vs
  Pseudo-Dirac Neutrinos at the ILC}},
  \href{https://doi.org/10.1140/epjc/s10052-019-6728-1}{\emph{Eur. Phys. J. C}
  {\bfseries 79} (2019) 220}
  [\href{https://arxiv.org/abs/1810.07210}{{\ttfamily 1810.07210}}].

\bibitem{Das:2018usr}
A.~Das, S.~Jana, S.~Mandal and S.~Nandi, \emph{{Probing right handed neutrinos
  at the LHeC and lepton colliders using fat jet signatures}},
  \href{https://doi.org/10.1103/PhysRevD.99.055030}{\emph{Phys. Rev. D}
  {\bfseries 99} (2019) 055030}
  [\href{https://arxiv.org/abs/1811.04291}{{\ttfamily 1811.04291}}].

\bibitem{FCC:2018vvp}
{\scshape FCC} collaboration, \emph{{FCC-hh: The Hadron Collider}: {Future
  Circular Collider Conceptual Design Report Volume 3}},
  \href{https://doi.org/10.1140/epjst/e2019-900087-0}{\emph{Eur. Phys. J. ST}
  {\bfseries 228} (2019) 755}.

\bibitem{Chakraborty:2018khw}
S.~Chakraborty, M.~Mitra and S.~Shil, \emph{{Fat Jet Signature of a Heavy
  Neutrino at Lepton Collider}},
  \href{https://doi.org/10.1103/PhysRevD.100.015012}{\emph{Phys. Rev. D}
  {\bfseries 100} (2019) 015012}
  [\href{https://arxiv.org/abs/1810.08970}{{\ttfamily 1810.08970}}].

\bibitem{Antusch:2019eiz}
S.~Antusch, O.~Fischer and A.~Hammad, \emph{{Lepton-Trijet and Displaced Vertex
  Searches for Heavy Neutrinos at Future Electron-Proton Colliders}},
  \href{https://doi.org/10.1007/JHEP03(2020)110}{\emph{JHEP} {\bfseries 03}
  (2020) 110} [\href{https://arxiv.org/abs/1908.02852}{{\ttfamily
  1908.02852}}].

\bibitem{Bolton:2019pcu}
P.~D. Bolton, F.~F. Deppisch and P.~S. Bhupal~Dev, \emph{{Neutrinoless double
  beta decay versus other probes of heavy sterile neutrinos}},
  \href{https://doi.org/10.1007/JHEP03(2020)170}{\emph{JHEP} {\bfseries 03}
  (2020) 170} [\href{https://arxiv.org/abs/1912.03058}{{\ttfamily
  1912.03058}}].

\bibitem{Das:2023tna}
A.~Das, S.~Mandal and S.~Shil, \emph{{Testing electroweak scale seesaw models
  at $e^{-} \gamma$ and $\gamma \gamma$ colliders}},
  \href{https://arxiv.org/abs/2304.06298}{{\ttfamily 2304.06298}}.

\bibitem{sterileorg}
``Sterile neutrino constraints.''
  \url{https://www.hep.ucl.ac.uk/~pbolton/index.html}.

\bibitem{Steingasser:2023ugv}
T.~Steingasser and D.~I. Kaiser, \emph{{Higgs Criticality beyond the Standard
  Model}},  \href{https://arxiv.org/abs/2307.10361}{{\ttfamily 2307.10361}}.

\bibitem{Giudice:2021viw}
G.~F. Giudice, M.~McCullough and T.~You, \emph{{Self-organised localisation}},
  \href{https://doi.org/10.1007/JHEP10(2021)093}{\emph{JHEP} {\bfseries 10}
  (2021) 093} [\href{https://arxiv.org/abs/2105.08617}{{\ttfamily
  2105.08617}}].

\bibitem{Denef:2017cxt}
F.~Denef, M.~R. Douglas, B.~Greene and C.~Zukowski, \emph{{Computational
  complexity of the landscape II\textemdash{}Cosmological considerations}},
  \href{https://doi.org/10.1016/j.aop.2018.03.013}{\emph{Annals Phys.}
  {\bfseries 392} (2018) 93}
  [\href{https://arxiv.org/abs/1706.06430}{{\ttfamily 1706.06430}}].

\bibitem{Khoury:2019yoo}
J.~Khoury and O.~Parrikar, \emph{{Search Optimization, Funnel Topography, and
  Dynamical Criticality on the String Landscape}},
  \href{https://doi.org/10.1088/1475-7516/2019/12/014}{\emph{JCAP} {\bfseries
  12} (2019) 014} [\href{https://arxiv.org/abs/1907.07693}{{\ttfamily
  1907.07693}}].

\bibitem{Khoury:2019ajl}
J.~Khoury, \emph{{Accessibility Measure for Eternal Inflation: Dynamical
  Criticality and Higgs Metastability}},
  \href{https://doi.org/10.1088/1475-7516/2021/06/009}{\emph{JCAP} {\bfseries
  06} (2021) 009} [\href{https://arxiv.org/abs/1912.06706}{{\ttfamily
  1912.06706}}].

\bibitem{Kartvelishvili:2020thd}
G.~Kartvelishvili, J.~Khoury and A.~Sharma, \emph{{The Self-Organized Critical
  Multiverse}},
  \href{https://doi.org/10.1088/1475-7516/2021/02/028}{\emph{JCAP} {\bfseries
  02} (2021) 028} [\href{https://arxiv.org/abs/2003.12594}{{\ttfamily
  2003.12594}}].

\bibitem{Khoury:2021grg}
J.~Khoury and S.~S.~C. Wong, \emph{{Early-time measure in eternal inflation}},
  \href{https://doi.org/10.1088/1475-7516/2022/05/031}{\emph{JCAP} {\bfseries
  05} (2022) 031} [\href{https://arxiv.org/abs/2106.12590}{{\ttfamily
  2106.12590}}].

\bibitem{Belfatto:2023tbv}
B.~Belfatto and S.~Trifinopoulos, \emph{{The remarkable role of the vector-like
  quark doublet in the Cabibbo angle and $W$-mass anomalies}},
  \href{https://arxiv.org/abs/2302.14097}{{\ttfamily 2302.14097}}.

\bibitem{Deppisch:2013jxa}
F.~F. Deppisch, J.~Harz and M.~Hirsch, \emph{{Falsifying High-Scale
  Leptogenesis at the LHC}},
  \href{https://doi.org/10.1103/PhysRevLett.112.221601}{\emph{Phys. Rev. Lett.}
  {\bfseries 112} (2014) 221601}
  [\href{https://arxiv.org/abs/1312.4447}{{\ttfamily 1312.4447}}].

\bibitem{Moffat:2018wke}
K.~Moffat, S.~Pascoli, S.~T. Petcov, H.~Schulz and J.~Turner,
  \emph{{Three-flavored nonresonant leptogenesis at intermediate scales}},
  \href{https://doi.org/10.1103/PhysRevD.98.015036}{\emph{Phys. Rev. D}
  {\bfseries 98} (2018) 015036}
  [\href{https://arxiv.org/abs/1804.05066}{{\ttfamily 1804.05066}}].

\bibitem{Davidson:2002qv}
S.~Davidson and A.~Ibarra, \emph{{A Lower bound on the right-handed neutrino
  mass from leptogenesis}},
  \href{https://doi.org/10.1016/S0370-2693(02)01735-5}{\emph{Phys. Lett. B}
  {\bfseries 535} (2002) 25}
  [\href{https://arxiv.org/abs/hep-ph/0202239}{{\ttfamily hep-ph/0202239}}].

\bibitem{Buchmuller:2002rq}
W.~Buchmuller, P.~Di~Bari and M.~Plumacher, \emph{{Cosmic microwave background,
  matter - antimatter asymmetry and neutrino masses}},
  \href{https://doi.org/10.1016/S0550-3213(02)00737-X}{\emph{Nucl. Phys. B}
  {\bfseries 643} (2002) 367}
  [\href{https://arxiv.org/abs/hep-ph/0205349}{{\ttfamily hep-ph/0205349}}].

\bibitem{Buchmuller:2004nz}
W.~Buchmuller, P.~Di~Bari and M.~Plumacher, \emph{{Leptogenesis for
  pedestrians}}, \href{https://doi.org/10.1016/j.aop.2004.02.003}{\emph{Annals
  Phys.} {\bfseries 315} (2005) 305}
  [\href{https://arxiv.org/abs/hep-ph/0401240}{{\ttfamily hep-ph/0401240}}].

\bibitem{Antusch:2003kp}
S.~Antusch, J.~Kersten, M.~Lindner and M.~Ratz, \emph{{Running neutrino masses,
  mixings and CP phases: Analytical results and phenomenological
  consequences}},
  \href{https://doi.org/10.1016/j.nuclphysb.2003.09.050}{\emph{Nucl. Phys. B}
  {\bfseries 674} (2003) 401}
  [\href{https://arxiv.org/abs/hep-ph/0305273}{{\ttfamily hep-ph/0305273}}].

\bibitem{Casas:1999cd}
J.~A. Casas, V.~Di~Clemente, A.~Ibarra and M.~Quiros, \emph{{Massive neutrinos
  and the Higgs mass window}},
  \href{https://doi.org/10.1103/PhysRevD.62.053005}{\emph{Phys. Rev. D}
  {\bfseries 62} (2000) 053005}
  [\href{https://arxiv.org/abs/hep-ph/9904295}{{\ttfamily hep-ph/9904295}}].

\bibitem{Elias-Miro:2011sqh}
J.~Elias-Miro, J.~R. Espinosa, G.~F. Giudice, G.~Isidori, A.~Riotto and
  A.~Strumia, \emph{{Higgs mass implications on the stability of the
  electroweak vacuum}},
  \href{https://doi.org/10.1016/j.physletb.2012.02.013}{\emph{Phys. Lett. B}
  {\bfseries 709} (2012) 222}
  [\href{https://arxiv.org/abs/1112.3022}{{\ttfamily 1112.3022}}].

\bibitem{Drewes:2021nqr}
M.~Drewes, Y.~Georis and J.~Klari\'c, \emph{{Mapping the Viable Parameter Space
  for Testable Leptogenesis}},
  \href{https://doi.org/10.1103/PhysRevLett.128.051801}{\emph{Phys. Rev. Lett.}
  {\bfseries 128} (2022) 051801}
  [\href{https://arxiv.org/abs/2106.16226}{{\ttfamily 2106.16226}}].

\bibitem{Brdar:2019iem}
V.~Brdar, A.~J. Helmboldt, S.~Iwamoto and K.~Schmitz, \emph{{Type-I Seesaw as
  the Common Origin of Neutrino Mass, Baryon Asymmetry, and the Electroweak
  Scale}}, \href{https://doi.org/10.1103/PhysRevD.100.075029}{\emph{Phys. Rev.
  D} {\bfseries 100} (2019) 075029}
  [\href{https://arxiv.org/abs/1905.12634}{{\ttfamily 1905.12634}}].

\end{thebibliography}\endgroup

\end{document}